\documentclass{amsart}
\usepackage{amssymb,latexsym}

\begin{document}

\newtheorem{conjecture}{Conjecture}
\newtheorem{definition}{Definition}
\newtheorem{result}{Result}

\title{A $U_q\bigl(\widehat{gl}(2\vert2)\bigr)_1$-vertex model:\\
Creation algebras and quasi-particles I}
\author{R.M. Gade}
\address{Lerchenfeldstra\ss e 12, 80538 Munich, Germany}
\email{renate.gade@t-online.de}
\thanks{This research project is supported by M. Krautg\"artner.}
\begin{abstract}
The infinite configuration space of an integrable vertex model based on
$U_q\bigl(\widehat{gl}(2\vert2)\bigr)_1$ is studied at $q=0$.
Allowing four particular boundary conditions, the infinite configurations are
mapped onto the semi-standard supertableaux of pairs of infinite border
strips. By means of this map, a weight-preserving one-to-one correspondence 
between the infinite configurations and the normal forms of a pair of creation
algebras is established for one boundary condition. A pair of type-II vertex operators
associated with an infinite-dimensional $U_q\bigl(gl(2\vert2)\bigr)$-module
$\mathring V$ and its dual $\mathring V^*$ is introduced. Their existence is
conjectured relying on a free boson realization. The realization allows to
determine the commutation relation satisfied by two vertex operators related to
the same $U_q\bigl(gl(2\vert2)\bigr)$-module. Explicit expressions are provided for
the relevant R-matrix elements. The formal $q\to0$ limit of
these commutation relations leads to the defining relations of the creation
algebras. Based on these findings it is conjectured that the type II vertex
operators associated with
$\mathring V$ and $\mathring V^*$ give rise to part of the eigenstates
of the row-to-row transfer matrix of the model. 
A partial discussion of the R-matrix elements introduced on $\mathring V\otimes
\mathring V^*$ is given.
\end{abstract}
\maketitle

\section{Introduction}

This paper continues the study of the $U_q\bigl(\widehat{gl}(2\vert2)
\bigr)$-vertex model started in \cite{gade1}. There an integrable vertex
model based on the vector representation of $U_q\bigl(gl(2\vert2)\bigr)$ and
its dual is investigated. In the limit $q\to0$, the action of
the corner transfer matrix Hamiltonian on the space of half-infinite
configurations takes a trigonal form provided that the configurations
obey a particular boundary condition. A one-to-one correspondence between
the half-infinite configurations and the weight states of a reducible
level-one module of $U_q\bigl(\widehat{sl}(2\vert2)\bigr)/\mathcal{H}$
with grade $-n$ is observed for $n\leq3$. Here the grade corresponds to
the diagonal element of the corner transfer matrix Hamiltonian. The
reducible module decomposes into one weakly integrable irreducible module
and one nonintegrable irreducible module of
$U_q\bigl(\widehat{sl}(2\vert2)\bigr)/\mathcal{H}$. Both are highest
weight modules.

Here the investigation of half-infinite configurations is supplemented
taking into account a second boundary condition. Various
choices of composite vertices and $U_q\bigl(gl(2\vert2)\bigr)$-weights
are considered. At $q=0$, a one-to-one correspondence
between the half-infinite
configurations and the weight states of reducible or irreducible level-one modules of
$U_q\bigl(\widehat{sl}(2\vert2)\bigr)/\mathcal{H}$ is found at the grades
$0,-1,-2$. This correspondence is assumed to hold true at all grades.
Similar as the modules relevant to the first boundary condition,
the reducible modules decompose into a weakly integrable and a nonintegrable highest-weight
module. These irreducible modules may be assembled in a reducible module
in various ways. In each case, one possibility is distinguished by a simple
free boson realization within the scheme employed in \cite{gade1}.

The arguments developed for the algebraic analysis of integrable
$U_q\bigl(\widehat{gl}(N)\bigr)$-vertex models (\cite{jm},\cite{dav},\cite{idz})
suggest that the full space of states is interpreted as 
a sum of tensor products each combining a highest weight module
with a suitable dual (and therefore lowest weight) module. The levels of both
constituents of the tensor product add to zero. According to
the method presented in these references, the row-to-row transfer matrix
should be described in terms of type I vertex operators associated with the
vector representation $W$ of $U_q\bigl(gl(2\vert2)\bigr)$
 and its dual $W^*$. Type II vertex operators are expected to create the
eigenstates of the row-to-row transfer matrix. In general, only a subset of all
type II vertex operators existing for the affine quantum algebra at fixed
level gives rise to such eigenstates. The correct choice of vertex operators
follows from the decomposition of the tensor products modelling
the space of states into irreducible components. For the 
$U_q\bigl(\widehat{gl}(N)\bigr)$-vertex models this decomposition can be
done at $q=0$ \cite{dav},\cite{naka}
making use of the crystal base theory \cite{aff}, \cite{kash}.
The path space of the model is described in terms of a creation algebra
whose generators can be seen as formal $q\to0$ limits of the appropriate
type II vertex operators. This picture is based on the existence of the
$q\to0$ limits of the creation operators acting on the true groundstate of the
model. The existence of these limits has been conjectured and checked
for the $XXZ$-model in \cite{dav}.

In this study, a similar description is proposed for an infinite configuration space
of the $U_q\bigl(\widehat{gl}(2\vert2)\bigr)$-model at $q=0$. The
second boundary condition is imposed in both directions. In the limit
$q\to0$, the diagonal elements of the corner transfer matrix Hamiltonian
decouple into a contribution depending only on the $W$-part of the
half-infinite configuration and into a second part depending only on the
$W^*$-part. For the boundary condition chosen here, the $W$-components and
the $W^*$-components of the infinite configurations can be described separately
in terms of two creation algebras $\mathcal{A}$ and $\mathcal{A}^*$. 
According to their defining relations, bases of both algebras are
given by their sets of normal forms. This has been proven in \cite{naka} for a quite
general type of creation algebras including $\mathcal{A}$ and
$\mathcal{A}^*$. 
The infinite $W$-components ($W^*$-components) are in one-to-one correspondence
with the normal
forms of $\mathcal{A}^*$ ($\mathcal{A}$) supplemented by the unity.

To demonstrate these correspondences, two types of
infinite border strips are employed. Border strips consisting of a finite number
of rows or columns of finite length framed by either two infinite
rows or two infinite columns are termed horizontal or vertical border strips,
respectively.
The semi-standard supertableaux of the horizontal (or vertical) border strips 
satisfying a suitable boundary condition are mapped onto the infinite
$W$-(or $W^*$-)components and vice versa. Then a weight-preserving one-to-one
map between the semi-standard supertableaux of the horizontal (vertical)
border strips and the set of normal forms of $\mathcal{A}^*$ ($\mathcal{A}$)
enlarged by the unity is given explicitely.
A similar description can be constructed for the infinite configuration space
restricted by the first boundary condition in both directions. If different
boundary conditions are imposed in the left and right direction, the simple
separation into the $W$-and $W^*$-components is lost. These cases will be
considered in a separate publication.

The defining relations of $\mathcal{A}$ and $\mathcal{A}^*$ are expected to
emerge from the
formal $q\to0$ limits of the commutation relations satisfied by particular type II
vertex operators. Suitable vertex operators should be associated with a reducible,
infinite-dimensional
highest weight module $\mathring{V}$ in the case of $\mathcal{A}$ and
with its dual module $\mathring{V}^*$ in the case of $\mathcal{A}^*$.
The module $\mathring V$ decomposes into a one-dimensional module and an
infinite-dimensional irreducible module. The latter coincides with the module $V$
introduced in \cite{gade1}. Based on a free boson realization for one
component, the existence of such vertex operators is conjectured.
The commutation relations of two vertex operators are governed by the R-matrices
related to $\mathring V$ and $\mathring V^*$. The free boson realizations allow
to determine the normalisation of the R-matrices defined on $\mathring V\otimes
\mathring V$ or $\mathring V^*\otimes \mathring V^*$. Evaluation of their 
$q\to0$ limits formally leads to the defining relations of $\mathcal A$ or
$\mathcal A^*$. It seems more difficult however to demonstrate the compatibility
of the commutation relations involving both types of vertex operators with the
separation into $W$- and $W^*$-components found at $q=0$. 
In the mixed case it is less obvious how to make use of
$q\to0$ limits of single R-matrix elements. The $q\to0$ limits of the R-matrix
elements on $\mathring V\otimes\mathring V$ or $\mathring V^*\otimes\mathring V^*$
reflect the structure of the irreducible components of these tensor products.
Hence this structure underlies the defining relations of $\mathcal A$ or
$\mathcal A^*$. This motivates
the search for an R-matrix action on
the irreducible components of the tensor products $\mathring V^*\otimes \mathring V$
or $\mathring V\otimes \mathring V^*$. This action is 
obtained on $\mathring V^*\otimes \mathring V$ only for a partial range of the
spectral parameter. Its formal $q\to0$ limit does not distinguish between
different irreducible components.

From the above findings it may be conjectured that a part of the
eigenstates of the row-to-row transfer matrix is generated by the
type II vertex operators associated with $\mathring{V}$ and $\mathring{V}^*$.
A full discussion of the mixed R-matrix elements is more conveniently given in context
with the R-matrix elements encountered in the case of mixed boundary conditions and
is therefore relegated to a forthcoming publication.

The main findings presented here
are formulated by conjecture \ref{C:c1} in section \ref{S:bound2},
conjectures \ref{C:c2} and \ref{C:c3}
in section \ref{S:vo} and by the results \ref{RR:r1}-\ref{RR:r3}
in section \ref{S:crealg}.

The paper is organised as follows. For the convenience of the reader,
subsection \ref{S:def} collects 
notations and recalls the free boson realization of $U_q\bigl(\widehat{gl}(2\vert2)
\bigr)$ at level-one used in \cite{gade1}. Subsection \ref{S:mod} gives
a short account of the vertex model and the structure of the half-infinite
configuration space subject to the boundary condition chosen in
\cite{gade1}. The second boundary condition is considered 
in section \ref{S:bound2}. Since the analysis is quite analogous to the one
presented in \cite{gade1}, the outline is kept short here.
In section \ref{S:bs}, the infinite configuration
space is mapped onto the semi-standard supertableaux. For one boundary
condition, a one-to-one correspondence between the semi-standard supertableaux
and the normal forms of two creation algebras is specified
in section \ref{S:crealg}. Section \ref{S:vo} deals with the
type II vertex operators related to the creation algebras. In section \ref{S:rmixed},
the mixed case is investigated.
Appendices \ref{A:map} and \ref{A:rmode} contain some details relegated from sections
\ref{S:crealg} and \ref{S:rmixed}.
Some properties of the R-matrix associated with the infinite-dimensional module
and the list of explicit expressions for the R-matrix elements are given in appendix
\ref{A:rmatrix}.

\section{The model}
\subsection{The quantum affine superalgebra $U_q\bigl(\widehat{gl}(2\vert2)\bigr)$}
\label{S:def}

The integrable vertex model investigated in \cite{gade1} is based on the quantum
affine superalgebra $U_q\bigl(\widehat{gl}(2\vert2)\bigr)$. Defining relations
in terms of the Chevalley or Drinfeld basis as well as references regarding the
representation theory of the algebra are given in \cite{gade1}.
$U_q\bigl(\widehat{gl}(2\vert2)\bigr)$ is an associative ${\mathbb{Z}}_2$-graded
algebra over ${\mathbb{C}}[[q-1]]$ with generators $E^{k,\pm}_n,\,k=1,2,3,\;n\in
{\mathbb{Z}}$ and $\Psi^{l,\pm}_{\pm n},\,l=1,2,3,4,\;n\in{\mathbb{Z}}_+$, 
the central element $c$ and the grading operator $d$. All simple roots are chosen odd.
A ${\mathbb{Z}}_2$ grading $\vert\cdot\vert$
is defined by $\vert
E^{k,\pm}_n\vert=1$ $\forall k,n$  and $\vert q^c\vert=\vert d\vert=
\vert\Psi^{l,\pm}_{\pm n}\vert=0$ $\forall l,n$.
The defining relations of 
$U_q\bigl(\widehat{gl}(2\vert2)\bigr)$
in terms of the Drinfeld basis can be found in \cite{gade1}.
$U_q\bigl(\widehat{sl}(2\vert2)\bigr)$
is the subalgebra with generators $E^{k,\pm}_n,\,n\in
{\mathbb{Z}}$ and $\Psi^{k,\pm}_{\pm n},\,n\in{\mathbb{Z}}_+$ with $k=1,2,3$, the
central element $q^c$ and the grading operator $d$.
$U'_q\bigl(\widehat{gl}(2\vert2)\bigr)$ and $U'_q\bigl(\widehat{sl}(2\vert2)\bigr)$
denote the superalgebras obtained by discarding the grading operator $d$ from
the set of generators of $U_q\bigl(\widehat{gl}(2\vert2)\bigr)$
or $U_q\bigl(\widehat{sl}(2\vert2)\bigr)$, respectively.

On any $U_q\bigl(\widehat{gl}
(2\vert2)\bigr)$-module considered here, $c$ acts as a scalar taking the value
$0$ or $1$. This value is referred to as the level of the module. Parts of the subsequent
analysis involve a free boson realisation of $U_q\bigl(\widehat{gl}(2\vert2)\bigr)$
applying to the level-one case. The generating functions
\begin{equation}\label{E:genf}
E^{k,\pm}(z)=\sum_{n\in{\mathbb{Z}}}E^{k,\pm}_nz^{-n-1}\qquad\qquad
\Psi^{l,\pm}(z)=\sum_{n\geq0}\Psi^{l,\pm}_{\pm n}z^{-n}
\end{equation}
and the grading operator $d$ can be expressed in terms of six sets $\{\varphi^l,\,
\varphi^l_0,\,\varphi^l_n,\;l=1,2,3,4;\,n\in{\mathbb{Z}}\}$ and $\{\beta^{\bar l},\,
\beta^{\bar l}_0,\,\beta^{\bar l}_n,\;\bar l=1,2;\,n\in{\mathbb{Z}}\}$ of bosonic
oscillators. These oscillators satisfy the commutation relations
\begin{equation}\label{E:osc1}
\begin{split}
\bigl[\varphi^l_n,\varphi^{l'}_m\bigr]&=\delta_{l,l'}\delta_{n+m,0}\frac{[n]^2}{n}
\qquad n,m\neq0\\
\bigl[\varphi^l,\varphi^{l'}_0\bigr]&=i\delta_{l,l'}
\end{split}
\end{equation}
and
\begin{equation}\label{E:osc2}
\begin{split}
\bigl[\beta^{\bar l}_n,\beta^{\bar l'}_m\bigr]&=-n\,\delta_{\bar l,\bar l'}
\delta_{n+m,0}\qquad n,m\neq0\\
\bigl[\beta^{\bar l},\beta^{\bar l'}_0\bigr]&=-i\delta_{\bar l,\bar l'}
\end{split}
\end{equation}
where $[n]\equiv\frac{q^n-q^{-n}}{q-q^{-1}}$.
The currents $\Psi^{l,\pm}(z)$ are realized by
\begin{equation}\label{E:psib1}
\Psi^{4,\pm}(z)=q^{\pm(\varphi^1_0-i\varphi^4_0)}\exp\Bigl(\pm(q-q^{-1})\sum_{n>0}
\bigl(\varphi^1_{\pm n}-i\varphi^4_{\pm n}\bigr)z^{\mp n}\Bigr)
\end{equation}
and
\begin{equation}\label{E:psib2}
\Psi^{l,\pm}(z)=q^{\mp i^l(\varphi^{l+1}_0+i\varphi^l_0)}\exp\Bigl(\mp(q-q^{-1})
\sum_{n>0}i^l\bigl(\varphi^{l+1}_{\pm n}+i\varphi^l_{\pm n}\bigr)z^{\mp n}\Bigr)
\end{equation}
for $l=1,2,3$. To express the remaining generating functions, it is convenient
to introduce the deformed free fields
\begin{equation}\label{E:freef1}
\varphi^{l,\pm}(z)=\varphi^l-i\varphi^l_0\ln z+i\sum_{n\neq0}\frac{q^{\mp
\frac{1}{2}\vert n\vert}}{[n]}\varphi^l_nz^{-n}
\end{equation}
and
\begin{equation}\label{E:freef2}
\beta^{\bar l}(z)=\beta^{\bar l}-i\beta^{\bar l}_0\ln z+i\sum_{n\neq0}\frac{1}{n}
\beta^{\bar l}_nz^{-n}
\end{equation}
Then the generating functions $E^{k\pm}(z)$ can be written
\begin{equation}\label{E:Ebos}
E^{k,\pm}(z)=:\exp\Bigl(\mp i^{k+1}\bigl(\varphi^{k+1,\pm}(z)+i\varphi^{k,\pm}
(z)\bigr)\Bigr):\exp\bigl(\pm i\pi\delta_{k,1}\varphi^3_0\bigr)X^{k,\pm}(z)
\end{equation}
with
\begin{equation}\label{E:Xbos}
\begin{split}
-X^{1,-}(z)=X^{2,+}(z)&=\frac{1}{z(q-q^{-1})}\Bigl(:\exp\bigl(\beta^1(q^{-1}z)\bigr)
:-:\exp\bigl(\beta^1(qz)\bigr):\Bigr)\\
X^{1,+}(z)=X^{2,-}(z)&=:\exp\bigl(-\beta^1(z)\bigr):\\
X^{3,+}(z)&=\frac{1}{z(q-q^{-1})}\Bigl(:\exp\bigl(\beta^2(q^{-1}z)\bigr):-:\exp
\bigl(\beta^2(qz)\bigr):\Bigr)\\
X^{3,-}(z)&=:\exp\bigl(-\beta^2(z)\bigr):
\end{split}
\end{equation}
The grading operator $d$ is characterised by the properties
\begin{equation}\label{E:dprop}
w^{-d}E^{k,\pm}(z)w^d=w\,E^{k,\pm}(wz)\qquad\qquad w^{-d}\Psi^{l,\pm}(z)w^d
=\Psi^{l,\pm}(wz)
\end{equation}
With \eqref{E:psib1}, \eqref{E:psib2} and \eqref{E:Ebos}, this implies
\begin{equation}\label{E:dbos}
d=-\frac{1}{2}\sum_{l=1}^4\bigl(\varphi^l_0\bigr)^2-\sum_{l=1}^4\sum_{n>0}
\frac{n^2}{[n]^2}\varphi^l_{-n}\varphi^l_n+\frac{1}{2}\sum_{\bar l=1,2}
\beta^{\bar l}_0(\beta^{\bar l}_0-i)
+\sum_{\bar l=1,2}\sum_{n>0}\beta^{\bar l}_{-n}\beta^{\bar l}_n
\end{equation}
Expressions \eqref{E:psib1},  \eqref{E:psib2}, \eqref{E:Ebos} and \eqref{E:dbos}
satisfy the defining relations of the Drinfeld
basis of $U_q\bigl(\widehat{gl}(2\vert2)\bigr)$ with $c$ replaced by the scalar $1$.
Later analysis of Fock spaces associated to the bosonic oscillators requires
four further fields $\eta^{\bar l}(z)$ and $\xi^{\bar l}(z)$ introduced by
\begin{equation}\label{E:etaxi}
\eta^{\bar l}(z)=\sum_{n\in\mathbb{Z}}\eta^{\bar l}_nz^{-n-1}=:e^{-\beta^{\bar l}
(z)}:\qquad\;\;
\xi^{\bar l}(z)=\sum_{n\in\mathbb{Z}}\xi^{\bar l}_nz^{-n}=:e^{\beta^{\bar l}(z)}:
\end{equation}
with $\bar l=1,2$.
The relations \eqref{E:osc2} imply $\xi^{\bar l}_n\eta^{\bar l'}_m+\eta^{\bar l'}_m
\xi^{\bar l}_n=\delta_{\bar l,\bar l'}\delta_{n+m,0}$
and
$\xi^{\bar l}_n\xi^{\bar l'}_m+\xi^{\bar l'}_m\xi^{\bar l}_n=
\eta^{\bar l}_n\eta^{\bar l'}_m+\eta^{\bar l'}_m\eta^{\bar l}_n=0$.
Both $\eta^1_0$ and $\eta^2_0$ commute
with all generators \eqref{E:psib1}, \eqref{E:psib2} and \eqref{E:Ebos}.
In section  \ref{S:vo}, the free boson realisation of the superalgebra will be employed
to establish the commutation relations of various type-II vertex operators.

For various purposes, it is convenient to write $\Psi^{l,\pm}(z)$ in terms
the generators $h_l$ and $H^l_n$ with $l=1,2,3,4$ and $n\in\mathbb{Z}\backslash0$:
\begin{equation}\label{E:h}
\Psi^{l,\pm}(z)=q^{\pm h_l}\exp\Bigl(\pm(q-q^{-1})\sum_{n>0}H^l_{\pm n}z^{\mp n}
\Bigr)\qquad l=1,2,3,4
\end{equation}
Equations \eqref{E:psib1} and \eqref{E:psib2} yield
\begin{alignat}{2}\label{E:hbos}
h_4&=\varphi^1_0-i\varphi^4_0\qquad\;\;&h_l&=-i^{l}\bigl(\varphi^{l+1}_0+i\varphi_0^l
\bigr)\\
H^4_n&=\varphi^1_n-i\varphi^4_n\qquad\;\;&H^l_n&=-i^l\bigl(\varphi^{l+1}_n
+i\varphi^l_n\bigr)
\end{alignat}
with $l=1,2,3,\,n\neq0$.
The quantum superalgebras generated by $E^{k,\pm}_0$ and $q^{\pm h_l}$ with
$k,l=1,2,3$ or $k=1,2,3$, $l=1,2,3,4$ are denoted by $U_q\bigl(sl(2\vert2)\bigr)$
or $U_q\bigl(gl(2\vert2)\bigr)$, respectively.
The  generators
$h_1+h_3$ and $H^1_n+H^3_n$ with
$n\neq0$ constitute the commutative algebra $\mathcal{H}$. All
generators of $U'_q\bigl(\widehat{sl}(2\vert2)\bigr)$ commute with $\mathcal{H}$.

In terms of the basis $\{\tau_l\}_{1\leq l\leq 4}$ with the bilinear form
$(\tau_l,\tau_{l'})=(-1)^{l+1}\delta_{l,l'}$, the classical simple roots $\bar{\alpha}_l$
are written $\bar{\alpha}_l=(-1)^{l+1}(\tau_l+\tau_{l+1})$ for $l=1,2,3$ and
$\bar{\alpha}_4=\tau_1-\tau_4$. The classical weights $\bar{\Lambda}_l$ with
$l=1,2,3,4$ are given by $\bar{\Lambda}_l=\sum_{l'=1}^l\tau _{l'}-\frac{1}{2}\sum_{l'=
1}^4\tau_l'$. In addition, an affine root $\delta$ and the corresponding
affine weight $\Lambda_0$ with the properties
$(\Lambda_0,\Lambda_0)=(\delta,\delta)=(\tau_l,\Lambda_0)=(\tau_l,\delta)=0$
and $(\Lambda_0,\delta)=1$ are introduced.
Then the set of simple roots is expressed by
$\alpha_0=\delta-\bar{\alpha}_1-\bar{\alpha}_2-\bar{\alpha}_3$ and $\alpha_l=
\bar{\alpha}_l$ for $l=1,2,3,4$. The weight lattice is the free Abelian group
$P=\oplus_{l=0}^4\mathbb{Z}\Lambda_l+\mathbb{Z}\delta$ with $\Lambda_l=\bar{\Lambda_l}
+\Lambda_0$ for $l=1,2,3$ and $\Lambda_4=\bar{\Lambda}_4$.
$P$ and its dual lattice $P^*=\oplus_{l=0}^4\mathbb{Z}h_l+\mathbb{Z}
d$ with $h_0=c-h_1-h_2-h_3$  can be identified via the bilinear form $(\,,)$ by setting
$\alpha_l=h_l$ and $d=\Lambda_0$.

\subsection{The vertex model}
\label{S:mod}

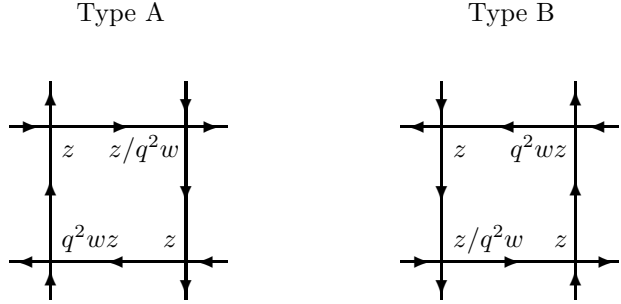
\begin{figure}
\begin{center}
\setlength{\unitlength}{1cm}
\begin{picture}(10,6)
\thicklines
\put(0.8,1.4){\line(1,0){0.1}}
\put(2.1,1.4){\vector(-1,0){1.2}}
\put(3.3,1.4){\vector(-1,0){1.2}}
\put(3.7,1.4){\vector(-1,0){0.4}}
\put(0.8,3.2){\vector(1,0){0.4}}
\put(1.2,3.2){\vector(1,0){1.2}}
\put(2.4,3.2){\vector(1,0){1.2}}
\put(3.6,3.2){\line(1,0){0.1}}
\put(1.35,0.9){\vector(0,1){0.4}}
\put(1.35,1.3){\vector(0,1){1.2}}
\put(1.35,2.5){\vector(0,1){1.2}}
\put(1.35,3.7){\line(0,1){0.1}}
\put(3.15,3.8){\vector(0,-1){0.45}}
\put(3.15,3.4){\vector(0,-1){1.2}}
\put(3.15,2.2){\vector(0,-1){1.25}}
\put(3.15,1){\line(0,-1){0.1}}
\put(1.5,1.6){$q^2wz$}
%\put(2.22,2.8){$w^{-1}z$}
\put(2.12,2.8){$z/q^2w$}
\put(1.5,2.8){$z$}
\put(2.85,1.6){$z$}
\put(1.7,4.6){Type A}

\put(6,1.4){\vector(1,0){0.4}}
\put(6.4,1.4){\vector(1,0){1.2}}
\put(7.6,1.4){\vector(1,0){1.25}}
\put(8.8,1.4){\line(1,0){0.1}}
\put(8.9,3.2){\vector(-1,0){0.4}}
\put(8.5,3.2){\vector(-1,0){1.2}}
\put(7.3,3.2){\vector(-1,0){1.2}}
\put(6.1,3.2){\line(-1,0){0.1}}
\put(6.55,3.8){\vector(0,-1){0.45}}
\put(6.55,3.4){\vector(0,-1){1.2}}
\put(6.55,2.2){\vector(0,-1){1.25}}
\put(6.55,1){\line(0,-1){0.1}}
\put(8.33,0.9){\vector(0,1){0.4}}
\put(8.33,1.3){\vector(0,1){1.2}}
\put(8.33,2.5){\vector(0,1){1.2}}
\put(8.33,3.7){\line(0,1){0.1}}
\put(6.7,1.6){$z/q^2w$}
\put(7.46,2.8){$q^2wz$}
\put(6.7,2.8){$z$}
\put(8.05,1.6){$z$}
\put(6.9,4.6){Type B}
\end{picture}\par

\caption{\textbf{Composite vertices and spectral parameters}\newline
Arrows pointing right or upwards (left or downwards) indicate lattice links related
to the module $W$ ($W^*$).}
\label{FF:spec}
\end{center}
\end{figure}

The vertex model is constructed from the four-dimensional $U'_q\bigl(\widehat{sl}(2
\vert2)\bigr)$-module $W$ with basis $\{w_k\}_{0\leq k\leq3}$ and the dual module
$W^*$ with basis $\{w^*_k\}_{0\leq k\leq3}$. Their $U'_q\bigl(\widehat{sl}(2
\vert2)\bigr)$-structures are given by
\begin{alignat}{2}\label{E:wmod2}
h_jw_k&=(-1)^{j+1}(\delta_{j,k+1}+\delta_{j,k})w_k&&\\
\Psi^{j,\pm}_{\pm n}w_k&=\pm(-1)^{j+1}(q-q^{-1})q^{\pm n(2-\delta_{j,2})}
\bigl(\delta_{j,k+1}+\delta_{j,k}\bigr)w_k\qquad &n>0&
\notag\\
h_jw^*_k&=(-1)^{j}(\delta_{j,k+1}+\delta_{j,k})w^*_k&\notag\\
\Psi^{j,\pm}_{\pm n}w^*_k&=\pm(-1)^{j}(q-q^{-1})q^{\pm n\delta_{j,2}}
\bigl(\delta_{j,k+1}+\delta_{j,k}\bigr)w^*_k\qquad& n>0\notag
\end{alignat}
with $j=1,2,3$ and
\begin{equation}\label{E:wmod1}
\begin{split}
E^{j,+}_nw_j&=(-1)^{j+1}q^{n(2-\delta_{j,2})}w_{j-1}\\
E^{j,-}_nw_{j-1}&=q^{n(2-\delta_{j,2})}w_j\\
E^{j,+}_nw^*_{j-1}&=-q^{(n+2)\delta_{j,2}-1}w^*_j\\
E^{j,-}_nw^*_j&=(-1)^{j+1}q^{(n-2)\delta_{j,2}+1}w^*_{j-1}
\end{split}
\end{equation}
$\forall n$.
Extension of \eqref{E:wmod1} and \eqref{E:wmod2} to 
$U'_q\bigl(\widehat{gl}(2\vert2)\bigr)$-structures is not unique. A convenient
choice is
\begin{alignat}{2}\label{E:wmod3}
h_4w_k&=(\delta_{k,0}-\delta_{k,3})w_k&\\
\Psi^{4,\pm}_{\pm n}w_k&=\pm(q-q^{-1})\bigl(q^{\pm 3n}\delta_{k,0}
-q^{\pm n}\delta_{k,3}\bigr)w_k\qquad &n>0\notag\\
h_4w^*_k&=-(\delta_{k,0}-\delta_{k,3})w^*_k&\notag\\
\Psi^{4,\pm}_{\pm n}w^*_k&=\mp(q-q^{-1})\bigl(q^{\mp n}\delta_{k,0}-q^{\pm n}
\delta_{k,3}\bigr)w^*_k\qquad &n>0\notag
\end{alignat}
The modules $W$ and $W^*$ are attributed alternately
to the horizontal as well as the vertical lines of the lattice
composing the integrable vertex
model. To each of the four types of elementary vertices, a
spectral parameter $z$ or $(q^2w)^{\pm 1}z$ is assigned as shown in figure \ref{FF:spec}.
Boltzmann weights are
assigned to the elementary vertices depending on the spectral parameter and on
 the configuration of basis elements $\{w_k\}_{0\leq k\leq 3}$
or $\{w^*_k\}_{0\leq k\leq3}$ on the joining links. Their values follow
from the R-matrices intertwining the action of $U_q\bigl(\widehat{gl}(2\vert2)
\bigr)$ on the tensor products of two evaluation modules \eqref{E:evmod} 
(see \cite{gade1}).
Four neighbouring vertices may be viewed as a composite vertex of type
A or B as illustrated in figure \ref{FF:spec}. In the limit $q\to0$, the Boltzmann
weights of a composite vertex provide a well-defined map $(W\otimes W^*)^{\otimes2}
\to(W\otimes W^*)^{\otimes2}$ (type A) or $(W^*\otimes W)^{\otimes 2}\to(W^*\otimes
W)^{\otimes2}$ (type B). Due to the particular spectral inhomogeneity chosen, the
maps are invertible. These properties allow to consider the Hamiltonian of a
corner transfer matrix at $q=0$. A northwest corner transfer matrix built
from composite vertices of type $A$ or type $B$ acts on the half-infinite configurations
$(\ldots\otimes v_{k_4}\otimes v_{k_3}\otimes v_{k_2}\otimes v_{k_1})$
subject to a suitable boundary condition. These configurations are specified by
$v_{2r}=w_{2r}$, $v_{2r-1}=w^*_{2r-1}$ for type $A$ and by $v_{2r}=w^*_{2r}$,
$v_{2r-1}=w_{2r-1}$ for type $B$.
The space of all configurations
$(\ldots\otimes v_{k_4}\otimes v_{k_3}\otimes v_{k_2}\otimes v_{k_1})$
with $k_r=k$ for almost all $r$ may be
called $\Omega_A^{(k)}$ in the type A case and $\Omega_B^{(k)}$ in the type B case.
In the following, the values $k=1$ and $k=3$ are considered. Analysis and structure
of the results for $\Omega_A^{(1)}$ and $\Omega_B^{(1)}$ are similar to the
procedure and findings described in
\cite{gade1} for the spaces $\Omega_A^{(3)}$ and $\Omega_B^{(3)}$. The action of
the CTM Hamiltonians becomes triangular in the limit $q\to0$.
Moreover, in
this limit the diagonal elements decouple into a part depending only on the
labels $k_{2r}$ and another part determined by the entries $k_{2r-1}$ only.

Denoting the diagonal element for the configuration
$(\ldots\otimes w_{k_4}\otimes w^*_{k_3}\otimes w_{k_2}\otimes w^*_{k_1})\in
\Omega_A^{(k)}$
by $h_{(\ldots,k_4,k_3,k_2,k_1);(\ldots,k_4,k_3,k_2,k_1)}$, the two contributions
are specified by
\begin{equation}\label{E:ctmh}
h_{(\ldots,k_4,k_3,k_2,k_1);(\ldots,k_4,k_3,k_2,k_1)}=-\sum_{r=1}^{\infty}
r\bigl(x_{k_{2r+1},k_{2r-1}}+y_{k_{2r+2},k_{2r}}\bigr)
\end{equation}
with
\begin{equation}\label{E:xy}
x_{k_1,k_2}=y_{k_2,k_1}=
\begin{cases}
0&\text{if $k_1>k_2$ or $k_1=k_2=1,3$};\\
1&\text{if $k_1<k_2$ or $k_1=k_2=0,2$}.
\end{cases}
\end{equation}
The diagonal element for the configuration $(\ldots\otimes w^*_{k_4}\otimes w_{k_3}
\otimes w^*_{k_2}\otimes w_{k_1})\in\Omega_B^{(k)}$ is given by
\begin{equation}\label{E:ctmh2}
-\sum_{r=1}^{\infty}
r\bigl(x_{k_{2r+2},k_{2r}}+y_{k_{2r+1},k_{2r-1}}\bigr)
\end{equation}
In view of the triangular action of the corner transfer matrix Hamiltonians,
$x_{k_1,k_2}$ and $y_{k_1,k_2}$ may be called the
quasi-energy functions of the vertex model. 

A $U_q\bigl(gl(2\vert2)\bigr)$-weight for a configuration $(\ldots\otimes w_{k_4}
\otimes w^*_{k_3}\otimes w_{k_2}\otimes w^*_{k_1})\in\Omega_A^{(k)}$ or
$(\ldots\otimes w^*_{k_4}\otimes w_{k_3}\otimes w^*_{k_2}
\otimes w_{k_1})\in\Omega_B^{(k)}$
follows unambiguously
from a $U_q\bigl(gl(2\vert2)\bigr)$-reference weight $(\bar h^{ref}_1,\bar h^{ref}_2,
\bar h^{ref}_3,\bar h^{ref}_4)$ introduced for the configuration
$(\ldots\otimes w_k\otimes w^*_k\otimes w_k\otimes w^*_k)$ or $(\ldots
\otimes w^*_k\otimes w_k\otimes w^*_k\otimes w_k)$, respectively. For
an arbitrary configuration in $\Omega_A^{(k)}$, the $U_q\bigl(gl(2\vert2)\bigr)$-weight
is given by
\begin{multline}\label{E:weight1}
h_l(\ldots\otimes w_{k_4}\otimes w^*_{k_3}\otimes w_{k_2}\otimes w^*_{k_1})=\\
\Bigl\{\bar h_l^{ref}+
\bar h^{A,k}_l(\ldots,k_4,k_3,k_2,k_1)\Bigr\}
(\ldots\otimes w_{k_4}\otimes w^*_{k_3}\otimes w_{k_2}\otimes w^*_{k_1})
\end{multline}
with
\begin{equation}\label{E:weight2}
\begin{split}
\bar h^{A,k}_1(\ldots,k_4,k_3,k_2,k_1)=&-\bar h_3^{A,k}(\ldots,k_4,k_3,k_2,k_1)=\\
&-\delta_{k,1}\sum_{r=1}^{\infty}\bigl(\delta_{k_{2r},2}+\delta_{k_{2r},3}-\delta_{k_{
2r-1},2}-\delta_{k_{2r-1},3}\bigr)\\
&+\delta_{k,3}\sum_{r=1}^{\infty}\bigl(\delta_{k_{2r},0}+\delta_{k_{2r},1}-\delta_{k_{
2r-1},0}-\delta_{k_{2r-1},1}\bigr)\\
\bar h^{A,k}_2(\ldots,k_4,k_3,k_2,k_1)=&\;
\delta_{k,1}\sum_{r=1}^{\infty}\bigl(\delta_{k_{2r},0}+\delta_{k_{2r},3}
-\delta_{k_{2r-1},0}-\delta_{k_{2r-1},3}\bigr)\\
&-\delta_{k,3}\sum_{r=1}^{\infty}\bigl(\delta_{k_{2r},1}+\delta_{k_{2r},2}
-\delta_{k_{2r-1},1}-\delta_{k_{2r-1},2}\bigr)\\
\bar h^{A,k}_4(\ldots,k_4,k_3,k_2,k_1)=&\;
\delta_{k,1}\sum_{r=1}^{\infty}\bigl(\delta_{k_{2r},0}
-\delta_{k_{2r},3}-\delta_{k_{2r-1},0}+\delta_{k_{2r-1},3}\bigr)
\\
+\delta_{k,3}\sum_{r=1}^{\infty}&\bigl(2\delta_{k_{2r},0}
+\delta_{k_{2r},1}+\delta_{k_{2r},2}-2\delta_{k_{2r-1},0}-\delta_{k_{2r-1},1}-
\delta_{k_{2r-1},2}\bigr)
\end{split}
\end{equation}
For any configuration $(\ldots\otimes w^*_{k_4}\otimes w_{k_3}\otimes
w^*_{k_2}\otimes w_{k_1})\in\Omega_B^{(k)}$, the weight reads
\begin{multline}\label{E:weight3}
h_l(\ldots\otimes w^*_{k_4}\otimes w_{k_3}\otimes w^*_{k_2}\otimes w_{k_1})=\\
\Bigl\{\bar h_l^{ref}-
\bar h^{A,k}_l(\ldots,k_4,k_3,k_2,k_1)\Bigr\}
(\ldots\otimes w^*_{k_4}\otimes w_{k_3}\otimes w^*_{k_2}\otimes w_{k_1})
\end{multline}
A suitable choice of reference weights for the spaces $\Omega_A^{(k)}$ and
$\Omega_B^{(k)}$ is
\begin{alignat}{2}\label{E:rw1}
h^{ref}_l&=\delta_{l,2}+\delta_{l,4}\qquad &\text{for}\;k=1,\\
h^{ref}_l&=0\;\;\forall l\qquad &\text{for}\;k=3.\notag
\end{alignat}
Another reference weight for $\Omega^{(k)}_A$ is provided by adding the
$U_q\bigl(gl(2\vert2)\bigr)$-weight of $w^*_k$ to the weight \eqref{E:rw1}:
\begin{alignat}{2}\label{E:rw2}
h^{ref}_l&=\delta_{l,4}+2\delta_{l,2}-\delta_{l,1}\qquad&\text{for}\;\;k=1,\\
h^{ref}_l&=\delta_{l,4}-\delta_{l,3}\qquad&\text{for}\;\;k=3.
\end{alignat}
The weight assignments \eqref{E:weight1} with \eqref{E:rw1} and \eqref{E:rw2}
will be called $\bar h^{A,k}$ and $\bar h'^{A,k}$, respectively.
Similarly, adding the $U_q\bigl(gl(2\vert2)\bigr)$-weight of $w_k$ to \eqref{E:rw1}
gives a second reference weight for $\Omega_B^{(k)}$:
\begin{equation}\label{E:rw3}
\begin{split}
h^{ref}_l&=\delta_{l,1}+\delta_{l,4}\qquad\text{for}\;\;k=1,\\
h^{ref}_l&=\delta_{l,3}-\delta_{l,4}\qquad\text{for}\;\;k=3.
\end{split}
\end{equation}
The assignments \eqref{E:weight3} with \eqref{E:rw1} and \eqref{E:rw3}
are denoted by $\bar h^{B,k}$ and $\bar h'^{B,k}$. In the following section,
a further reference weight
\begin{alignat}{2}\label{E:rw4}
h_l^{ref}&=-s\delta_{l,1}+(1+s)\delta_{l,2}-(1+2s')\delta_{l,4}\qquad&\text{for}\;\;k=1,\\
h_l^{ref}&=s\delta_{l,3}+s'\delta_{l,4}\qquad&\text{for}\;\;k=3\phantom{.}\notag
\end{alignat}
with arbitrary $s'$ and $s\notin \mathbb Z$ will be taken into account both for
$\Omega^{(1)}_A$ and $\Omega^{(1)}_B$. The assignments \eqref{E:weight3} with
\eqref{E:rw4} will be referred to as $\bar h^{C,k}$.

In \cite{gade1}, the half-infinite configurations of the spaces
$\Omega_A^{(3)}$ and $\Omega_B^{(3)}$ are
compared to the weight vectors of reducible level-one modules
of $U_q\bigl(\widehat{sl}(2\vert2)\bigr)/{\mathcal H}$ denoted by
$\tilde V(\Lambda_0)$,
$\tilde V(\Lambda_1+\Lambda_4)$ and $\tilde V(2\Lambda_0-\Lambda_3+\Lambda_4)$.
The eigenvalues of the grading operator $d$ and the generators
$h_l$, $l=1,2,3,4,$ are called the grade and the $U_q\bigl(gl(2\vert2)\bigr)$-weight
of a weight vector. For a fixed weight assignment 
$\bar h^A,\,\bar h'^A, \bar h^B$ or $\bar h'^B$, the $U_q\bigl(
gl(2\vert2)\bigr)$-weights of all configurations with the diagonal element of the
CTM Hamiltonian given by $-n$ are collected. At $n=0,1,2,3$, they are in one-to-one
correspondence with the $U_q\bigl(gl(2\vert2)\bigr)$-weights of all vectors with grade
$-n$ found in one of the reducible level-one modules. This correspondence may be assumed to
hold true at any grade. Then the character of the reducible module can be expressed
in terms of the quasi-energy functions. Due to the
decomposition of the diagonal elements \eqref{E:ctmh}, \eqref{E:ctmh2}, the character expressions
factorise into two parts, each of them depending only on one quasi-energy function.
Each of the reducible module can be decomposed into two
irreducible level-one modules. One them is weakly integrable \cite{kac1}, the other
nonintegrable. 
Table \ref{T:t1} specifies the relevant level-one modules for the four
choices of composite vertices and assignment of weights.
The weakly integrable irreducible module is listed left of the nonintegrable module.

\noindent
\begin{table}[h]
%\begin{center}
\renewcommand{\arraystretch}{1.5}
\begin{tabular}{|l|l|l|l|l|}\hline
space&weights&reducible module&\multicolumn{2}{c |}{irreducible modules}\\
\hline
$\Omega_A^{(3)}$&$\bar h^{A,3}$&$\tilde V(\Lambda_0)$&$V(\Lambda_0)$&$V(\Lambda_2
-\Lambda_4)$\\
\hline
$\Omega_A^{(3)}$&$\bar h'^{A,3}$&$\tilde V(2\Lambda_0-\Lambda_3-\Lambda_4)$&$V(2\Lambda_0-
\Lambda_3-\Lambda_4)$&$V(-\Lambda_1+2\Lambda_2-\Lambda_4)$\\
\hline
$\Omega_B^{(3)}$&$\bar h^{B,3}$&$\tilde V(\Lambda_0)$&$V(\Lambda_0)$&$V(\Lambda_2
-\Lambda_4)$\\
\hline
$\Omega_B^{(3)}$&$\bar h'^{B,3}$&$\tilde V(\Lambda_1+\Lambda_4)$&$V(\Lambda_1+\Lambda_4)$&$
V(\Lambda_3-\Lambda_4)$\\
\hline
\end{tabular}
\vspace{0.5cm}
\caption{Assignment of $U_q\bigl(gl(2\vert2)\bigr)$-weights
and level-one modules for the configuration spaces $\Omega_A^{(3)}$
and $\Omega_B^{(3)}$} 
\label{T:t1}
%\end{center}
\end{table}

A similar analysis suggests that the weight states of the irreducible, nonintegrable
level-one module $V\bigl((1-s)\Lambda_0+s\Lambda_3+s'\Lambda_4\bigr)$-module
of $U_q\bigl(\widehat{sl}(2\vert2)\bigr)/\mathcal H$ correspond to the configurations
if the reference weight $\bar h^{C,3}$ is adopted.

\section{The second boundary condition}
\label{S:bound2}

The configurations in $\Omega_A^{(1)}$ and $\Omega_B^{(1)}$ can be related to the weight
vectors of level-one modules of $U_q\bigl(\widehat{sl}(2\vert2)
\bigr)/{\mathcal H}$. 
An assignment of weights $\bar h^{A,1}$, $\bar h'^{A,1}$, $\bar h^{B,1}$,
$\bar h'^{B,1}$ or $\bar h^{C,1}$ introduced in the previous section is fixed. Then
the $U_q\bigl(
gl(2\vert2)\bigr)$-weights of all configurations with a fixed value $-n$ of the
diagonal elements \eqref{E:ctmh} or \eqref{E:ctmh2} are compared to the $U_q\bigl(
gl(2\vert2)\bigr)$-weights of all vectors with the associated grade present in a
suitable level-one
module. The value of the associated grade may equal the diagonal element or differ
from it by a constant value. Consideration of the three lowest values $n=0,1,2$
indicates the appropriate modules.
Three reducible level-one modules of $U_q\bigl(
\widehat{sl}(2\vert2)\bigr)/{\mathcal H}$ denoted by $\mathring{V}(\Lambda_0)$,
$\mathring{V}(2\Lambda_0-\Lambda_3+\Lambda_4)$ and $\mathring{V}(\Lambda_1+
\Lambda_4)$ account for the first four choices of composite vertices and
$U_q\bigl(gl(2\vert2)\bigr)$-weights. Similar as the  reducible modules related to
$\Omega_A^{(3)}$ and $\Omega_B^{(3)}$, each of them decomposes
into an irreducible weakly integrable and an irreducible nonintegrable module.
In table \ref{T:t2},
the level-one modules related to each case are listed.
There the weakly integrable module appears left of the nonintegrable irreducible
module. The irreducible, nonintegrable level-one module $V\bigl(-s\Lambda_1+(s+1)\Lambda_2-(
1+2s')\Lambda_4\bigr)$ of $U_q\bigl(\widehat{sl}(2\vert2)\bigr)/{\mathcal H}$
accounts for the choice $\bar h^{C,1}$.
\noindent
\begin{table}[h]
%\begin{center}
\renewcommand{\arraystretch}{1.5}
\begin{tabular}{|l|l|l|l|l|}\hline
space&weights&reducible module&\multicolumn{2}{c |}{irreducible modules}\\
\hline
$\Omega_A^{(1)}$&$\bar h^{A,1}$&$\mathring{V}(\Lambda_0)$&$V(\Lambda_0)$&$
V(\Lambda_2+\Lambda_4)$\\
\hline
$\Omega_A^{(1)}$&$\bar h'^{A,1}$&$\mathring{V}(2\Lambda_0-\Lambda_3+\Lambda_4)$&$
V(2\Lambda_0-\Lambda_3+\Lambda_4)$&$V(-\Lambda_1+2\Lambda_2+\Lambda_4)$\\
\hline
$\Omega_B^{(1)}$&$\bar h^{B,1}$&$\mathring{V}(\Lambda_0)$&$V(\Lambda_0)$&$V(
\Lambda_2+\Lambda_4)$\\
\hline
$\Omega_B^{(1)}$&$\bar h'^{B,1}$&$\mathring{V}(\Lambda_1+\Lambda_4)$&$V(
\Lambda_1+\Lambda_4)$&$V(\Lambda_3+\Lambda_4)$\\
\hline
\end{tabular}
\vspace{0.5cm}
\caption{Assignment of $U_q\bigl(gl(2\vert2)\bigr)$-weights
and level-one modules for the configuration spaces $\Omega_A^{(1)}$ and
$\Omega_B^{(1)}$}
\label{T:t2}
%\end{center}
\end{table}

A pair of irreducible modules can be assembled in a reducible module in
various ways. In each case found in table \ref{T:t2}, a particular reducible
module is distinguished by a simple free boson realization in terms of the
scheme given in section \ref{S:mod}. These modules are obtained from the
vectors
\begin{equation}\label{E:vec}
\kappa_0=e^{\beta^1}\vert0\rangle\qquad\kappa_1=e^{i\varphi^1+\beta^1}\vert0
\rangle\qquad \kappa_3=\beta^2_{-1}e^{-\varphi^4+\beta^1+\beta^2}\vert0\rangle
\end{equation}
with the boson Fock vacuum $\vert0\rangle$ characterised by the properties
\begin{equation}\label{E:vac}
\varphi^l_0\vert0\rangle=\beta^{\bar l}_0\vert0\rangle=0\qquad \varphi^l_n\vert0\rangle
=\beta^{\bar l}_n\vert0\rangle=0\qquad\forall n>0,\;l=1,2,3,4,\;\bar l=1,2
\end{equation}
For the irreducible module $V\bigl(-s\Lambda_1+(1+s)\Lambda_2-(1+2s')\Lambda_4
\bigr)$, the corresponding vector reads
\begin{equation}
\kappa_4=e^{-s'i\varphi^1-(s'-s)\varphi^2+(1+s')(i\varphi^3+\varphi^4)-s\beta^1}
\,\vert0\rangle
\end{equation}
The action of the grading operator on $\kappa_I$ reads $d\,\kappa_I=
\tfrac{1}{2}(\delta_{I,0}-1)\kappa_I$ for $I=0,1,3$
and $d\,\kappa_4=-\tfrac{1}{2}s(1+2s')\kappa_4$. Arbitrary polynomials of the
$U_q\bigl(\widehat{gl}(2\vert2)\bigr)$-generators \eqref{E:psib1},\eqref{E:psib2}
and \eqref{E:Ebos} applied on $\kappa_I$ give rise to vectors with the
maximal value of their grades given by $\tfrac{1}{2}(\delta_{I,0}+2\delta_{I,3}
-1)$ for $I=0,1,3$ and by $-\tfrac{1}{2}s(1+2s')$ for $I=4$.
If $I=0$, the vectors with grade $0$ form a reducible $U_q\bigl(gl(2
\vert2)\bigr)$-module which decomposes into the infinite-dimensional, irreducible
module $V(\bar{\Lambda}_2+\bar{\Lambda}_4)$ and the one-dimensional module
with weight $(0,0,0,0)$. For $I=1$, the vectors with grade $-\tfrac{1}{2}$
furnish a reducible $U_q\bigl(gl(2
\vert2)\bigr)$-module which is decomposed into the four-dimensional module
$V(\bar{\Lambda}_1+\bar{\Lambda}_4)$ and the infinite-dimensional module
$V(\bar{\Lambda}_3+\bar{\Lambda}_4)$. The latter are both irreducible.
In the case $I=3$, the vectors with grade $\tfrac{1}{2}$ form the
infinite-dimensional, irreducible $U_q\bigl(gl(2\vert2)\bigr)$-module
$V(-\bar{\Lambda}_1+2\bar{\Lambda}_2+\bar{\Lambda}_4)$. Similarly,
for $I=4$, the vectors with grade $-\tfrac{1}{2}s(1+2s')$ constitute
the infinite-dimensional, irreducible $U_q\bigl(gl(2\vert2)\bigr)$-module
$V\bigl(-s\bar{\Lambda}_1+(1+s)\bar{\Lambda}_2-(1+2s')\bar{\Lambda}_4\bigr)$.
The $U_q\bigl(gl(2\vert2)\bigr)$-modules obtained in this way
can be viewed as the maximal-grade subspaces of
level-one modules of $U_q\bigl(\widehat{sl}(2\vert2)\bigr)/
{\mathcal H}$. This suggests to examine the Fock spaces $\mathring{{\mathcal F}}_I$
with $I=0,1,3,4$  defined by
\begin{multline}\label{E:fock1}
\mathring{\mathcal F}_I=\mathbb{C}\bigl[\tilde{\varphi}^{\bar l_1}_{-1},
\beta^{\bar l_2}_{-1},
\tilde{\varphi}^{\bar l_3}_{-2},\beta^{\bar l_4}_{-2},\ldots\bigr]\\
\otimes\Bigl(\oplus_{S_1,
S_2,S_3\in\mathbb{Z}}e^{S_1(i\varphi^1+\varphi^2-\beta^1)+S_2(\varphi^2-i
\varphi^3-\beta^1)+S_3(i\varphi^3+\varphi^4-\beta^2)+\beta^1+\varpi_I}\vert0
\rangle\Bigr)
\end{multline}
with $\bar l_i=1,2$,
\begin{equation}\label{E:fock3}
\begin{split}
\tilde{\varphi}^1_{-n}&\equiv \varphi^1_{-n}-i\varphi^2_{-n}+\varphi^3_{-n}-i
\varphi^4_{-n}\\
\tilde{\varphi}^2_{-n}&\equiv i\varphi^2_{-n}+\varphi^3_{-n}
\end{split}
\end{equation}
and
\begin{equation}\label{E:fock2}
\begin{split}
\varpi_0&=0\qquad\varpi_1=-\varphi^2+\beta^1\qquad\varpi_3=\varphi^2-\beta^1\\
\varpi_4&=s(\varphi^2-\beta^1)-(1+s')\bigl(i\varphi^1+\varphi^2-i\varphi^3-\varphi^4
\bigr)
\end{split}
\end{equation}
The remaining two linear combinations in $\varphi^l_{-n}$ are not included in
\eqref{E:fock1} since the Fock space $\mathring{{\mathcal F}}_I$ should accommodate
a module of $U_q\bigl(\widehat{sl}(2\vert2)\bigr)/{\mathcal H}$ rather than a
$U_q\bigl(\widehat{gl}(2\vert2)\bigr)$-module.
Investigation of the next lower grades indicates that the required reducible
modules are realized as restricted Fock spaces:
\begin{equation}\label{E:modb1}
\begin{split}
\mathring{V}(\Lambda_0)=Ker_{\eta^2_0}\mathring{\mathcal F}_0\qquad
\mathring{V}(\Lambda_1+\Lambda_4)&=Ker_{\eta^2_0}\mathring{\mathcal F}_1\qquad
\mathring{V}(2\Lambda_0-\Lambda_3+\Lambda_4)=Ker_{\eta^2_0}\mathring{\mathcal F}_3\\
V\bigl(-s\Lambda_1+(1+s)\Lambda_2-(1+2s')\Lambda_4\bigr)&=Ker_{\eta^2_0}
\mathring{\mathcal F}_4
\end{split}
\end{equation}
The expression \eqref{E:dbos} for the grading operator $d$ and the properties
\eqref{E:vac} yield
\begin{multline}\label{E:dact}
d\,e^{S_1(i\varphi^1+\varphi^2-\beta^1)+S_2(\varphi^2-i
\varphi^3-\beta^1)+S_3(i\varphi^3+\varphi^4-\beta^2)+\beta^1+\varpi_I}\vert0
\rangle=\\
-\frac{1}{2}\Bigl(S_1(S_1-1)+(S_2-S_3)(S_2-S_3-1)+\delta_{I,1}-
\delta_{I,3}+s(1+2s')\delta_{I,4}\Bigr)\cdot\\
\cdot e^{S_1(i\varphi^1+\varphi^2-\beta^1)+S_2(\varphi^2-i
\varphi^3-\beta^1)+S_3(i\varphi^3+\varphi^4-\beta^2)+\beta^1+\varpi_I}\vert0
\rangle
\end{multline}
for $I=0,1,3,4$.
With \eqref{E:hbos} and \eqref{E:dact}, the $U_q\bigl(gl(2\vert2)\bigr)$-weights
of all vectors in $Ker_{\eta^2_0}\mathring{{\mathcal F}}_I$ with grade $-n$ are easily
collected for small $n$. Then the one-to-one correspondence between the
$U_q\bigl(gl(2\vert2)\bigr)$-weights $\bar h^{A,1}$, $\bar h'^{A,1}$, $\bar h^{B,1}$,
$\bar h'^{B,1}$ or $\bar h^{C,1}$
of the configurations in $\Omega_A^{(1)}$ or $\Omega_B^{(1)}$ and the weights of the vectors
in the modules \eqref{E:modb1} is readily verified for $n=0,1,2$.

In the cases $I=0,1,3$,
the direct sum decomposition $Ker_{\eta^2_0}\mathring{\mathcal F}_I=
\xi^1_0\eta^1_0Ker_{\eta^2_0}\mathring{\mathcal F}_I\oplus\eta^1_0\xi^1_0
Ker_{\eta^2_0}\mathring{\mathcal F}_I$ allows to separate the
irreducible components of $Ker_{\eta^2_0}\mathring{\mathcal F}_I$.
Boson realizations for the irreducible $U_q\bigl(\widehat{sl}(2\vert2)\bigr)/
{\mathcal H}$-modules in table \ref{T:t2} are provided by
\begin{alignat}{2}\label{E:modb2}
V(\Lambda_0)&=\eta^1_0Ker_{\eta^2_0}\mathring{\mathcal F}_0&V(\Lambda_2+\Lambda_4)
&=Ker_{\eta^1_0}Ker_{\eta^2_0}\mathring{\mathcal F}_0\\
V(\Lambda_1+\Lambda_4)&=\eta^1_0Ker_{\eta^2_0}\mathring{\mathcal F}_1
&V(\Lambda_3+\Lambda_4)
&=Ker_{\eta^1_0}Ker_{\eta^2_0}\mathring{\mathcal F}_1\notag\\
V(2\Lambda_0-\Lambda_3+\Lambda_4)&=\eta^1_0Ker_{\eta^2_0}\mathring{\mathcal F}_3
\qquad\;\;&V(-\Lambda_1+2\Lambda_2+\Lambda_4)
&=Ker_{\eta^1_0}Ker_{\eta^2_0}\mathring{\mathcal F}_3\notag
\end{alignat}
Hence, the zero mode $\eta^1_0$ annihilates the nonintegrable components.
Some features of the reducible modules can be read from \eqref{E:vec} and
\eqref{E:modb2}. Since $\eta^1_0\kappa_I\neq0\forall I$, the vectors $\kappa_I$
belong to the weakly integrable irreducible submodules. The coupling of both
irreducible submodules is described by
\begin{alignat}{2}\label{E:coup}
E^{1,+}_0\kappa_I&\neq0\qquad &\eta^1_0E^{1,+}_0\kappa_I=E^{1,+}_0\eta^1_0\kappa_I
=0\qquad\qquad& I=0,3\\
E^{2,-}_0\kappa_J&\neq0\qquad &\eta^1_0E^{2,-}_0\kappa_J=E^{2,-}_0\eta^1_0\kappa_J
=0\qquad\qquad&J=0,1\notag
\end{alignat}
Thus $E^{1,+}_0\kappa_0$, $E^{2,-}_0\kappa_1$ and $E^{1,+}_0\kappa_3$ are
contained in the nonintegrable submodules. A vector with maximal grade in
$V(2\Lambda_0-\Lambda_3+\Lambda_4)$ is given by $E^{2,-}_1E^{3,-}_0\kappa_3$,
for example.
In the following, the notations 
$\mathring{V}(\Lambda_0)$, $\mathring{V}(\Lambda_1+\Lambda_4)$ and
$\mathring{V}(2\Lambda_0-\Lambda_3+\Lambda_4)$ refer to the reducible modules
 realized by \eqref{E:modb1}.

Provided that the correspondence between half-infinite configurations and
weight states found for $n=0,1,2$ remains valid for all $n\geq0$, the
characters of the reducible modules can be written in terms of the energy
functions \eqref{E:xy}. For the nonintegrable modules described above,
well defined characters are introduced by
\begin{equation}\label{E:chardef}
ch_{{V}(\cdot)}(\varrho,\rho_0,\rho_2,\rho_3)\equiv tr_{{V}(\cdot
)}\varrho^d\rho_0^{\frac{1}{2}(h_1+h_2+h_3+h_4)}
\rho_2^{-\frac{1}{2}(h_1+h_2-h_3-h_4)}\rho_3^{\frac{1}{2}(h_1+h_2+h_3-h_4)}
\end{equation}
where $\vert \rho_3\vert<1$ and ${V}(\cdot)$ denotes any of the modules
listed in \eqref{E:modb1}.

\begin{conjecture}\label{C:c1}
The characters of the $U_q\bigl(\widehat{sl}(2\vert2)\bigr)/{\mathcal H}$-modules
$\mathring{V}(\Lambda_0)$, $\mathring{V}(\Lambda_1+\Lambda_4)$,
$\mathring{V}(2\Lambda_0-\Lambda_3+\Lambda_4)$ and $V\bigl(-s\Lambda_1+(1+s)
\Lambda_2-(1+2s')\Lambda_4\bigr)$ 
are given by
\begin{equation}\label{E:char1}
\begin{split}
ch_{\mathring{V}(\Lambda_0)}(\varrho,\rho_0,\rho_2,\rho_3)
&=\sum_{\{\ldots,k_6,k_4,k_2\}}\varrho^{-\sum_{r>0}r
y_{k_{2r+2},k_{2r}}}\prod_{l=0,2,3}\rho_l^{\sum_{r>0}\delta_{k_{2r},l}}\\
&\qquad\;\;\cdot\sum_{\{\ldots,k_5,k_3,k_1\}}\varrho^{-
\sum_{\bar r>0}\bar rx_{k_{2\bar r+1},k_{2\bar r-1}}}\prod_{\bar l=0,2,3}
\rho_{\bar l}^{-\sum_{r>0}\delta_{k_{2\bar r-1},\bar l}}
\end{split}
\end{equation}
and
\begin{equation}\label{E:char2}
\begin{split}
ch_{\mathring{V}(\Lambda_1+\Lambda_4)}(\varrho,\rho_0,\rho_2,\rho_3)&=
\varrho^{-\frac{1}{2}}
\,ch_{\mathring{V}(\Lambda_0)}(\varrho,\rho_0,\rho_2,\rho_3)\\
ch_{\mathring{V}(2\Lambda_2-\Lambda_3+\Lambda_4)}(\varrho,\rho_0,\rho_2,\rho_3)&=
\varrho^{\frac{1}{2}}\,ch_{\mathring{V}(\Lambda_0)}(\varrho,\rho_0,\rho_2,\rho_3)\\
ch_{\mathring{V}\bigl(-s\Lambda_1+(1+s)\Lambda_2-(1+2s')\Lambda_4\bigr)}
(\varrho,\rho_0,\rho_2,\rho_3)&=
\varrho^{-\frac{1}{2}s(1+2s')}
\,ch_{\mathring{V}(\Lambda_0)}(\varrho,\rho_0,\rho_2,\rho_3)
\end{split}
\end{equation}
with $\vert \rho_3\vert<1$. The sums in \eqref{E:char1} are restricted by the
requirement that $k_{r}=1$ for almost all $r>0$.
\end{conjecture}

In most of the remainder, the assignment $\bar h^{A,1}$ is
considered.

\section{Infinite border strips}
\label{S:bs}

Within the framework proposed in \cite{jm}, \cite{dav}, \cite{idz},
the eigenstates of the row-to-row transfer matrix of the vertex model are
created on the ground state by means of type II vertex operators. This general
feature can be expected to apply to the $U_q\bigl(\widehat{gl}(2\vert2)
\bigr)$-model as well. Unfortunately, there is no simple method to
single out the appropriate vertex operators among all those existing at
the given level. In \cite{dav} and \cite{naka}, the
space of states of the $XXZ$-model and its higher spin generalisation
has been decomposed at $q=0$. The analysis involves a creation algebra 
whose defining relations can be viewed as formal $q\to0$ limits of the
commutation relations satisfied by the appropriate type II vertex operators.
A relation between the space of states at $q=0$ and the creation algebra
is established by means of the domain wall picture and the crystal theory
associated with the paths. The resulting expressions in terms of the
generators of the creation algebra are interpreted as the
$q\to0$-limits of the $n$-particle eigenstates of the model. In the next
section, two creation algebras relevant to the present model
will be considered.
Infinite border strips prove a useful tool for setting up the
relation between their generators and the configuration space.

All following considerations refer to infinite configurations
$(\ldots\otimes w_{j_2}\otimes w^*_{j_1}\otimes w_{j_0}\otimes w^*_{j_{-1}}
\otimes w_{j_{-2}}\otimes w^*_{j_{-3}}\otimes\ldots)$ subject to suitable
boundary conditions. The set of all infinite configurations satisfying
$k_r=k\,\forall r>r_+>0$ and $k_r=k'\,\forall r<r_-<0$ with $k,k'=1,3$ is
denoted by $\mathcal{K}_{k,k'}$. Where convenient, the infinite components $(\ldots
\otimes w_{j_2}\otimes w_{j_0}\otimes w_{j_{-2}}\otimes\ldots)$ and $(\ldots
\otimes w^*_{j_1}\otimes w^*_{j_{-1}}\otimes w^*_{j_{-3}}\otimes\ldots)$
will be abbreviated by $(\ldots,j_2,j_0,j_{-2},\ldots)$ and $(\ldots,j_1,j_{-1},
j_{-3},\ldots)$, respectively. All four choices of $k,k'$ allow for a well-defined
generalisation of the expression \eqref{E:ctmh}:
\begin{equation}\label{E:hinf2}
h_{(\ldots,k_2,k_0,k_{-2},k_{-4},\ldots);(\ldots,k_2,k_0,k_{-2},
k_{-4},\ldots)}=-\sum_{r\in\mathbb{Z}}ry_{k_{2r+2},k_{2r}}
\end{equation}
\begin{equation}\label{E:hinf1}
h_{(\ldots,k_3,k_1,k_{-1},k_{-3},\ldots);(\ldots,k_3,k_1,k_{-1},
k_{-3},\ldots)}=-\sum_{r\in\mathbb{Z}}rx_{k_{2r+1},k_{2r-1}}
\end{equation}
with $x_{k_{2r+1},k_{2r-1}}$ and $y_{k_{2r+2},k_{2r}}$ defined by
\eqref{E:xy}. 
$U_q\bigl(gl(2\vert2)\bigr)$-weights $(\bar h_1,\bar h_2,\bar h_3,\bar h_4)$
compatible with the assignment $\bar h^{A,1}$ defined by
\eqref{E:weight1} and \eqref{E:rw1} are introduced
for the infinite components of $\mathcal{K}_{1,1}$ by
\begin{equation}\label{E:infw1}
\begin{split}
\bar h_1(\ldots,k_2,k_0,k_{-2},k_{-4},\ldots)=&-
\bar h_3(\ldots,k_2,k_0,k_{-2},k_{-4},\ldots)\\
&=-\sum_{r\in\mathbb{Z}}
(\delta_{k_{2r},2}+\delta_{k_{2r},3}) (\ldots,k_2,k_0,k_{-2},k_{-4},\ldots)\\
\bar h_2(\ldots,k_2,k_0,k_{-2},k_{-4},\ldots)=&\sum_{r\in\mathbb{Z}}
(\delta_{k_{2r},0}+\delta_{k_{2r},3})(\ldots,k_2,k_0,k_{-2},k_{-4},\ldots)\\
\bar h_4(\ldots,k_2,k_0,k_{-2},k_{-4},\ldots)=&\sum_{r\in\mathbb{Z}}
(\delta_{k_{2r},0}-\delta_{k_{2r},3})(\ldots,k_2,k_0,k_{-2},k_{-4},\ldots)
\end{split}
\end{equation}
and
\begin{equation}\label{E:infw2}
\begin{split}
\bar h_1(\ldots,k_3,k_1,k_{-1},k_{-3},\ldots)=&-
\bar h_3(\ldots,k_3,k_1,k_{-1},k_{-3},\ldots)\\
&=\sum_{r\in\mathbb{Z}}
(\delta_{k_{2r-1},2}+\delta_{k_{2r-1},3}) (\ldots,k_3,k_1,k_{-1},k_{-3},\ldots)\\
\bar h_2(\ldots,k_3,k_1,k_{-1},k_{-3},\ldots)=&-\sum_{r\in\mathbb{Z}}
(\delta_{k_{2r-1},0}+\delta_{k_{2r-1},3})(\ldots,k_3,k_1,k_{-1},k_{-3},\ldots)\\
\bar h_4(\ldots,k_3,k_1,k_{-1},k_{-3},\ldots)=&-\sum_{r\in\mathbb{Z}}
(\delta_{k_{2r-1},0}-\delta_{k_{2r-1},3})(\ldots,k_3,k_1,k_{-1},k_{-3},\ldots)
\end{split}
\end{equation}

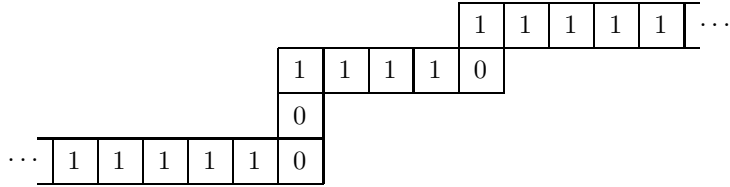
\begin{figure}[ht]
\begin{center}
\setlength{\unitlength}{1cm}
\begin{picture}(10,3)
%\thicklines
\put(0.5,0.3){\line(1,0){3.8}}
\put(0.5,0.9){\line(1,0){3.8}}
\put(3.7,1.5){\line(1,0){3}}
\put(3.7,2.1){\line(1,0){5.6}}
\put(6.1,2.7){\line(1,0){3.2}}
\put(3.7,0.3){\line(0,1){1.8}}
\put(4.3,0.3){\line(0,1){1.8}}
\put(6.1,1.5){\line(0,1){1.2}}
\put(6.7,1.5){\line(0,1){1.2}}
\multiput(0.7,0.3)(0.6,0){5}{\line(0,1){0.6}}
\put(4.9,1.5){\line(0,1){0.6}}
\multiput(7.3,2.1)(0.6,0){4}{\line(0,1){0.6}}
\multiput(0.9,0.5)(0.6,0){5}{$1$}
\multiput(3.9,1.7)(0.6,0){4}{$1$}
\multiput(6.3,2.3)(0.6,0){5}{$1$}
\put(3.9,0.5){$0$}
\put(3.9,1.1){$0$}
\put(6.3,1.7){$0$}
\put(0.1,0.6){$\ldots$}
\put(9.3,2.4){$\ldots$}
%\linethickness{1.5pt}
\put(5.5,1.5){\line(0,1){0.6}}
\end{picture}\par

\caption{The reference labelling 
for the horizontal border strip with $R=2$ and parameters $p_1=1$, $p_2=2$,
$\bar p_1=4$.}
\label{FF:infbs}
\end{center}
\end{figure}

For homogeneous vertex models related to quantum affine algebras, a one-to-one
correspondence between the spin configurations and semi-standard super tableaux
of skew Young diagrams has been demonstrated in \cite{kuniba}. A similar one-to-one
correspondence between the half-infinite spin configurations of the present
model and two types of semi-standard super tableaux of finite and half-infinite
border strips is pointed out in \cite{gade1}. 
Making use of this correspondence, pairs of infinite border strips can be
related to the components $(\ldots\otimes w_{k_4}\otimes w_{k_2}\otimes w_{k_0}
\otimes w_{k_{-2}}\otimes w_{k_{-4}}\otimes\ldots)$ and 
 $(\ldots\otimes w^*_{k_3}\otimes w^*_{k_1}\otimes w^*_{k_{-1}}
\otimes w^*_{k_{-3}}\otimes w^*_{k_{-5}}\otimes\ldots)$ of the infinite
configurations $(\ldots w_{k_2}\otimes w^*_{k_1}\otimes w_{k_0}\otimes w^*_{k_{-1}}
\otimes w_{k_{-2}}\otimes w^*_{k_{-3}}\otimes\ldots)$.
These border strips consist of finitely many rows and columns of finite length
assembled between either two half-infinite rows or two half-infinite columns.
In the following, they will be referred to as horizontal or vertical border strips,
respectively. As an example, figure \ref{FF:infbs} shows a horizontal border
strip.
The horizontal border strips are related to the components 
$(\ldots\otimes w_{k_4}\otimes w_{k_2}\otimes w_{k_0}
\otimes w_{k_{-2}}\otimes w_{k_{-4}}\otimes\ldots)$ and the vertical border strips
to the components
$(\ldots \otimes w^*_{k_3}\otimes w^*_{k_1}\otimes w^*_{k_{-1}}
\otimes w^*_{k_{-3}}\otimes w^*_{k_{-5}}\otimes\ldots)$.
A semi-standard super tableau of a horizontal or vertical
border strip is obtained by assigning one of the
numbers $0,1,2,3$ to each box such that the numbers of each two neighbouring boxes
satisfy two rules:
\begin{enumerate}
\item
If the side common to both boxes is vertical, then the
number $k_1$ in the left box and the number $k_2$ in the right
box fulfil
\begin{equation}\label{E:label1}
k_1>k_2\quad\text{or}\quad k_1=k_2=\begin{cases}
0,2&\text{vertical strip}\\
1,3&\text{horizontal strip}
\end{cases}
\end{equation}
\item
If the side common to both boxes is horizontal, then the number $k_1$ in
the upper box and the number $k_2$ in the lower box fulfil
\begin{equation}\label{E:label2}
k_1>k_2\quad\text{or}\quad k_1=k_2=\begin{cases}
1,3&\text{vertical strip}\\
0,2&\text{horizontal strip}
\end{cases}
\end{equation}
\end{enumerate}
Almost all numbers attributed to the infinite strip are fixed by a
boundary condition. The set of all semi standard super tableaux of
horizontal (vertical) border strips with the
number $k$ given to
almost all boxes of the lower half-infinite row (left half-infinite column)
and the number $k'$
given to almost all boxes in the upper half-infinite row (right half-infinite column)
will be called $\mathcal{B}^h_{k,k'}$ ($\mathcal{B}^v_{k,k'}$).
Each of the four choices $k,k'=1,3$ is consistent
with the rules \eqref{E:label1}, \eqref{E:label2}. The set
$\mathcal{B}^h_{k,k}$ ($\mathcal{B}^v_{k,k}$) includes exactly one semi standard
super tableau for the infinite border strip consisting of one single row (column).
This tableau attributes the number $k$ to each box. Excluding this tableau
from $\mathcal{B}^h_{k,k}$ $(\mathcal{B}^v_{k,k}$) yields a set called
$\mathcal{B}^{h\backslash 0}_{k,k}$ $\bigl(\mathcal{B}^{v\backslash 0}_{k,k}\bigr)$.

A horizontal border strip with at least one finite row
is characterised by the set $R,(p_1,p_2,\ldots,p_R;\,\bar p_1,\bar p_2,\ldots
,\bar p_{R-1})$ with $R>1$ and $p_i,\bar p_i\in\mathbb{N}$. This border strip contains
$R$ finite columns with more than one box. A vertical border strip with at least
one finite column is described by the set $R,(p_1,p_2,\ldots,p_{R-1};\,\bar p_1,
\bar p_2,\ldots,\bar p_R)$ with $R>1$ and $p_i,\bar p_i\in\mathbb{N}$. In both cases,
the parameter $p_i+1$ ($\bar p_i+1$) specifies the number of the boxes
contained in the $i$-th finite column (finite row) composed of at least two boxes.
Here the counting proceeds from the right to the left end of a horizontal border
strip or from the upper to the lower end of a vertical border strip. In the
remainder, the counting of rows or columns refers only to rows or
columns consisting of more than one box.
The set $R=1,(p_1;\emptyset)$ with $p_1\in\mathbb{N}$ specifies a horizontal
border strip composed of the lower and upper half-infinite row and a column
containing $p_1+1\geq2$ boxes. Similarly, the set $R=1,(\emptyset;\,\bar p_1)$
characterises
a vertical border strip built from the left and right half-infinite columns
and a row with $\bar p_1+1\geq2$ boxes. Finally, the value $R=0$
refers to the border strips consisting of one single infinite row or column.

A horizontal (vertical) border strip with $R\geq1$
has exactly one semi-standard super tableau in $\mathcal{B}^h_{1,1}$
$(\mathcal{B}^v_{1,1}$) involving only the numbers $0$ and $1$. In a horizontal
border strip, the $p_i$ lower boxes of each column 
with $p_i+1\geq2$ boxes receive the number $0$. All other boxes obtain the number $1$.
In case of a vertical border strip, the rightmost $\bar p_i$ boxes in a row with
$\bar p_i+1\geq2$ boxes receive the value $0$. The number $1$ is given to all other boxes.
These tableaux will be called the reference labellings in the following.
An example is illustrated by figure \ref{FF:infbs}.

In order to define a mapping from a semi standard super tableau onto a
component of an infinite configuration, a counting of the boxes needs to be
specified. The leftmost box of the upper half-infinite row of a horizontal
border strip or the lowest box in the right half-infinite column of a
vertical border strip is counted as the $r_0$-th box for some $r_0\in
\mathbb{Z}$. The $(r+1)$-th box is left of or below the $r$-th box for any
$r\in\mathbb{Z}$. Then for a horizontal border strip with parameters $R=0$,
$R=1,(p_1;\emptyset)$ or $R>1,(p_1,\ldots,p_R;\bar p_1,\ldots,\bar p_{R-1})$
an arbitrary semi-standard
super tableau in $\mathcal{B}^h_{k,k'}$ may be considered. 
Given a fixed value of $r_0$, it is convenient to introduce numbers $s_i$, $1
\leq i\leq R$, by
\begin{equation}\label{E:sdef}
\begin{split}
s_1&=r_0\\
s_i&=r_0+\sum_{i_<=1}^{i-1}(p_{i_<}+\bar p_{i_<})\qquad 2\leq i\leq R
\end{split}
\end{equation}
and
\begin{equation}\label{E:dform1}
d^h(p_1,\ldots,p_R;\,\bar p_1,\ldots,\bar p_{R-1};\,r_0)=
\sum_{i=1}^Rp_i\Bigl(s_i+\frac{1}{2}(p_i-1)\Bigr)
\end{equation}
A component $(\ldots\otimes w_{k_4}\otimes w_{k_2}
\otimes w_{k_{0}}\otimes w_{k_{-2}}\otimes w_{k_{-4}}\otimes\ldots)$
is associated with the semi-standard super tableau by identifying the number
attributed to the $(r+1)$-th box with $k_{-2r}$. 
With \eqref{E:xy} and \eqref{E:hinf2} it is easily verified that the contribution
of this component to the diagonal element of the CTM Hamiltonian coincides with
\eqref{E:dform1}.

For a vertical border strip with parameters $R'=0$, $R'=1,(\emptyset;\,
\bar p'_1)$ or $R'>1,(p'_1,\ldots,p'_{R'-1};\,\bar p'_1,\ldots,
\bar p'_{R'})$ and the box counting fixed by the number $r'_0$, the number
\begin{equation}\label{E:dform2}
d^v(p'_1,\ldots,p'_{R'-1};\,\bar p'_1,\ldots,\bar p'_{R'};\,r_0)=
\sum_{i=1}^{R'}\bar p'_i\Bigl(s'_i+\frac{1}{2}
(\bar p'_i-1)\Bigr)
\end{equation}
is introduced.
Here $s'_i$ is defined by \eqref{E:sdef} with $r_0$, $p_{i_<}$, $\bar p_{i_<}$
replaced by $r'_0$, $p'_{i_<}$ and $\bar p'_{i_<}$, respectively.
A semi standard super tableau in $\mathcal{B}^v_{k,k'}$
is related to a component $(\ldots\otimes w^*_{k_3}\otimes w^*_{k_1}\otimes w^*_{k_{-1}}
\otimes w^*_{k_{-3}}\otimes w^*_{k_{-5}}\otimes\ldots)$
by identifying the number attributed to the $(r+1)$-th box with $k_{-(2r-1)}$. 
According to \eqref{E:xy} and \eqref{E:hinf1}, the contribution of this component
to the diagonal element of the CTM Hamiltonian equals \eqref{E:dform2}.

Analogous statements apply to the components $(\ldots\otimes w_1\otimes w_1
\otimes w_1\otimes\ldots)$ and $(\ldots\otimes w^*_1\otimes w^*_1\otimes w^*_1
\otimes\ldots)$ provided that
\begin{equation}\label{E:dform3}
d^h(0,\emptyset)=d^v(\emptyset,0)=0
\end{equation}

This is readily verified comparing the definitions \eqref{E:xy} for
$y_{k_{2r+2},k_{2r}}$ and $x_{k_{2r+1},k_{2r-1}}$
with the rules \eqref{E:label1}, \eqref{E:label2}. A $U_q\bigl(gl(2\vert2)
\bigr)$-weight is assigned to each semi standard super tableau in
$\mathcal{B}^h_{1,1}$  or $\mathcal{B}^v_{1,1}$ via \eqref{E:infw2} or \eqref{E:infw1}
and the above identifications. 

The above prescription gives a one-to-one correspondence between the components
$(\ldots\otimes w_{k_4}\otimes w_{k_2}
\otimes w_{k_{0}}\otimes w_{k_{-2}}\otimes w_{k_{-4}}\otimes\ldots)$
and $(\ldots\otimes w^*_{k_3}\otimes w^*_{k_1}\otimes w^*_{k_{-1}}
\otimes w^*_{k_{-3}}\otimes w^*_{k_{-5}}\otimes\ldots)$
of the configurations in $\mathcal{K}_{k,k'}$ and the sets $B^{h}_{k,k'}$ and
$B^{v}_{k,k'}$ of semi-standard super tableaux associated with infinite
border strips. This correspondence will be used in the next section to
describe the configuration space in terms of two creation algebras.

\section{The creation algebras}
\label{S:crealg}

\begin{definition}\label{D:d1}
The creation algebra $\mathcal{A}^*$ 
is generated by $\{\phi^*_{j,t;\,m}\vert 0\leq j\leq3,\,
t\in\mathbb{N}_0,\,m\in\mathbb{Z}\}$ over $\mathcal{Z}$
subject to the defining
relations
\begin{equation}\label{E:credef1}
\phi^*_{j_2,t_2;\,m_2}\,\phi^*_{j_1,t_1;\,m_1}=-\phi^*_{j_2,t_2;\,m_1+t_2
+\theta_{j_2,j_1}}\phi^*_{j_1,t_1;\,m_2-t_2-\theta_{j_2,j_1}}
\end{equation}
with
\begin{equation}\label{E:theta}
\theta_{j_2,j_1}=
\begin{cases}
0,&\text{if $j_2=2$};\\
0,&\text{if $j_2=0,3$ and $j_1=0,1$;}\\
1,&\text{otherwise.}
\end{cases}
\end{equation}
\end{definition}
A special case of the defining relations \eqref{E:credef1} is
\begin{equation}\label{E:spec1}
\phi^*_{j_2,t_2;\,m+t_2+\theta_{j_2,j_1}}\,\phi^*_{j_1,t_1;\,m}=0\qquad\forall
t_1\in\mathbb{N}_0,\,m\in\mathbb{Z}
\end{equation}
For the subsequent analysis, it is useful to introduce the notion of normal
forms \cite{naka}.
%\begin{definition}\label{E:normdef}
The product
\begin{equation}\label{E:prod}
\phi^*_{j_n,t_n;\,m_n}\ldots\phi^*_{j_2,t_2;\,m_2}\phi^*_{j_1,t_1;\,m_1}
\end{equation}
is called a normal form (of $\mathcal{A}^*$) iff
\begin{equation}\label{E:cond}
m_{i+1}>m_i+t_{i+1}+\theta_{j_{i+1},j_i}\qquad \text{for}\;\;1\leq i< n
\end{equation}
%\end{definition}
The set 
\begin{equation}\label{E:Bdef}
B^*=\cup_{n\in\mathbb{N}}\bigl\{\phi^*_{j_n,t_n;\,m_n}\ldots\phi^*_{j_1,t_1;\,m_1}
\vert j_i=0,1,2,3,\,t_i\in\mathbb{N}_0,\,m_i\in\mathbb{Z}\; \text{satisfy
\eqref{E:cond}}\}
\end{equation}
provides a $\mathbb{Z}$-linear base of the algebra
$\mathcal{A}^*$.
This statement is a special case of corollary 2 proven in
\cite{naka}. 
A $U_q\bigl(gl(2\vert2)\bigr)$-weight $(h^{\phi^*}_1,h^{\phi^*}_2,
h^{\phi^*}_3,h^{\phi^*}_4)$ is introduced for the generators $\phi^*_{j,t;\,m}$ via
\begin{equation}\label{E:algw}
\begin{split}
\bigl[h_l,\phi^*_{j,t;\,m}\bigr]&=h^{\phi^*}_l(j,t)\phi^*_{j,t;\,m},
\qquad l=1,2,3,4
\\
h_1^{\phi^*}(j,t)&=-h_3^{\phi^*}(j,t)=
-(t+1-\delta_{j,2})\\
h_2^{\phi^*}(j,t)&=t+1-\delta_{j,0}\\
h_4^{\phi^*}(j,t)&=-(t-1+\delta_{j,0}+2\delta_{j,1})
\end{split}
\end{equation}
The commutator with the grading operator $d$ is defined by
\begin{equation}\label{E:algd}
\bigl[d,\phi^*_{j,t;\,m}\bigr]=m\phi^*_{j,t;\,m}
\end{equation}
Taking into account \eqref{E:algw}, \eqref{E:algd} and \eqref{E:infw2},
the set $B^*$ of normal forms may be
compared to the set $\mathcal{B}^{h\backslash 0}_{1,1}$ of semi standard super tableaux.

For fixed values of $r_0\in\mathbb{Z}$ and $p_1\in\mathbb{N}$, the set of all normal forms
\begin{equation}\label{E:phiset1}
\phi^*_{j_{p_1},t_{p_1};\,r_0+p_1-1}\ldots\phi^*_{j_2,t_2;\,r_0+1}\phi^*_{j_1,t_1;
\,r_0}
\end{equation}
is called $B^*(p_1;\emptyset;\,r_0)$.

For $r_0\in\mathbb{Z}$ and $(p_1,p_2,\ldots,p_R;\,\bar p_1,\bar p_2,\ldots,
\bar p_{R-1})$ with $p_i,\bar p_i\in\mathbb{N}$ and $R>1$, 
the set of all normal forms
$\phi^*_{j_N,t_N;\,m_N}\ldots\phi^*_{j_2,t_2;\,m_2}\phi^*_{j_1,t_1;\,m_1}$
with 
\begin{equation}\label{E:phiset2}
\begin{split}
N&=\sum_{i=1}^Rp_i,\quad m_1=r_0,\\
m_{i+1}-m_i&=\begin{cases}
\bar p_{\hat R}+1&\text{if $i=p_1+p_2+\ldots+p_{\hat R}$ for $1\leq \hat R<R$},\\
1&\text{otherwise.}
\end{cases}
\end{split}
\end{equation}
is denoted by $B^*(p_1,p_2,\ldots,p_R;\,\bar p_1,\bar p_2,\ldots,\bar p_{R-1};\,r_0)$.

The set of all semi-standard super tableaux associated with the horizontal
border strip $R>1,(p_1,\ldots,p_R;\,\bar p_1,\ldots,\bar p_{R-1})$ or
$R=1,(p_1,\emptyset)$ with the box counting fixed by $r_0$ can be mapped onto 
$B^*(p_1,\ldots,p_R;\,\bar p_1,\ldots,\bar p_{R-1};\,r_0)$ or $B^*(p_1;\emptyset;\,r_0)$,
respectively. In particular, the reference labelling introduced in the previous
section is mapped onto the normal form 
with $j_r=2,\,t_r=0$ $\forall r$. This amounts to attributing the generator
$\phi^*_{2,0;\,r}$ to the $r+1$-th box if the reference labelling assigns the
number $0$ to this box. 
According to \eqref{E:infw2} and the correspondence
between configurations and semi-standard super tableaux,
the reference labelling is the only tableau of the border strip with the
$U_q\bigl(gl(2\vert2)\bigr)$-weight given by $(0,N,0,N)$. Equation
\eqref{E:algw} specifies the only normal form in 
$B^*(p_1,\ldots,p_R;\,\bar p_1,\ldots,\bar p_{R-1};\,r_0)$ or $B^*(p_1;\emptyset;r_0)$
with the same values.

An arbitrary semi-standard super tableau in $\mathcal{B}^{h\backslash0}_{1,1}$
with $R\geq1$ is mapped onto a normal form
\eqref{E:phiset2} in 
three steps. It is convenient to rewrite the normal form
\eqref{E:phiset2} as
\begin{equation}\label{E:nform2}
\begin{split}
&\phi^*_{j_{p_1},t_{p_1};\,s_1+p_1-1}\ldots\phi^*_{j_2,t_2;\,s_1+1}\phi^*_{j_1,t_1;
\,s_1}\\
&\cdot\phi^*_{j_{p_1+p_2},t_{p_1+p_2};\,s_2+p_2-1}\ldots\phi^*_{j_{p_1+2},
t_{p_1+2};s_2+1}\phi^*_{j_{p_1+1},t_{p_1+1};\,s_2}\\
&\cdot\phi^*_{j_{p_1+p_2+p_3},t_{p_1+p_2+p_3};\,s_3+p_3-1}\ldots
\phi^*_{j_{p_1+p_2+2},t_{p_1+p_2+2};\,s_3+1}\phi^*_{j_{p_1+p_2+1},t_{p_1+p_2+1};,s_3}\\
&\ldots\\
&\cdot\phi^*_{j_{N},t_{N};s_R+p_R-1}\ldots\phi^*_{j_{p_1
+\ldots+p_{R-1}+2},t_{p_1+\ldots+p_{R-1}+2};s_R+1}
\phi^*_{j_{p_1+\ldots+p_{R-1}+1},t_{p_1+\ldots+p_{R-1}+1};s_R}
\end{split}
\end{equation}
with the numbers $s_i$ defined in  \eqref{E:sdef}. The factors in the $i$-th
line in \eqref{E:nform2} are related to the numbers given to the boxes in the
$i$-th column and the right neighbouring row as follows.

\begin{figure}[ht]
\begin{center}
\setlength{\unitlength}{1cm}
\begin{picture}(12,6)

\put(0.4,1.2){\line(1,0){4.4}}
\put(0.8,0.8){\line(1,0){4}}
\multiput(0.4,0.8)(0.4,0){3}{\line(0,1){0.4}}
\multiput(1.9,0.8)(0.4,0){8}{\line(0,1){0.4}}
\put(1.35,1){$\ldots$}
\put(4.82,1){$\ldots$}
\put(0.52,0.9){$3$}
\put(0.92,0.9){$3$}
\put(2.02,0.9){$3$}
\put(2.42,0.9){$2$}
\multiput(2.82,0.9)(0.4,0){5}{$1$}
\put(0.52,0.48){$k$}
\put(0.42,1.1){$\overbrace{\phantom{3333333333\,\,}}^{t\geq0}$}
\put(5.5,0.9){$\longleftrightarrow$}
\put(7,0.9){$\begin{cases}
k=0:&\phi^*_{3,t;\,s_1}\\k=1,2:&\phi^*_{0,t;\,s_1}
\end{cases}$}

\put(0.4,4.2){\line(1,0){4}}
\put(0.8,3.8){\line(1,0){3.6}}
\multiput(0.4,3.8)(0.4,0){3}{\line(0,1){0.4}}
\multiput(1.9,3.8)(0.4,0){7}{\line(0,1){0.4}}
\put(1.35,4){$\ldots$}
\put(4.42,4){$\ldots$}
\put(0.52,3.9){$3$}
\put(0.92,3.9){$3$}
\put(2.02,3.9){$3$}
\multiput(2.42,3.9)(0.4,0){5}{$1$}
\put(0.52,3.48){$k$}
\put(0.42,4.1){$\overbrace{\phantom{3333333333\,\,}}^{t\geq0}$}
\put(5.5,3.9){$\longleftrightarrow$}
\put(7,3.9){$\begin{cases}
k=0:&\phi^*_{2,t;\,s_1}\\k=1,2,\,t>0:&\phi^*_{1,t-1;\,s_1}
\end{cases}$}

\linethickness{1.5pt}  
\put(0.4,0.4){\line(1,0){0.4}}
\put(0.4,0.8){\line(1,0){0.4}}
\put(0.4,0.4){\line(0,1){0.4}}
\put(0.8,0.4){\line(0,1){0.4}}
\put(0.4,3.4){\line(1,0){0.4}}
\put(0.4,3.8){\line(1,0){0.4}}
\put(0.4,3.4){\line(0,1){0.4}}
\put(0.8,3.4){\line(0,1){0.4}}
\end{picture}\par

\caption{Mapping of the upper half-infinite row}
\label{FF:map3}
\end{center}
\end{figure}
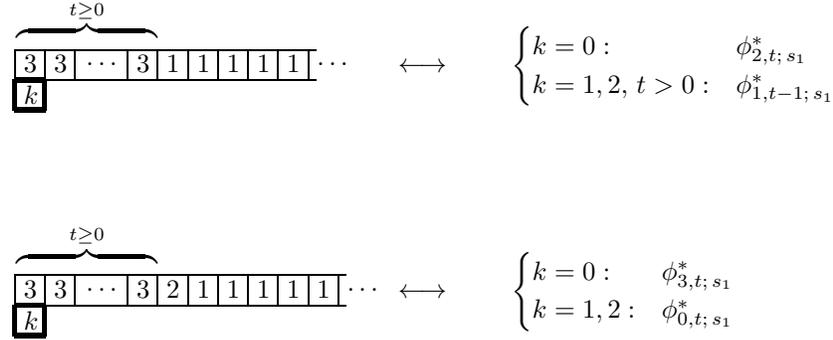

\begin{enumerate}

\item For $1\leq i\leq R$ and $p_i\geq2$, the factors
\begin{multline}
%\label{E:fact}
\qquad\;\;\phi^*_{j_{p_1+\ldots+p_i},t_{p_1+\ldots+p_i};s_i+p_i-1}\ldots
\phi^*_{j_{p_1+\ldots+p_{i-1}+3},t_{p_1+\ldots+p_{i-1}+3};s_i+2}\notag\\
\cdot\phi^*_{j_{p_1+\ldots+p_{i-1}+2},t_{p_1+\ldots+p_{i-1}+2};s_i+1}
\end{multline}
are determined according to the numbers attributed to the $p_i$ lower boxes of
the $i$-th column. Figure \ref{FF:map1} specifies these factors 
for all cases allowed by the rules \eqref{E:label1} and \eqref{E:label2}
for the tableaux of horizontal strips.\label{R:step1}

\item To obtain the factor $\phi^*_{j_{p_1+\ldots+p_{i-1}+1},t_{p_1+\ldots+
p_{i-1}+1};\,s_i}$  
with $i>1$ and $p_i\geq1$, the $i$-th column and the $(i-1)$-th finite row with length
$\bar p_{i-1}+1$ is taken into account. Figure \ref{FF:map2} collects all possible
cases  together with the associated factors.
\label{R:step2}

\item Depending on the numbers given to the boxes of the upper infinite column,
the rightmost factor $\phi^*_{j_1,t_1;\,s_1}$ is determined. All cases are listed in
figure \ref{FF:map3}. \label{R:step3}

\end{enumerate}

For a semi-standard super tableau of the border strip $(p_1,\emptyset)$ with
$p_1>1$, the normal form \eqref{E:phiset1} is determined by steps (\ref{R:step2})
and (\ref{R:step3}). In the case $(1;\emptyset)$, only step (\ref{R:step3})
is required. A box on the lhs of figures \ref{FF:map1}-\ref{FF:map3}
is drawn boldly if it is the $(r+1)$th box in the complete border strip 
and the corresponding expression on the rhs contains a generator $\phi^*_{j,t;r}$.

\begin{figure}[ht]
\begin{center}
\setlength{\unitlength}{1cm}
\begin{picture}(16,16)

\put(1.23,13.25){$\left.\phantom{\line(0,1){1.1}}\right\}$}
\put(1.6,13.25){$p_i-1$}
\put(0.96,13){$\vdots$}
\put(0.4,12.4){\line(1,0){0.4}}
\put(0.4,12.4){\line(0,1){0.4}}
\put(0.4,12.8){\line(1,0){0.4}}
\put(0.8,14.3){\line(0,1){0.8}}
\put(1.2,14.3){\line(0,1){0.8}}
\put(1.6,14.7){\line(0,1){0.4}}
\put(0.8,15.1){\line(1,0){0.8}}
\put(0.8,14.7){\line(1,0){0.8}}
\put(0.93,12.5){$0$}
\put(0.93,13.6){$0$}
\put(0.93,14){$0$}
\put(0.93,14.4){$k$}
\put(3,13.25){$\longleftrightarrow\qquad k=0,1:\;\;
\phi^*_{2,0;\,s_i+p_i-1}\ldots\phi^*_{2,0;\,s_i+2}\phi^*_{2,0;\,s_i+1}$}

\put(1.23,1.25){$\left.\phantom{\line(0,1){1.1}}\right\}$}
\put(1.6,1.25){$p_i-t-1$}
\put(0.96,1){$\vdots$}
\put(1.23,3.55){$\left.\phantom{\line(0,1){1.1}}\right\}$}
\put(1.6,3.55){$t>0$}
\put(0.96,3.3){$\vdots$}

\put(1.23,7.25){$\left.\phantom{\line(0,1){1.1}}\right\}$}
\put(1.6,7.25){$p_i-t$}
\put(0.96,7){$\vdots$}
\put(1.23,9.15){$\left.\phantom{\line(0,1){1.1}}\right\}$}
\put(1.6,9.45){$t>0$}
\put(0.96,8.9){$\vdots$}

\put(0.4,0.4){\line(1,0){0.4}}
\put(0.4,0.4){\line(0,1){0.4}}
\put(0.4,0.8){\line(1,0){0.4}}

\put(0.4,6.4){\line(1,0){0.4}}
\put(0.4,6.4){\line(0,1){0.4}}
\put(0.4,6.8){\line(1,0){0.4}}

\put(0.8,5){\line(1,0){0.8}}
\put(0.8,4.6){\line(1,0){0.8}}
\put(0.8,4.2){\line(0,1){0.8}}
\put(1.2,4.2){\line(0,1){0.8}}
\put(1.6,4.6){\line(0,1){0.4}}

\put(0.8,10.6){\line(1,0){0.8}}
\put(0.8,10.2){\line(1,0){0.8}}
\put(0.8,9.8){\line(0,1){0.8}}
\put(1.2,9.8){\line(0,1){0.8}}
\put(1.6,10.2){\line(0,1){0.4}}

\put(0.93,0.5){$0$}
\put(0.93,1.6){$0$}
\put(0.93,2){$0$}
\put(0.93,2.4){$1$}
\put(0.93,2.8){$2$}
\put(0.93,3.9){$2$}
\put(0.93,4.3){$2$}

\put(0.93,6.5){$0$}
\put(0.93,7.6){$0$}
\put(0.93,8){$0$}
\put(0.93,8.4){$2$}
\put(0.93,9.5){$2$}
\put(0.93,9.9){$2$}

\put(1.7,2.4){$\longleftrightarrow
\begin{cases}
p_i=t+1:&\phi^*_{0,0;\,s_i+p_i-1}\ldots\phi^*_{0,0;\,s_i+2}\phi^*_{0,0;\,s_i+1}\\
p_i>t+1:&\phi^*_{2,0;\,s_i+p_i-1}\ldots\phi^*_{2,0;\,s_i+t+2}\phi^*_{2,0;\,s_i+t+1}\\
&\qquad\qquad\;\;\cdot
\phi^*_{0,0;\,s_i+t}\ldots\phi^*_{0,0;\,s_i+2}\phi^*_{0,0;\,s_i+1}
\end{cases}$}

\put(1.8,8.4){$\longleftrightarrow
\begin{cases}
t=1,p_i=2:&\phi^*_{3,0;\,s_i+1}\\
t=1,
p_i>2:&\phi^*_{2,0;\,s_i+p_i-1}\ldots\phi^*_{2,0;\,s_i+3}\phi^*_{2,0;\,s_i+2}\,
\phi^*_{3,0;\,s_i+1}\\
t>1,&\\
p_i=t+1:&\phi^*_{3,0;\,s_i+t}\phi^*_{0,0;\,s_i+t-1}\ldots\phi^*_{0,0;\,s_i+2}
\phi^*_{0,0;\,s_i+1}\\
t>1,&\\
p_i>t+1:&\phi^*_{2,0;\,s_i+p_i-1}\ldots\phi^*_{2,0;\,s_i+t+2}\phi^*_{2,0;\,s_i+t+1}\\
&\cdot
\phi^*_{3,0;\,s_i+t}\phi^*_{0,0;\,s_i+t-1}\ldots\phi^*_{0,0;\,s_i+2}\phi^*_{0,0;\,s_i+1}
\end{cases}$}

\linethickness{1.5pt}
\put(0.8,0.4){\line(0,1){3.8}}
\put(1.2,0.4){\line(0,1){3.8}}
\multiput(0.8,0.4)(0,0.4){2}{\line(1,0){0.4}}
\multiput(0.8,1.5)(0,0.4){5}{\line(1,0){0.4}}
\multiput(0.8,3.8)(0,0.4){2}{\line(1,0){0.4}}
\put(0.8,6.4){\line(0,1){3.4}}
\put(1.2,6.4){\line(0,1){3.4}}
\multiput(0.8,6.4)(0,0.4){2}{\line(1,0){0.4}}
\multiput(0.8,7.5)(0,0.4){4}{\line(1,0){0.4}}
\multiput(0.8,9.4)(0,0.4){2}{\line(1,0){0.4}}
\put(0.8,12.4){\line(0,1){1.9}}
\put(1.2,12.4){\line(0,1){1.9}}
\multiput(0.8,12.4)(0,0.4){2}{\line(1,0){0.4}}
\multiput(0.8,13.5)(0,0.4){3}{\line(1,0){0.4}}

\end{picture}\par

\caption{Mapping of the columns}
\label{FF:map1}
\end{center}
\end{figure}
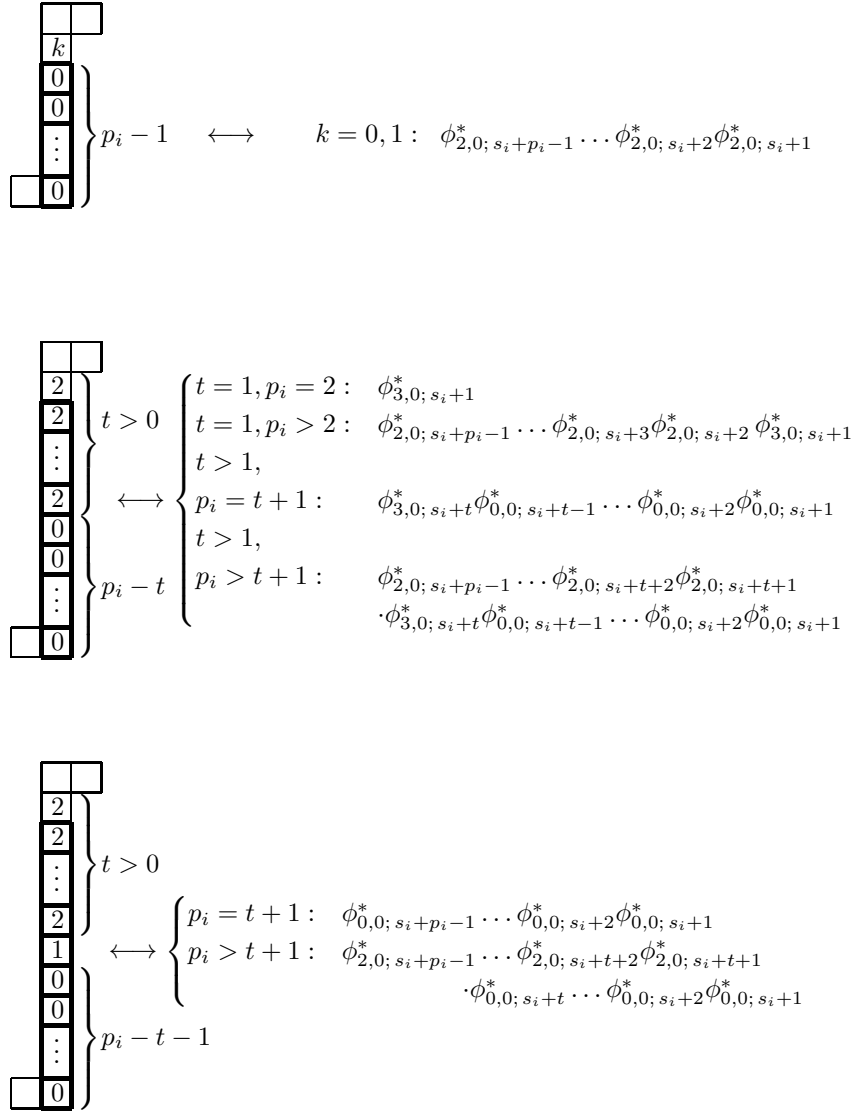

\begin{figure}[ht]
\begin{center}
\setlength{\unitlength}{1cm}
\begin{picture}(12,8)
\put(0.4,1.2){\line(1,0){4.6}}
\put(0.8,0.8){\line(1,0){4.2}}
\multiput(0.4,0.8)(0.4,0){3}{\line(0,1){0.4}}
\multiput(1.9,0.8)(0.4,0){5}{\line(0,1){0.4}}
\put(4.2,0.8){\line(0,1){0.4}}
\put(4.6,0.8){\line(0,1){0.8}}
\put(5,0.8){\line(0,1){0.8}}
\put(4.6,1.6){\line(1,0){0.4}}
\put(1.35,1){$\ldots$}
\put(3.65,1){$\ldots$}
\put(0.52,0.9){$3$}
\put(0.92,0.9){$3$}
\put(2.02,0.9){$3$}
\put(2.42,0.9){$2$}
\put(2.82,0.9){$1$}
\put(3.22,0.9){$1$}
\put(4.32,0.9){$1$}
\put(4.72,0.9){$0$}
\put(0.52,0.48){$k$}
\put(0.42,1.1){$\overbrace{\phantom{3333333333\,\,}}^{t\geq0}$}
\put(2.72,0.65){$\underbrace{\phantom{3333333333\,\,}}_{\bar p_{i-1}-t-1\geq0}$}
%\put(2.72,1.1){$\overbrace{\phantom{3333333333\,\,}}^{s_i-t-1\geq0}$}
\put(5.5,0.9){$\longleftrightarrow$}
\put(7,0.9){$\begin{cases}
k=0:&\phi^*_{3,t;\,s_i}\\k=1,2:&\phi^*_{0,t;\,s_i}
\end{cases}$}

\put(0.4,4.2){\line(1,0){4.2}}
\put(0.8,3.8){\line(1,0){3.8}}
\multiput(0.4,3.8)(0.4,0){3}{\line(0,1){0.4}}
\multiput(1.9,3.8)(0.4,0){5}{\line(0,1){0.4}}
%\put(3.8,3.8){\line(0,1){0.4}}
\put(4.2,3.8){\line(0,1){0.8}}
\put(4.6,3.8){\line(0,1){0.8}}
\put(4.2,4.6){\line(1,0){0.4}}
\put(1.35,4){$\ldots$}
\put(3.65,4){$\ldots$}
\put(0.52,3.9){$3$}
\put(0.92,3.9){$3$}
\put(2.02,3.9){$3$}
\put(2.42,3.9){$2$}
\put(2.82,3.9){$1$}
\put(3.22,3.9){$1$}
\put(4.32,3.9){$1$}
\put(0.52,3.48){$k$}
\put(0.42,4.1){$\overbrace{\phantom{3333333333\,\,}}^{t\geq0}$}
\put(2.72,3.65){$\underbrace{\phantom{3333333333\,\,}}_{\bar p_{i-1}-t\geq0}$}
\put(5.5,3.9){$\longleftrightarrow$}
\put(7,3.9){$\begin{cases}
k=0:&\phi^*_{3,t;\,s_i}\\k=1,2:&\phi^*_{0,t;\,s_i}
\end{cases}$}

\put(0.4,7.2){\line(1,0){4.2}}
\put(0.8,6.8){\line(1,0){3.8}}
\multiput(0.4,6.8)(0.4,0){3}{\line(0,1){0.4}}
\multiput(1.9,6.8)(0.4,0){4}{\line(0,1){0.4}}
\put(3.8,6.8){\line(0,1){0.4}}
\put(4.2,6.8){\line(0,1){0.8}}
\put(4.6,6.8){\line(0,1){0.8}}
\put(4.2,7.6){\line(1,0){0.4}}
\put(1.35,7){$\ldots$}
\put(3.25,7){$\ldots$}
\put(0.52,6.9){$3$}
\put(0.92,6.9){$3$}
\put(2.02,6.9){$3$}
\put(2.42,6.9){$1$}
\put(2.82,6.9){$1$}
\put(3.92,6.9){$1$}
\put(0.52,6.48){$k$}
\put(0.42,7.1){$\overbrace{\phantom{3333333333\,\,}}^{t}$}
\put(2.32,6.65){$\underbrace{\phantom{3333333333\,\,}}_{\bar p_{i-1}-t\geq0}$}
%\put(2.32,7.1){$\overbrace{\phantom{3333333333\,\,}}^{s_i-t\geq0}$}
\put(5.5,6.9){$\longleftrightarrow$}
\put(7,6.9){$\begin{cases}
k=0;\,t\geq0:&\phi^*_{2,t;\,s_i}\\k=1,2;\,t>0:&\phi^*_{1,t-1;\,s_i}
\end{cases}$}

\linethickness{1.5pt}  
\put(0.4,0.4){\line(1,0){0.4}}
\put(0.4,0.8){\line(1,0){0.4}}
\put(0.4,0.4){\line(0,1){0.4}}
\put(0.8,0.4){\line(0,1){0.4}}
\put(0.4,3.4){\line(1,0){0.4}}
\put(0.4,3.8){\line(1,0){0.4}}
\put(0.4,3.4){\line(0,1){0.4}}
\put(0.8,3.4){\line(0,1){0.4}}
\put(0.4,6.4){\line(1,0){0.4}}
\put(0.4,6.8){\line(1,0){0.4}}
\put(0.4,6.4){\line(0,1){0.4}}
\put(0.8,6.4){\line(0,1){0.4}}
\end{picture}\par

\caption{Mapping of the finite rows}
\label{FF:map2}
\end{center}
\end{figure}
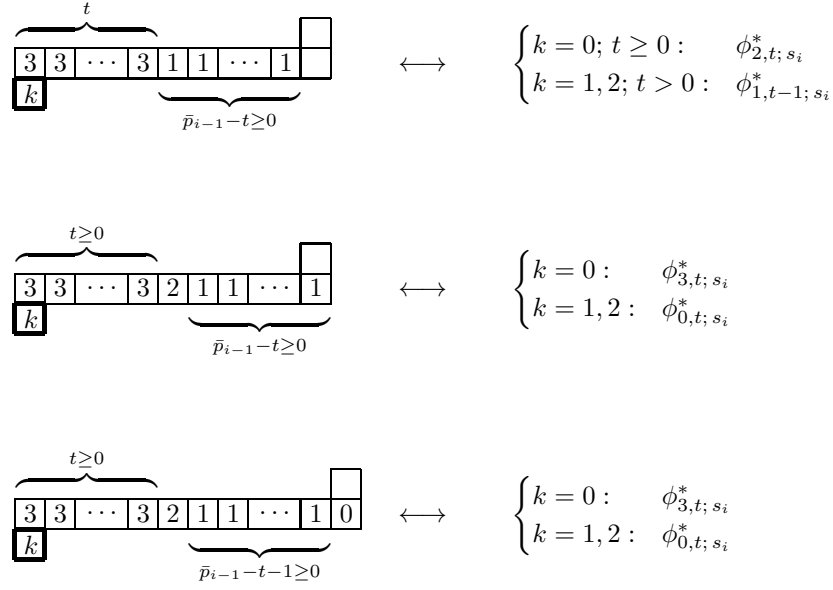

Each correspondence listed in figures \ref{FF:map1}-\ref{FF:map3} matches the
$U_q\bigl(gl(2\vert2)\bigr)$-weights introduced in \eqref{E:infw2} and
\eqref{E:algw}. Moreover, the sum $m_1+m_2+\ldots+m_N$ coincides with the
number $d^h(p_1,\ldots,p_R;\,\bar p_1,\ldots,\bar p_{R-1};\,r_0)$ introduced
in \eqref{E:dform1}. 
As a consequence of the rules \eqref{E:label1}-\eqref{E:label2},
an arbitrary tableau in $\mathcal{B}^{h\backslash0}_{1,1}$
assigns the number $1$ to all boxes of the lower half-infinite row except the
rightmost of them.
Thus the prescriptions given in figures \ref{FF:map1}-\ref{FF:map3} specify
one normal form of $B^*(p_1,\ldots,p_R;\bar p_1,\ldots,\bar p_{R-1};r_0)$
$\bigl(\text{or}\,B^*(p_1;\emptyset;r_0)\bigr)$
for each tableau of the border strip with the parameters
$(p_1,\ldots,p_R;
\bar p_1,\ldots,\bar p_{R-1})$
$\bigl(\text{or}\,(p_1;
\emptyset)\bigr)$ and $r_0$. Due to
the definition \eqref{E:cond}, the same prescriptions attribute exactly
one semi-standard super tableau related for the parameters
$(p_1,\ldots,p_R;\bar p_1,\ldots,\bar p_{R-1})$
$\bigl(\text{or}\,(p_1;
\emptyset)\bigr)$ and $r_0$ to
each normal form in 
$B^*(p_1,\ldots,p_R;\bar p_1,\ldots,\bar p_{R-1};r_0)$
$\bigl(\text{or}\,B^*(p_1;\emptyset;r_0)\bigr)$.

\begin{result}\label{RR:r1}
The semi-standard super tableaux in $\mathcal{B}^{h\backslash0}_{1,1}$
 related a horizontal border strip with the parameters $r_0$,
$(p_1,\ldots,p_R;\bar p_1,\ldots,\bar p_{R-1})\;\bigl(\text{or}\,(p_1;
\emptyset)\bigr)$, 
 and the normal forms in $B^*(p_1,\ldots,p_R;\bar p_1,\ldots,\bar p_{R-1};r_0)
\bigl(\text{or}\,B^*(p_1;\emptyset;r_0)\bigr)$ are in one-to-one
correspondence. Hence, the tableaux in $\mathcal{B}^{h\backslash0}_{1,1}$
and the normal forms in $B^*$ are in one-to-one correspondence.
\end{result}

The set $\mathcal{B}^{v\backslash0}_{1,1}$ of semi-standard super tableaux
associated with the vertical border strips is related to the set of
normal forms of the creation algebra $\mathcal{A}$.

\begin{definition}\label{D:d2}
The creation algebra $\mathcal{A}$ 
is generated by $\{\phi_{j,t;\,m}\vert 0\leq j\leq3,\,
t\in\mathbb{N}_0,\,m\in\mathbb{Z}\}$ over $\mathcal{Z}$
subject to the defining
relations
\begin{equation}\label{E:credef2}
\phi_{j_2,t_2;\,m_2}\,\phi_{j_1,t_1;\,m_1}=-\phi_{j_2,t_2;\,m_1+t_1
+\theta_{j_1,j_2}}\phi_{j_1,t_1;\,m_2-t_1-\theta_{j_1,j_2}}
\end{equation}
with $\theta_{j,j'}$ defined by \eqref{E:theta}.
\end{definition}
 
The procedure applying to the case of vertical border strips is quite analogous
to the one outlined above for the horizontal border strips.
A product
\begin{equation}\label{E:prod2}
\phi_{j_n,t_n;\,m_n}\ldots\phi_{j_2,t_2;\,m_2}\phi_{j_1,t_1;\,m_1}
\end{equation}
is referred to as a normal form (of $\mathcal{A}$) iff
\begin{equation}\label{E:cond2}
m_{i+1}>m_i+t_{i}+\theta_{j_{i},j_{i+1}}\qquad \text{for}\;\;1\leq i< n
\end{equation}
According to corollary 2 in \cite{naka}, the set
\begin{equation}\label{E:Bstardef}
B=\cup_{n\in\mathbb{N}}\bigl\{\phi_{j_n,t_n;\,m_n}\ldots\phi_{j_1,t_1;\,m_1}
\vert j_i=0,1,2,3,\,t_i\in\mathbb{N}_0,\,m_i\in\mathbb{Z}\; \text{satisfy
\eqref{E:cond2}}\}
\end{equation}
provides a $\mathbb{Z}$-linear basis of $\mathcal{A}$.
A $U_q\bigl(gl(2\vert2)\bigr)$-weight is defined for the generators
$\phi_{j,t;\,m}$ by
\begin{equation}\label{E:algstarw}
\bigl[h_l,\phi_{j,t;\,m}\bigr]=-h^{\phi^*}_l(j,t)\phi_{j,t;\,m},
\qquad l=1,2,3,4
\end{equation}
with $h^{\phi^*}_l$ given by \eqref{E:algw}. 
A mapping of the set $\mathcal{B}^{v\backslash0}_{1,1}$ of semi-standard
super tableaux onto the set $B$ similar to the mapping of
$\mathcal{B}^{h\backslash0}_{1,1}$ on $B^*$ is specified in Appendix
\ref{A:map}.
The set of all normal forms $\phi_{j_{\bar p_1},t_{\bar p_1};\,r_0+\bar p_1-1}
\ldots\phi_{j_1,t_2;\,r_0+1}\phi_{j_1,t_1;r_0}$ with fixed $r_0\in
\mathbb{Z}$ and $\bar p_1\in\mathbb{N}$ is denoted by $B(\emptyset;\bar p_1)$.
For $r_0\in\mathbb{Z}$ and $(p_1,p_2,\ldots,p_R;\,\bar p_1,\bar p_2,\ldots,
\bar p_{R-1})$ with $p_i,\bar p_i\in\mathbb{N}$ and $R>1$, 
the set of all normal forms
$\phi_{j_{\bar N},t_{\bar N};\,m_{\bar N}}\ldots\phi_{j_2,t_2;\,m_2}
\phi_{j_1,t_1;\,m_1}$
with 
\begin{equation}\label{E:phiset3}
\begin{split}
\bar N&=\sum_{i=1}^R\bar p_i,\quad m_1=r_0,\\
m_{i+1}-m_i&=\begin{cases}
p_{\hat R}+1&\text{if $i=\bar p_1+\bar p_2+\ldots+\bar p_{\hat R}$ for
$1\leq \hat R<R$},\\
1&\text{otherwise.}
\end{cases}
\end{split}
\end{equation}
is denoted by $B(p_1,p_2,\ldots,p_{R-1};\,\bar p_1,\bar p_2,\ldots,
\bar p_{R};\,r_0)$.

\begin{result}\label{RR:r2}
The semi-standard super tableaux in $\mathcal{B}^{v\backslash0}_{1,1}$ related
to a vertical border strip with parameters $r_0$, $(p_1,\ldots,p_{R-1};\bar p_1,
\ldots,\bar p_R)$ $\bigl(\text{or} \,(\emptyset,\bar p_1)\bigr)$  and the normal
forms in $B(p_1,\ldots,p_{R-1};\bar p_1,\ldots,\bar p_R;r_0)$ $\bigl(\text{or}\,
B(\emptyset,\bar p_1)\bigr)$ are in one-to-one correspondence .
\end{result}

For a normal form $\phi_{j_{\bar N},t_{\bar N};\,m_{\bar N}}\ldots
\phi_{j_1,t_1;\,m_1}$ in $B(p_1,\ldots,p_{R-1};\bar p_1,\ldots,
\bar p_R;r_0)$ or $B(\emptyset,\bar p_1)$, the sum $m_1+\ldots+m_{\bar N}$
equals the number $d^v(p_1,\ldots,p_{R-1};\,\bar p_1,\ldots,\bar p_R;\,r_0)$
introduced in \eqref{E:dform2}.

In the following section, the definition 
of the algebra $\mathcal{A^*}$ (or $\mathcal{A}$) is motivated by a naive limit
of the commutation
relations of type II vertex operators. Besides the components related to
the generators $\phi^*_{j,t;m}$ (or $\phi_{j,t;m}$), the vertex operators have
a further component equal to the unit. Each of the sets $B^*$,  $B$, 
may be supplemented by the unit. The resulting sets are
denoted by $B^*_{1}$ and $B_{1}$, respectively.
Formally, the unit may be assigned to the tableau in
$\mathcal{B}^h_{1,1}$ or $\mathcal{B}^v_{1,1}$
attributing the number one to each box of the
border strip given by the single infinite or half-infinite row  or column.
Application of the map between the components and the semi-standard
supertableaux specified in section \ref{S:bs} yields a relation between the
enlarged sets and the infinite configurations.

\begin{result}\label{RR:r3}
The infinite components $(\ldots\otimes w_{k_4}\otimes w_{k_2}
\otimes w_{k_{0}}\otimes w_{k_{-2}}\otimes w_{k_{-4}}\otimes\ldots)$
with $k_r=1$ for almost all $r$ are in
one-to-one correspondence with the set $B^*_1$.
The infinite components
$(\ldots\otimes w^*_{k_3}\otimes w^*_{k_1}\otimes w^*_{k_{-1}}
\otimes w^*_{k_{-3}}\otimes w^*_{k_{-5}}\otimes\ldots)$ with $k_r=1$ for almost
all $r$ are in one-to-one correspondence with the set $B_1$.
\end{result}

\section{Type II vertex operators}
\label{S:vo}

The type II vertex operators considered below
 are intertwiners of $U$-modules of the form
\begin{equation}\label{E:vodef1}
\Phi_{V_I}^{V\,V_J}(z):V_I\longrightarrow V_z\otimes V_J
\end{equation}
where $V_I$ and $V_J$ are level-one $U_q\bigl(\widehat{sl}(2\vert2)\bigr)
/{\mathcal H}$-modules. $V$ with basis $\{v_j\}$ denotes a $U'_q\bigl(\widehat{sl}
(2\vert2)\bigr)$-module obtained from
an infinite-dimensional $U_q\bigl(gl(2\vert2)\bigr)$-module 
by means of the evaluation homomorphism \cite{jimbo},
\cite{zhang1}, \cite{gade2}. 
The evaluation module $V_z=V\otimes\mathbb{C}
[z,z^{-1}]$ is endowed with a $U_q\bigl(\widehat{sl}(2\vert2)\bigr)$-structure via
\begin{alignat}{2}\label{E:evmod}
E^{k,\pm}_n\bigl(v_j\otimes z^m\bigr)&=E^{k,\pm}_nv_j\otimes z^{m+n}\qquad
&\forall n\\
\Psi^{l,\pm}_{\pm n}\bigl(v_j\otimes z^m\bigr)&=\Psi^{l,\pm}_{\pm n}v_j
\otimes z^{m\pm n}\qquad&n\geq0\notag\\
d\bigl(v_j\otimes z^m\bigr)&=mv_j\otimes z^m&&\notag\\
c\bigl(v_j\otimes z^m
\bigr)&=0&&\notag
\end{alignat}
The vertex operator \eqref{E:vodef1} is introduced as the formal series
\begin{equation}\label{E:vodef2}
\begin{split}
\Phi_{V_I}^{V\,V_J}(z)&=\sum_jv_j\otimes\left(\Phi_{V_I}^{V\,V_J}\right)_j(z)\\
\left(\Phi_{V_I}^{V\,V_J}\right)_j(z)&=\sum_{m\in\mathbb{Z}}
\left(\Phi_{V_I}^{V\,V_J}\right)_{j;m}z^{-m}
\end{split}
\end{equation}
In terms of the maps $\left(\Phi_{V_I}^{V\,V_J}\right)_{j;m}:V_I\longrightarrow
V_J$ the intertwining property reads
\begin{multline}\label{E:intprop}
\Delta(a)\Biggl\{\sum_j\sum_{m\in\mathbb{Z}}\bigl(v_j\otimes z^{-m}\bigr)\otimes
\left(\Phi_{V_I}^{V\,V_J}\right)_{j;m}\Biggr\}\\
=\sum_j\sum_{m\in\mathbb{Z}}(-1)^{\vert a\vert\cdot(\vert v_j\vert+
\vert\Phi_j\vert)}
\bigl(v_j\otimes z^{-m}\bigr)\otimes\left(\Phi_{V_I}^{V\,V_J}\right)_{j;m}a
\end{multline}
$\forall a\in U_q\bigl(\widehat{gl}(2\vert2)\bigr)$. Here $\vert v_j\vert$ and
$\vert \Phi_j\vert$ denote the $\mathbb{Z}_2$-gradings of $v_j$ and
the component $\left(\Phi_{V_I}^{V\,V_J}\right)_{j;m}$. In the remainder, their
relation is fixed by $\vert \Phi_j\vert=\vert v_j\vert$.
For the generators of $U_q\bigl(gl(2\vert2)\bigr)$,
$q^{\pm\frac{1}{2}c}$ and the grading operator $d$, the coproduct $\Delta$ is
defined by
\begin{alignat}{2}\label{E:co1}
\Delta\bigl(E^{k,+}_0\bigr)&=E^{k,+}_0\otimes 1+q^{h_k}\otimes E^{k,+}_0\qquad&
\Delta\bigl(E^{k,-}_0\bigr)&=E^{k,-}_0\otimes q^{-h_k}+1\otimes E^{k,-}_0\\
\Delta\bigl(q^{\pm h_l}\bigr)&=q^{\pm h_l}\otimes q^{\pm h_l}&k=1,2,3,\;
&l=1,2,3,4\notag
\end{alignat}
and
\begin{equation}\label{E:co2}
\Delta\bigl(q^{\pm\frac{1}{2}c}\bigr)=q^{\pm\frac{1}{2}c}\otimes q^{\pm\frac{1}{2}
c}\qquad\;\;\Delta(d)=d\otimes1+1\otimes d
\end{equation}
A partial information on the coproduct of the
remaining Drinfeld generators of $U_q\bigl(\widehat{gl}(2\vert2)\bigr)$
is provided by the formulae \cite{zhang2}
\begin{alignat}{2}\label{E:co3}
\Delta\bigl(E^{k,+}_n\bigr)&=E^{k,+}_n\otimes q^{nc}+q^{2nc+h_k}\otimes
E^{k,+}_n&&\\
&+\sum_{n'=0}^{n-1}q^{\frac{1}{2}(n+3n')c}\Psi^{k,+}_{n-n'}\otimes q^{(n-n')c}
E^{k,+}_{n'}\quad\mod{N_-\otimes N^2_+}&&\notag\\
\Delta\bigl(E^{k,+}_{-n}\bigr)&=E^{k,+}_{-n}\otimes q^{-nc}+q^{-h_k}\otimes E^{k,+}_{
-n}&&\notag\\
&+\sum_{n'=1}^{n-1}q^{\frac{1}{2}(n-n')c}\Psi^{k,-}_{-n+n'}\otimes q^{-(n-n')
c}E^{k,+}_{-n'}\quad\mod{N_-\otimes N^2_+}&&\notag\\
\Delta\bigl(E^{k,-}_n\bigr)&=E^{k,-}_n\otimes q^{h_k}+q^{nc}\otimes E^{k,-}_n
&&\notag\\
&+\sum_{n'=1}^{n-1}q^{(n-n')c}E^{k,-}_{n'}\otimes q^{-\frac{1}{2}(n-n')c}
\Psi^{k,+}_{n-n'}\quad\mod{N^2_-\otimes N_+}&&\notag\\
\Delta\bigl(E^{k,-}_{-n}\bigr)&=E^{k,-}_{-n}\otimes q^{-2mc-h_k}
+q^{-nc}\otimes E^{k,-}_{-m}&&\notag\\
&+\sum_{n'=0}^{n-1}q^{-(n-n')c}E^{k,-}_{-n'}\otimes q^{-\frac{1}{2}(n+3n')}
\Psi^{k,-}_{-n+n'}\quad\mod{N^2_-\otimes N_+}&&\notag\\
\Delta\bigl(H^l_n\bigr)&=H^l_n\otimes q^{\frac{1}{2}nc}+q^{\frac{3}{2}nc}
\otimes H^l_n\quad\mod{N_-\otimes N_+}&&\notag\\
\Delta\bigl(H^l_{-n}\bigr)&=H^l_{-n}\otimes q^{-\frac{3}{2}nc}+q^{-\frac{1}{2}
nc}\otimes H^l_{-n}\quad\mod{N_-\otimes N_+}&&\notag
\end{alignat}
for $k=1,2,3$, $l=1,2,3,4$,
and $n>0$. $N_{\pm}$ and $N^2_{\pm}$ are left $\mathbb{Q}(q)\bigl[
q^{\pm c},q^{\pm h_l},\Psi^{l,\pm}_{\pm n}\bigr]$-modules generated by $\bigl\{E^{k,\pm}_m
\bigr\}_{k=1,2,3,\,m\in\mathbb{Z}}$
and $\bigl\{E^{k,\pm}_mE^{k',\pm}_{m'}\bigr\}_{k,k'=1,2,3,\,m,m'\in\mathbb{Z}}$,
respectively. Equations \eqref{E:co3} are readily verified using the Hopf algebra
structure of $U_q\bigl(\widehat{gl}(2\vert2)\bigr)$ given in \cite{gade3}. In
context with vertex models based on the quantum affine algebra $U_q\bigl(
\widehat{gl}(N)\bigr)$, type II vertex operators associated with finite-dimensional
evaluation modules of the algebra are considered. Expressions analogous to
\eqref{E:co3} allow to derive free boson expressions for a particular component
of a vertex operator (see \cite{jm}, \cite{kato},\cite{konno} and \cite{koyama},
for example). Similar results have been obtained in \cite{zhang2}
for $U_q\bigl(\widehat{gl}(2\vert2)
\bigr)$-vertex operators related to the four-dimensional modules $W_z$ and $W^*_z$
defined in section \ref{S:mod}.

For two $U_q\bigl(gl(2\vert2)\bigr)$-modules $V^{(1)}$ and $V^{(2)}$ with
bases $\{v^{(1)}\}_j$ and $\{v^{(2)}\}_{j'}$,
the $R$-matrix $R_{V^{(1)}V^{(2)}}\left(\tfrac{z}{w}\right)\in End(V^{(1)}_z
\otimes V^{(2)}_w)$ intertwines
the action of $U_q\bigl(\widehat{sl}(2\vert2)\bigr)$ on the tensor product of the
two evaluation modules $V^{(1)}_z$ and $V^{(2)}_w$:
\begin{equation}\label{E:rdef}
R_{V^{(1)}V^{(2)}}\left(\tfrac{z}{w}\right)\Delta(a)=\Delta'(a)
R_{V^{(1)}V^{(2)}}\left(\tfrac{z}{w}\right)
\qquad\;\;\forall a\in U_q\bigl(\widehat{sl}(2\vert2)\bigr)
\end{equation}
where $\Delta'(a)=P^{gr}\circ\Delta(a)$ with $P^{gr}(x_1\otimes x_2)=(-1)^{\vert x_1
\vert\cdot\vert x_2\vert}x_2\otimes x_1$. It is convenient to introduce a
second matrix $\bar R_{V^{(1)}V^{(2)}}\left(\tfrac{z}{w}\right)$ by
\begin{equation}\label{E:rel}
\begin{split}
R_{V^{(1)}V^{(2)}}\left(\tfrac{z}{w}\right)\bigl(v^{(1)}_{j_1}\otimes
v^{(2)}_{j_2}\bigr)&=
\sum_{j_3,j_4}(-1)^{\vert v^{(1)}_{j_3}\vert\cdot\vert v^{(2)}_{j_4}\vert}
\Bigl(\bar R_{V^{(1)}V^{(2)}}\Bigr)_{j_1,j_2}^{j_3,j_4}\left(\tfrac{z}{w}\right)
v^{(1)}_{j_3}\otimes v^{(2)}_{j_4}\\
\bar R_{V^{(1)}V^{(2)}}\left(\tfrac{z}{w}\right)\bigl(v^{(1)}_{j_1}\otimes
v^{(2)}_{j_2}\bigr)&=
\sum_{j_3,j_4}\Bigl(\bar R_{V^{(1)}V^{(2)}}\Bigr)_{j_1,j_2}^{j_3,j_4}\left(
\tfrac{z}{w}\right)v_{j_3}\otimes v_{j_4}
\end{split}
\end{equation}
Uniqueness of the normalised vertex operators \eqref{E:vodef1} provided,
their commutation relation reads 
\begin{multline}\label{E:vocom1}
\left(\Phi_{V_J}^{V^{(1)}\,V_K}\right)_{j_1}(z)\left(\Phi_{V_I}^{V^{(2)}\,V_J}
\right)_{j_2}(w)=\\
c\left(\tfrac{z}{w}\right)\cdot\sum_{j_3,j_4}
\Bigl(\bar R_{V^{(1)}V^{(2)}}\Bigr)_{j_1,j_2}^{j_3,j_4}\left(\frac{z}{w}\right)
\left(\Phi_{V_J}^{V^{(2)}\,V_K}\right)_{j_4}(w)\left(\Phi_{V_I}^{V^{(1)}\,V_J}
\right)_{j_3}(z)
\end{multline}
with a scalar function $c\left(\tfrac{z}{w}\right)$.
The relation \eqref{E:vocom1} follows from the argument given in chapter 6
of \cite{jm} for a pair of evaluation modules $V^{(1)}_z$ and $V^{(2)}_w$ with
an intertwiner $R_{V^{(1)}V^{(2)}}\bigl(\tfrac{z}{w}\bigr)$.
For quantum affine algebras, the normalisation of the
$R$-matrix governing the analogous commutations relations
has been determined via free boson realizations
\cite{jm} or from the related two-point functions obtained as solutions
of certain $q$-difference equations \cite{idzumi}. A free boson realization
has also been used in \cite{zhang2} to fix the commutation relations of two
$U_q\bigl(\widehat{gl}(2\vert2)\bigr)$-vertex operators related to $W_z$ or $W^*_z$.

The $U'_q\bigl(\widehat{sl}(2\vert2)\bigr)$-module relevant to
the configuration spaces $\Omega_A^{(1)}$ and $\Omega_B^{(1)}$ of the present vertex
model is constructed from the infinite-dimensional, irreducible $U_q\bigl(gl(2
\vert2)\bigr)$-module $V_{\bar{\Lambda}_2+\bar{\Lambda}_4}$ with basis
$\{v_{j,t}\}_{0\leq j\leq3,\,t\in\mathbb{N}_0}$. Its $U_q\bigl(gl(2\vert2)
\bigr)$-weights are
\begin{alignat}{2}\label{E:mod1}
h_1v_{j,t}&=-(t+1-\delta_{j,2})v_{j,t}\qquad\qquad&h_2v_{j,t}&=(t+1-\delta_{j,0})
v_{j,t}\\
h_3v_{j,t}&=(t+1-\delta_{j,2})v_{j,t}\qquad\qquad&h_4v_{j,t}&=-(t-1+\delta_{j,0}
+2\delta_{j,1})v_{j,t}\notag
\end{alignat}
A $\mathbb{Z}_2$-grading is defined by $\vert v_{0,t}\vert=\vert v_{2,t}\vert=1$
and $\vert v_{1,t}\vert=\vert v_{3,t}\vert=0$ $\forall t$. The evaluation
homomorphism found in \cite{gade2} yields a $U'_q\bigl(\widehat{sl}(2\vert2)
\bigr)$-structure given by the action of $h_1,\,h_2, h_3$ specified in
\eqref{E:mod1} and
\begin{equation}\label{E:mod2}
\begin{split}
H^1_mv_{j,t}&=-\tfrac{1}{m}q^{m(t+1-\delta_{j,2})}\bigl[m(t+1-\delta_{j,2})\bigr]
v_{j,t}\\
H^2_mv_{j,t}&=\tfrac{1}{m}q^{m(t+\delta_{j,0}+2\delta_{j,1})}\bigl[m(t+1-
\delta_{j,0})\bigr]v_{j,t}\\
H^3_mv_{j,t}&=-H^1_mv_{j,t}
\end{split}
\end{equation}
for $m\neq0$ and
\begin{alignat}{2}\label{E:mod3}
E^{1,-}_mv_{2,t}&=v_{1,t-1}\qquad&E^{1,+}_mv_{1,t}&=-[t+1]v_{2,t+1}\\
E^{1,-}_mv_{3,t}&=v_{0,t}\qquad&E^{1,+}_mv_{0,t}&=-[t+1]v_{3,t}\notag\\
E^{2,-}_mv_{1,t}&=-q^{m(2t+3)}\tfrac{[t+1]}{[t+2]}v_{0,t+1}\qquad&
E^{2,+}_mv_{0,t+1}&=-q^{m(2t+3)}[t+2]v_{1,t}\notag\\
E^{2,-}_mv_{2,t}&=q^{m(2t+1)}v_{3,t}\qquad&E^{2,+}_mv_{3,t}&=q^{m(2t+1)}
[t+1]v_{2,t}\notag\\
E^{3,-}_mv_{0,t}&=-q^{2m(t+1)}v_{1,t}\qquad&E^{3,+}_mv_{1,t}&=-q^{2m
(t+1)}[t+1]v_{0,t}\notag\\
E^{3,-}_mv_{3,t}&=q^{2m(t+1)}v_{2,t+1}\qquad&E^{3,+}_mv_{2,t+1}&=
q^{2m(t+1)}[t+1]v_{3,t}\notag
\end{alignat}
$\forall m\in\mathbb{Z}$. The basis $\{v_{j,t}\}_{0\leq j\leq 3}$ coincides
with the basis of the module $V$ introduced in \cite{gade1}.
A reducible, infinite-dimensional $U'_q\bigl(\widehat{sl}(2\vert2)
\bigr)$-module $\mathring{V}$ with basis $v_{1,-1}\cup\{v_{j,t}\}_{0\leq j\leq3,\,
t\in\mathbb{N}_0}$
is defined by \eqref{E:mod1}-\eqref{E:mod3} and
\begin{equation}\label{E:vred1}
\begin{split}
h_lv_{1,-1}&=0\qquad l=1,2,3,4\\
H^k_mv_{1,-1}&=0\qquad k=1,2,3,\,m\neq0
\end{split}
\end{equation}
and
\begin{equation}\label{E:vred2}
\begin{split}
E^{1,-}_mv_{1,-1}&=E^{2,+}_mv_{1,-1}=E^{3,\pm}_mv_{1,-1}=0\\
E^{1,+}_mv_{1,-1}&=\frac{[m]}{m}v_{2,0}\\
E^{2,-}_mv_{1,-1}&=q^m\frac{[m]}{m}
v_{0,0}
\end{split}
\end{equation}
$\forall m$. Any action of Drinfeld generators of
$U'_q\bigl(\widehat{sl}(2\vert2)\bigr)$ on $\mathring{V}$ not listed in
\eqref{E:mod1}-\eqref{E:vred2} vanishes. Specifying the action of all
$H^4_m$ on $v_{j,t}$ for one pair $j,t$ 
determines a $U'_q\big(\widehat{gl}(2\vert2)\bigr)$-structure on $\mathring{V}$.  
The appropriate choice will be given below.

For shorter notation, the components
$\left(\Phi_{V_I}^{\mathring{V}\,V_J}\right)_{j,t}(z)$ 
may be written $\Phi_{j,t}(z)$.
The intertwining property \eqref{E:intprop} with $a=e_k$, $k=1,2,3$,
relates all components 
$\left(\Phi_{V_I}^{\mathring{V}\,V_J}\right)_{j,t}(z)$ with $j=0,1,2,3$ and $t\geq0$:
\begin{equation}\label{E:intbos3}
\begin{split}
[t+1]\Phi_{3,t}(z)&=q^{t+1}e_2\Phi_{2,t}(z)+\Phi_{2,t}(z)e_2\\
[t+1]\Phi_{2,t+1}(z)&=-q^{t+1}e_3\Phi_{3,t}(z)+\Phi_{3,t}(z)e_3\\
[t+1]\Phi_{0,t}(z)&=q^{-t-1}e_1\Phi_{3,t}(z)-\Phi_{3,t}(z)e_1\\
[t+1]\Phi_{1,t}(z)&=-q^{-t-1}e_1\Phi_{2,t+1}(z)-\Phi_{2,t+1}(z)e_1
\end{split}
\end{equation}
According to \eqref{E:co1}-\eqref{E:co3}, the intertwining property \eqref{E:intprop}
imposes the following  conditions on the component $\Phi_{2,0}(z)$: 
%\begin{subequations}\label{E:intc1}
\begin{equation}\label{E:intc1a}
\begin{split}
E^{k,-}_m\Phi_{2,0}(z)-(-1)^{\vert\Phi_{2,0}\vert}\Phi_{2,0}(z)E^{k,-}_m&=0\qquad
k=1,2,3\\
E^{3,+}_m\Phi_{2,0}(z)-(-1)^{\vert\Phi_{2,0}\vert}\Phi_{2,0}(z)E^{3,+}_m
&=0
\end{split}
\end{equation}
\begin{equation}\label{E:intc1b}
E^{1,+}_m\Phi_{2,0}(z)-(-1)^{\vert\Phi_{2,0}\vert}\Phi_{2,0}(z)E^{1,+}_m
=-(-1)^{\vert v_{2,0}\vert}q^mz^m\Phi_{1,-1}(z)
\end{equation}
%\end{subequations}
$\forall m$ and
\begin{equation}\label{E:intc2}
\begin{split}
\bigl[h_l,\Phi_{2,0}(z)\bigr]&=-(\delta_{l,2}+\delta_{l,4})\Phi_{2,0}(z)
\qquad l=1,2,3,4\\
w^{-d}\Phi_{2,0}(z)w^d&=\Phi_{2,0}(wz)\\
%\end{split}
%\end{equation}
%\begin{equation}\label{E:intc3}
%\begin{split}
\bigl[H^1_{\pm n},\Phi_{2,0}(z)\bigr]&=\bigl[H^3_{\pm n},\Phi_{2,0}(z)\bigr]=0\\
\bigl[H^2_n,\Phi_{2,0}(z)\bigr]&=-q^{\frac{1}{2}n}\frac{[n]}{n}z^n\Phi_{2,0}(z)\\
\bigl[H^2_{-n},\Phi_{2,0}(z)\bigr]&=-q^{-\frac{3}{2}n}\frac{[n]}{n}z^{-n}
\Phi_{2,0}(z)
\end{split}
\end{equation}
for $n>0$. 
The conditions \eqref{E:intc1a} and \eqref{E:intc2} are fulfilled by
\begin{equation}\label{E:intbos1}
\begin{split}
&\Phi_{2,0}(z)=\\
&:\exp\Biggl\{\beta^1(qz)
-i\varphi^1-\varphi^2-\ln (qz)(\varphi^1_0-i
\varphi^2_0)-i\pi\varphi^3_0+C\bigl(\varphi^1_0-i\varphi^2_0-\varphi^3_0
+i\varphi^4_0\bigr)\\
&\;\;+\sum_{m\neq0}C_m\bigl(\varphi^1_m-i\varphi^2_m-\varphi^3_m+
i\varphi^4_m\bigr)
z^{-m}+\sum_{m\neq0}\frac{1}{[m]}q^{-\frac{1}{2}\vert m\vert-m}\bigl(\varphi^3_m-i
\varphi^4_m\bigr)z^{-m}\Biggr\}:
\end{split}
\end{equation}
and
\begin{multline}\label{E:intbos2}
\Phi_{1,-1}(z)=:\exp\Biggl\{C\bigl(\varphi^1_0-i\varphi^2_0-\varphi^3_0
+i\varphi^4_0\bigr)\\
+\sum_{m\neq0}\bigl(C_m-\tfrac{1}{[m]}q^{-\frac{1}{2}
\vert m\vert-m}\bigr)\bigl(\varphi^1_m-i\varphi^2_m-\varphi^3_m+i\varphi^4_m\bigr)
z^{-m}\Biggr\}:
\end{multline}
for $C,C_{m}\in\mathbb{C}$. Here the free boson realizations \eqref{E:psib2},
\eqref{E:Ebos} and \eqref{E:dbos} for the Drinfeld generators have been employed. The
expressions \eqref{E:intbos1} and \eqref{E:intbos2} satisfy
\begin{equation}\label{E:com1}
\Phi_{2,0}(z)\Phi_{2,0}(w)=-\Phi_{2,0}(w)\Phi_{2,0}(z)
\end{equation}
and
\begin{equation}\label{E:com2}
\Phi_{1,-1}(z)\Phi_{1,-1}(w)=\Phi_{1,-1}(w)\Phi_{1,-1}(z)\qquad
\Phi_{1,-1}(z)\Phi_{2,0}(w)=\Phi_{2,0}(w)\Phi_{1,-1}(z)
\end{equation}
As required by \eqref{E:intprop},
$\Phi_{1,-1}(z)$ commutes with $h_l$, $H^k_{\pm n}$ and $E^{k,\pm}_m$
with $k=1,2,3,\,l=1,2,3,4,\,n>0,\,m\neq0$.
Due to \eqref{E:co3}, the property \eqref{E:intprop} is satisfied 
for $a=H^4_m$ provided that
\begin{equation}\label{E:H4a}
H^4_mv_{2,0}=-\frac{[m]}{m}\Bigl(2q^{-m+\frac{1}{2}\vert m\vert}[m]C_{-m}+1
\Bigr)z^m\,v_{2,0}\qquad m\neq0
\end{equation}
The components $\Phi_{1,-1;m}$ and
$\Phi_{2,0;m}$ defined by $\Phi_{1,-1}(z)=\sum_{m\in\mathbb{Z}}\Phi_{1,-1;m}z^{-m}$
and $\Phi_{2,0}(z)=\sum_{m\in\mathbb{Z}}\Phi_{2,0;m}z^{-m}$
may be applied to the level-one modules
$\mathring{V}(\Lambda_0)$, $\mathring{V}(\Lambda_1+\Lambda_4)$,
$\mathring{V}(2\Lambda_2-\Lambda_3+\Lambda_4)$ or $V\bigl(-s\Lambda_1+(1+s)
\Lambda_2-(1+2s')\Lambda_4\bigr)$ introduced in section
\ref{S:bound2}. Use of the realizations \eqref{E:modb1} and taking into
account the property \eqref{E:vac} reveals that
the components $\Phi_{1,-1;m}$ and $\Phi_{2,0;m}$ map each of these level-one
modules onto itself if
\begin{equation}\label{E:h4cond}
[m]C_m=q^{-\frac{1}{2}\vert m\vert-m}\qquad\mbox{for}\;m<0
\end{equation}
This suggests the existence of intertwiners of the form \eqref{E:vodef1}
with $V=\mathring{V}$ and $V_I=V_J$ given by any of the four level-one
modules. The validity of the entire set of equations \eqref{E:intprop}
remains to be demonstrated.
\begin{conjecture}\label{C:c2}
There exist type II vertex operators
\begin{equation}\label{E:intex1}
\Phi_{{V}_I}^{\mathring{V}\;{V}_I}(z):
{V}_I\longrightarrow\mathring{V}_z\otimes{V}_I\qquad\;\;I=0,1,3,4
\end{equation}
for
${V}_0=\mathring{V}(\Lambda_0)$, ${V}_1=\mathring{V}(\Lambda_1
+\Lambda_4)$, ${V}_3=\mathring{V}(2\Lambda_0-\Lambda_3+\Lambda_4)$ and
$V_4=V\bigl(-s\Lambda_1+(1+s)\Lambda_2-(1+2s')\Lambda_4\bigr)$.
They can be normalised such that
\begin{multline}\label{E:intnorm}
\left(\Phi_{{V}_I}^{\mathring{V}\;{V}_I}
\right)_{1,-1}(z)=:\exp\biggl\{C\bigl(\varphi^1_0-i\varphi^2_0-\varphi^3_0
+i\varphi^4_0\bigr)\\
+\sum_{m>0}\bigl(C_m-\tfrac{1}{[m]}q^{-\frac{1}{2}
\vert m\vert-m}\bigr)\bigl(\varphi^1_m-i\varphi^2_m-\varphi^3_m+i\varphi^4_m\bigr)
z^{-m}\biggr\}:
\end{multline}
with arbitrary complex numbers $C$ and $C_m$ with $m>0$.
A further component is given by
\begin{multline}\label{E:intcomp1}
\left(\Phi_{{V}_I}^{\mathring{V}\;{V}_I}
\right)_{2,0}(z)=\\:\exp\biggl\{\beta^1(qz)-i\varphi^{1,+}(qz)-\varphi^{2,+}(qz)
-i\pi\varphi^3_0+C\bigl(\varphi^1_0-i\varphi^2_0-\varphi^3_0
+i\varphi^4_0\bigr)\\
+\sum_{m>0}\bigl(C_m-\tfrac{1}{[m]}q^{-\frac{1}{2}
\vert m\vert-m}\bigr)\bigl(\varphi^1_m-i\varphi^2_m-\varphi^3_m+i\varphi^4_m\bigr)
z^{-m}\biggr\}:
\end{multline}
All the remaining components $\left(\Phi_{{V}_I}^{\mathring{V}\;
{V}_I}\right)_{j,t}(z)$, $0\leq j\leq3$, $t-\delta_{j,2}\geq0$,
follow from \eqref{E:intbos3} and \eqref{E:intcomp1}.
The vertex operators satisfy
the commutation relation
\begin{equation}\label{E:vocom2}
\Phi_{{V}_I}^{\mathring{V}\;{V}_I}(z)
\Phi_{{V}_I}^{\mathring{V}\;{V}_I}(w)=-
\bar R_{\mathring{V}\mathring{V}}\left(\tfrac{z}{w}\right)\,
\Phi_{{V}_I}^{\mathring{V}\;{V}_I}(w)
\Phi_{{V}_I}^{\mathring{V}\;{V}_I}(z)
\end{equation}
with the $R$-matrix $R_{\mathring{V}\mathring{V}}
\left(\tfrac{z}{w}\right):\mathring{V}_z\otimes
\mathring{V}_w\longrightarrow\mathring{V}_z\otimes\mathring{V}_w$ normalised
by
\begin{equation}\label{E:rnorm}
\Bigl(\bar R_{\mathring V\mathring V}\Bigr)_{2,0;\,2,0}^{2,0;\,2,0}
\left(\tfrac{z}{w}\right)=1
\end{equation}
\end{conjecture}

The normalisation \eqref{E:rnorm} stems from the relation \eqref{E:com1}.
All $R$-matrix elements involving the index pair $1,-1$ are listed by
$\bigl(\bar R_{\mathring V\mathring V}\bigr)_{1,-1;\,1,-1}^{1,-1;\,1,-1}(z)=1$ and
$\bigl(\bar R_{\mathring V\mathring V}\bigr)_{1,-1;\,j,t}^{1,-1;\,j,t}(z)=
\bigl(\bar R_{\mathring V\mathring V}\bigr)_{j,t;\,1,-1}^{j,t;\,1,-1}(z)=1\;
\forall j,t$.

The action of $H^4_m$ on $v_{j,t}$ follows from the values of $C_m$.
With
\begin{equation}\label{Cchoice1}
C=0,\qquad [m]C_m=q^{-\frac{\vert m\vert}{2}-m}\quad\forall m\neq0
\end{equation}
equation \eqref{E:H4a} yields
\begin{alignat}{2}\label{E:H4b}
H^4_mv_{1,-1}&=0&\qquad&\\
H^4_mv_{0,t}&=-\tfrac{1}{m}q^{m(t+1)}[mt]v_{0,t}\qquad&H^4_mv_{1,t}&=
-\tfrac{1}{m}q^{m(t+2)}[m(t+1)]v_{1,t}\notag\\
H^4_mv_{2,t}&=\tfrac{1}{m}\Bigl([m]-q^{m(t+1)}[tm]\Bigr)v_{2,t}\qquad&
H^4_mv_{3,t}&=\tfrac{1}{m}\Bigl([m]-q^{m(t+1)}[tm]\Bigr)v_{3,t}\notag
\end{alignat}
$\forall m\neq0$ and $t\in\mathbb{N}_0$. \eqref{Cchoice1} is a convenient
choice kept for the remainder of this section.

A similar analysis indicates the existence of type II vertex operators
\begin{equation}\label{E:intex2}
\Phi_{{V}_I}^{\mathring{V}^*\;{V}_I}(z):
{V}_I\longrightarrow\mathring{V}^*_z\otimes{V}_I\qquad\;\;I=0,1,3,4
\end{equation}
for the dual module $\mathring{V}^*$. In terms of the dual basis $v^*_{1,-1}\cup
\{v^*_{j,t}\}_{0\leq j\leq3,\,t\in\mathbb{N}_0}$ with
$\langle v^*_{j_1,t_1}\vert v_{j_2,t_2}
\rangle=\delta_{j_1,j_2}\delta_{t_1,t_2}$, the $U'_q\bigl(\widehat{gl}(2\vert2
)\bigr)$-structure on $\mathring{V}^*$ is introduced by
\begin{equation}\label{E:dualmod}
\langle av^*\vert v\rangle=(-1)^{\vert a\vert\cdot\vert v^*\vert}\langle v^*
\vert S(a)v\rangle\qquad\forall a\in U'_q\bigl(\widehat{gl}(2\vert2)\bigr)
\end{equation}
where $S$ denotes the antipode. On $U_q\bigl(gl(2\vert2)\bigr)$, the antipode is
defined by
\begin{equation}\label{E:antip}
S(E^{k,+}_0)=-q^{-h_k}E^{k,+}_0\qquad S(E^{k,-}_0)=-E^{k,-}_0q^{h_k}\qquad
S(h_l)=-h_l
\end{equation}
for $k=1,2,3$ and $l=1,2,3,4$. Its definition on $U_q\bigl(\widehat{gl}(2\vert
2)\bigr)$ can be found in \cite{gade3}.
For two vertex operators \eqref{E:intex2}, the commutation relation reads
\begin{equation}\label{E:vocom3}
\Phi_{{V}_I}^{\mathring{V}^*\;{V}_I}(z)
\Phi_{{V}_I}^{\mathring{V}^*\;{V}_I}(w)=-
\bar R_{\mathring{V}^*\mathring{V}^*}\left(\tfrac{z}{w}\right)\,
\Phi_{{V}_I}^{\mathring{V}^*\;{V}_I}(w)
\Phi_{{V}_I}^{\mathring{V}^*\;{V}_I}(z)
\end{equation}
In the following, the index pairs of the R-matrix elements will be indicated
by $j_i,t_i$ if associated with $\mathring V$ and by $j_i,t_i^*$ if
associated to $\mathring V^*$. Then the subscript specifying $V^{(1)}V^{(2)}$ in
\eqref{E:rel} and \eqref{E:vocom1} can be skipped.
The R-matrix elements of $\bar R_{\mathring V^*\mathring V^*}(z)$  can be obtained
from the matrix elements of $\bar R_{\mathring{V}\mathring{V}}(z)$
by means of the relation
\begin{equation}\label{E:rrel}
{{\bar R}}_{j_1,t_1^*;\,j_2,t_2^*}^{j_3,t_3^*;\,j_4,t_4^*}(z)=
\bar R_{j_3,t_3;\,j_4,t_4}^{j_1,
t_1;\,j_2,t_2}(z)
\end{equation}

Explicite expressions for the matrix elements of $\bar R_{\mathring{V}
\mathring{V}}(z)$ are found in appendix \ref{A:rmatrix}. In order to obtain
well-defined limits for the R-matrix elements in the limit $q\to0$,
the basis of the modules $V$ and $V^*$ needs to be changed slightly. A new basis
$\{\tilde v_{j,t}\}_{0\leq j\leq3}$ or $\{\tilde v^*_{j,t}\}_{0\leq j\leq3}$ is
introduced by
\begin{equation}\label{E:basech}
\tilde v_{j,t}=q^{t(t+\alpha_j)}v_{j,t}\qquad\;\;
\tilde v^*_{j,t}=q^{-t(t+\alpha_j)}v^*_{j,t}
\end{equation}
with $\alpha_j=\tfrac{1}{2}+\delta_{j,1}-\delta_{j,2}$.
Then the expressions collected in
formulae \eqref{E:r1}-\eqref{E:r13} yield
\begin{equation}\label{E:rlim1}
\lim_{q\to0}\tilde R_{j_1,t_1;\,j_2,t_2}^{j_3,t_3;\,j_4,t_4}(z)=
\delta_{j_1,j_4}\delta_{j_2,j_3}\,\delta_{t_1,t_4}\delta_{t_2,t_3}\;
z^{t_2+\theta_{j_2,j_1}}
\end{equation}
Here $\tilde R_{j_1,t_1;\,j_2,t_2}^{j_3,t_3;\,j_4,t_4}(z)$ denotes the
R-matrix elements in the basis $\{\tilde v_{j,t}\}_{0\leq j\leq3}$
as given by \eqref{E:r14} and
$\theta_{j_2,j_1}$ is defined by \eqref{E:theta}.
Changing the basis
according to \eqref{E:basech} does not affect the expression \eqref{E:intcomp1}.
With respect to the basis $\{\tilde v_{j,t}\}_{0\leq j\leq3,t\in\mathbb{N}_0}$
or $\{\tilde v^*_{j,t}\}_{0\leq j\leq3,t\in\mathbb{N}_0}$,
the commutation relations \eqref{E:vocom1} for $V=\mathring{V}$ or $V=\mathring{V}^*$
 with $0\leq j\leq3$, $t\in\mathbb{N}_0$
may be viewed as equations for objects with small $q$ expansions
$\left(\Phi_{{V}_I}^{\mathring{V}\;{V}_I}\right)_{j,t}(z^{-1})
=\phi_{j,t}(z)+O(q)$ or
$\left(\Phi_{{V}_I}^{\mathring{V}^*\;{V}_I}\right)_{j,t}(z^{-1})
=\phi^*_{j,t}(z)+O(q)$.
Then in the limit of vanishing $q$ the commutation relations reads
\begin{equation}\label{E:limcom1}
\begin{split}
\phi_{j_1,t_1}(z)\phi_{j_2,t_2}(w)&=-\left(\frac{z}{w}\right)^{-t_2-
\theta_{j_2,j_1}}\phi_{j_1,t_1}(w)\phi_{j_2,t_2}(z)\\
\phi^*_{j_1,t_1}(z)\phi^*_{j_2,t_2}(w)&=-\left(\frac{z}{w}\right)^{-t_1-
\theta_{j_1,j_2}}\phi^*_{j_1,t_1}(w)\phi^*_{j_2,t_2}(z)
\end{split}
\end{equation}
With
the mode expansions $\phi_{j,t}(z)=\sum_{m\in\mathbb{Z}}\phi_{j,t;\,m}z^{-m}$,
$\phi^*_{j,t}(z)=\sum_{m\in\mathbb{Z}}\phi^*_{j,t;\,m}z^{-m}$, the relations
\eqref{E:limcom1} yield the defining equations \eqref{E:credef1},
\eqref{E:credef2} for the creation algebras given in section \ref{S:crealg}:
\begin{equation}\label{E:limcom2}
\begin{split}
\phi_{j_1,t_1;\,m_1}\,\phi_{j_2,t_2;\,m_2}&=-\phi_{j_1,t_1;\,m_2+t_2
+\theta_{j_2,j_1}}\phi_{j_2,t_2;\,m_1-t_2-\theta_{j_2,j_1}}\\
\phi^*_{j_1,t_1;\,m_1}\,\phi^*_{j_2,t_2;\,m_2}&=-\phi^*_{j_1,t_1;\,m_2+t_1
+\theta_{j_1,j_2}}\phi^*_{j_2,t_2;\,m_1-t_1-\theta_{j_1,j_2}}
\end{split}
\end{equation}
The properties \eqref{E:algw} and \eqref{E:algstarw} follow from
\eqref{E:intc2}. $\Phi^*_{1,-1;0}$  $\bigl(\Phi_{1,-1;0}\bigr)$ can be
viewed as the component corresponding to the unique semi-standard
super tableau in $\mathcal{B}^{h}_{1,1}$  $(\mathcal{B}^v_{1,1})$
associated with the border strip consisting of a single row (column).

A free boson realization for 
$\Phi_{{V}_I}^{\mathring{V}^*\;{V}_I}(z)$ is obtained
employing the coproduct of the antipode of the $U_q\bigl(\widehat{gl}(2\vert2)
\bigr)$-generators. Replacing $q$ by $q^{-1}$ on the rhs of
\eqref{E:hbos} gives a free boson realization of $-q^mS\bigl(H^l_m\bigr)$.
On the lhs of \eqref{E:Xbos}, $X^{1,\pm}(z)$ is replaced by $X^{2,\pm}(z)$
and vice versa. Then substitution of $q$ by $q^{-1}$ on the rhs of 
\eqref{E:Ebos} yields a suitable free boson realization of
$-q^{\pm h_k}S\bigl(E^{k,\pm}(q^{-1}z)\bigr)$. 
Coproduct and antipode of the algebra satisfy
$\Delta S(a)=(S\otimes S)\Delta'(a)$. Due to this property, the coproduct formulae
\eqref{E:co3} give sufficient information to provide a free boson expression
for $\left(\Phi_{{V}_I}^{\mathring{V}^*\;{V}_I}\right)_{2,0}(z)$
analogous to \eqref{E:intcomp1}. 
The result yields the commutation relation
\begin{equation}
\left(\Phi_{{V}_I}^{\mathring{V}^*\;{V}_I}\right)_{2,0}(z)
\left(\Phi_{{V}_I}^{\mathring{V}^*\;{V}_I}\right)_{2,0}(w)=-
\left(\Phi_{{V}_I}^{\mathring{V}^*\;{V}_I}\right)_{2,0}(w)
\left(\Phi_{{V}_I}^{\mathring{V}^*\;{V}_I}\right)_{2,0}(z)
\end{equation}
Together with \eqref{E:rrel}, this leads to \eqref{E:vocom3}.

The relation with the creation algebras
suggests that the particular vertex operators discussed above create
eigenstates of the row-to-row transfer matrix.
\begin{conjecture}\label{C:c3}
The type II vertex operators
$\Phi_{{V}_I}^{\mathring{V}\;{V}_I}(z)$ and
$\Phi_{{V}_I}^{\mathring{V}^*\;{V}_I}(z)$ provide a part
of the quasi-particle structure of the vertex model.
\end{conjecture}
Further details will be given together with a more complete account in a
forthcoming publication.

The limiting behaviour of the R-matrix elements $\tilde R_{j_1,t_1;\,j_2,t_2}^{
j_3,t_3;\,j_4,t_4}(z)$
specified in \eqref{E:rlim1} reflects the eigenvalue structure of the intertwiner
$\check R_{VV}(z)=P^{gr}R_{VV}(z)$.
The decomposition of the tensor product $V\otimes V$ into
irreducible $U_q\bigl(gl(2\vert2)\bigr)$-modules is given by
\begin{equation}\label{E:VVdec}
V\otimes V=V_{2\bar{\Lambda}_2+2\bar{\Lambda}_4}\oplus\Biggl(\,\bigoplus_{n=0}^{
\infty}V_{-(n+1)\bar{\Lambda}_1+(n+2)\bar{\Lambda}_2+(n+1)\bar{\Lambda}_3
+(2-n)\bar{\Lambda}_4}\Biggr)
\end{equation}
Using the explicit expressions for the matrix elements listed in appendix
\ref{A:rmatrix}, the eigenvalues of $\check R_{VV}(z)$ are readily evaluated.
On any vector $\sum_{j_1,j_2,t_1,t_2}\varkappa_{j_1,t_1;j_2,t_2}v_{j_1,
t_1}\otimes v_{j_2,t_2}$ contained in the component 
$V_{-(n+1)\bar{\Lambda}_1+(n+2)\bar{\Lambda}_2+(n+1)\bar{\Lambda}_3
+(2-n)\bar{\Lambda}_4}$, the linear operator $\check R_{VV}(z)$ acts as the scalar
\begin{equation}\label{E:rn}
r_n(z)=\frac{z-q^{2(n+1)}}{1-q^{2(n+1)}z}\cdot\ldots\cdot\frac{z-q^4}{1-q^4z}
\frac{z-q^2}{1-q^2z}
\end{equation}
Here the normalisation is chosen such that $\check R_{VV}(z)$ acts as the unit on
the component $V_{2\bar{\Lambda}_2+2\bar{\Lambda}_4}$. The  $U_q\bigl(gl(
2\vert2)\bigr)$-weight space of the component 
$V_{-(n+1)\bar{\Lambda}_1+(n+2)\bar{\Lambda}_2+(n+1)\bar{\Lambda}_3 +(2-n)\bar{\Lambda}_4}$
is in one-to-one
correspondence with the pairs $(j_1,t_1;j_2,t_2)$ attributed to the power
$z^{n}$ by \eqref{E:rlim1}.
Hence the $U_q\bigl(gl(2\vert2)\bigr)$-weights attributed to all normal forms
$\phi_{j_2,t_2;\,m+n+1}\phi_{j_1,t_1;\,m}$ with $n>0$ and $m$ fixed 
exactly corresponds to the $U_q\bigl(gl(2\vert2)\bigr)$-weights of the
components $V_{2\bar{\Lambda}_2+2\bar{\Lambda}_4}$ and
$V_{-(n'+1)\bar{\Lambda}_1+(n'+2)\bar{\Lambda}_2+(n'+1)\bar{\Lambda}_3 +(2-n')
\bar{\Lambda}_4}$ with $n'<n$. The $U_q\bigl(gl(2\vert2)\bigr)$-weights of
$\phi_{j_r,t_r;\,m+r-1}\ldots\phi_{j_2,t_2;\,t_2}\phi_{j_1,t_1;\,m_1}$ correspond
to the weights of $V_{r\bar{\Lambda}_2+r\bar{\Lambda}_4}$. Generally, a finite
set of infinite-dimensional $U_q\bigl(gl(2\vert2)\bigr)$-modules can be
associated with each infinite vertical or horizontal border strip
 with parameters $r_0$, $(p_1,\ldots,p_{R-1};\bar p_1,\ldots,\bar p_{R})$ such that
the weights found in the modules coincide with the weights of the normal forms
in the set $B(p_1,\ldots,p_{R-1};\bar p_1,\ldots,\bar p_{R};r_0)$.
Analogous statements apply to horizontal border strips.

In contrast,
the infinite configuration space of the vertex model does not seem to indicate
the structure of $V\otimes V^*$ or $V^*\otimes V$ in an analogous
way. Within a certain region of the spectral parameter $z$, the action of the R-matrix
on the tensor product $V\otimes V^*$ is analysed
in the following section.

\section{The mixed case}
\label{S:rmixed}

The description of the space of infinite configurations outlined in sections
\ref{S:bs} and \ref{S:crealg} is achieved by means of two sets
of normal forms. Each set is provided by a creation algebra and applies to the
description of one space of components. The two creation algebras can be taken
independent. As fare as described by the previous section,
the limiting procedure connecting the type II vertex operators to the creation algebras 
treats both pairs $\bigl(\Phi^{V\,V_{J_1}}_{V_{I_1}}(z), \mathcal A\bigr)$ and
$\bigl(\Phi^{V^*\,V_{J_2}}_{V_{I_2}}(w),\mathcal A^*\bigr)$
separately. This raises
the questions whether combining both vertex operators for the description of
the entire configuration space is compatible with a description by two
independent creation algebras.
Assuming that $V_z\otimes V^*_w$ and $V^*_w\otimes V_z$ are intertwined by
$R_{VV^*}\bigl(\tfrac{z}{w}\bigr)$ according to \eqref{E:rdef}, the commutation
relations involving both types of vertex operators can be written in the
form \eqref{E:vocom1} with $V^{(1)}=V$ and $V^{(2)}=V^*$. However, this choice of
evaluation modules does not allow for the introduction of an intertwiner in
the general sense. Though the intertwining conditions
\eqref{E:rdef} can be solved formally for the single R-matrix elements \eqref{E:rel},
the corresponding action of the R-matrix on the $U_q\bigl(gl(2\vert2)\bigr)$-irreducible
components of
$V\otimes V^*$ is well defined only for $\vert\tfrac{q^2w}{z}\vert<1$.
In the following, it will be shown that within this region, 
this action does not depend on the irreducible component in the limit $q\to0$.

In certain cases,
consideration of single matrix elements of $\check R_{V^{(1)}V^{(2)}}(z)
=P^{gr}R_{V^{(1)}V^{(2)}}(z)$ is less
instructive in the case $V^{(1)}\neq V^{(2)}$. 
The normalisation of a vector in an irreducible
component of $V^{(1)}\otimes V^{(2)}$ can be chosen independently of the
normalisation of a vector in the corresponding component of $V^{(2)}\otimes V^{(1)}$.
Choices of the normalisations depend on the purpose, e.g. taking a certain
limit, and is not apparent from a single matrix element.
In fact, it seems hard to extract some formal limit of vanishing $q$ from
the equations \eqref{E:vocom1} with $V^{(1)}=V$ and $V^{(2)}=V^*$.

The decomposition of $V\otimes V^*$
can be found studying the action of the quadratic Casimir $C^{(2)}$ on 
the tensor products.

The operator $C^{(2)}$ given by
\begin{equation}\label{E:cform}
\begin{split}
&\frac{C^{(2)}}{(q-q^{-1})^2}=\\
&-E^{1,-}_0E^{1,+}_0q^{h_2+h_3+h_4}-q^{-2}
E^{2,-}_0E^{2,+}_0q^{-h_1+h_3+h_4}-E^{3,-}_0E^{3,+}_0q^{-h_1-h_2+h_4}\\
&+q^{-1}[E^{1,-}_0,E^{2,-}_0]_{q^{-1}}[E^{1,+}_0,E^{2,+}_0]_qq^{h_3+h_4}-q^{-1}
[E^{2,-}_0,E^{3,-}_0]_{q^{-1}}[E^{2,+}_0,E^{3,+}_0]_qq^{-h_1+h_4}\\
&-\bigl(E^{3,-}_0[E^{1,-}_0,E^{2,-}_0]_{q^{-1}}-q^{-1}[E^{1,-}_0,E^{2,-}_0]_{q^{-1}}
E^{3,-}_0\bigr)\cdot\\
&\qquad\qquad\qquad\qquad\qquad\qquad\qquad\bigl(E^{3,+}_0[E^{1,+}_0,
E^{2,+}_0]_q-q[E^{1,+}_0,E^{2,+}_0]_qE^{3,+}_0\bigr)q^{h_4}\\
&+\frac{[2]}{(q-q^{-1})^2}-\frac{q^{h_4-1}}{(q-q^{-1})^2}\Bigl(q^{h_1+h_2+h_3}
-q^{-h_1+h_2+h_3}+q^{-h_1-h_2+h_3}-q^{-h_1-h_2-h_3}\Bigr)
\end{split}
\end{equation}
is a central element of $U_q\bigl(gl(2\vert2)\bigr)$. Here the $q$-deformed
commutators are defined by
\begin{equation}
\begin{split}
[E^{i,+}_0,E^{i+1,+}_0]_q&=E^{i,+}_0E^{i+1,+}_0+q^{(-1)^{i+1}}E^{i+1,+}_0E^{i,+}_0
\\
[E^{i,-}_0,E^{i+1,-}_0]_{q^{-1}}&=E^{i,-}_0E^{i+1,-}_0+q^{(-1)^i}E^{i+1,-}_0E^{i,-}_0
\end{split}
\end{equation}
Generalised
eigenvectors of $\Delta\bigl(C^{(2)}\bigr)$ with eigenvalue $2y$ are provided by
\begin{alignat}{2}\label{E:ceig1}
&\sum_{t=0}^{\infty}q^{t(2r-1)}\frac{(q^2;q^2)_{t+1}(q^2;q^2)_r}{(q^2;q^2)_{t+r}
(q^2;q^2)_1}\tilde P_t\bigl(y;q^{2r+1},q,q
\vert q^2\bigr)\,v_{2,t+r}\otimes v^*_{0,t}\qquad\;\;&&r\geq0\\
&\sum_{t=0}^{\infty}q^{-t(2r+1)}\frac{(q^2;q^2)_{t+r+1}}{(q^2;q^2)_t(q^2;q^2)_{r+1}}
\tilde P_t\bigl(y;q^{2r+1},q,q
\vert q^2\bigr)\,v_{2,t}\otimes v^*_{0,t+r}\qquad\;\;&&r>0
\end{alignat}
and
\begin{alignat}{2}\label{E:ceig2}
&\sum_{t=0}^{\infty}q^{t(2r-1)}\frac{(q^2;q^2)_{t+1}(q^2;q^2)_r}{(q^2;q^2)_{t+r}(q^2;
q^2)_1}\tilde P_t\bigl(y;q^{-2r-1},q^{-1},
q^{-1}\vert q^{-2}\bigr)\,v^*_{0,t}\otimes v_{2,t+r}\qquad\;\;&&r\geq0\\
&\sum_{t=0}^{\infty}q^{-t(2r+1)}\frac{(q^2;q^2)_{t+r+1}}{(q^2;q^2)_t(q^2;q^2)_{r+1}}
\tilde P_t\bigl(y;q^{-2r-1},q^{-1},
q^{-1}\vert q^{-2}\bigr)\,v^*_{0,t+r}\otimes v_{2,t}\qquad\;\;&&r>0
\end{alignat}
where $r\geq0$, $(q^2;q^2)_n=(1-q^2)(1-q^4)\ldots(1-q^{2n})$ and
$\tilde P_t\bigl(y;a,c,d\vert q^2\bigr)\equiv \tilde P_t(y)$ denote orthonormal
continuous dual Hahn polynomials in base $q^2$ \cite{askwi}.
They satisfy the recurrence relation
\begin{equation}\label{E:recur1}
2y\tilde P_t(y)=a_t\tilde P_{t+1}(y)+b_t\tilde P_t(y)+a_{t-1}\tilde P_{t-1}(y)
\end{equation}
where
\begin{equation}\label{E:recur2}
\begin{split}
a_t&=\sqrt{\bigl(1-q^{2(t+1)}\bigr)\bigl(1-acq^{2t}\bigr)\bigl(1-adq^{2t}
\bigr)\bigl(1-cdq^{2t}\bigr)}\\
b_t&=a+a^{-1}-a^{-1}\bigl(1-acq^{2t}\bigr)\bigl(1-adq^{2t}\bigr)-a\bigl(1-
q^{2t}\bigr)\bigl(1-cdq^{2(t-1)}\bigr)
\end{split}
\end{equation}
and $t\geq -1$, $\tilde P_0(y)=1$ and $\tilde P_{-1}(y)=0$.
Replacing $q$ by $q^{-1}$ in the coefficients \eqref{E:recur2}
yields the recursion relations for the polynomials
$\tilde P_t\bigl(y;a,c,d\vert q^{-2}\bigr)$. They are supplemented by
the initial condition $\tilde P_{-1}\bigl(y;a,c,d\vert q^{-2}\bigr)=0$
and a convenient choice of the value of
$\tilde P_{0}\bigl(y;a,c,d\vert q^{-2}\bigr)$ to be specified below. 
The polynomials $\tilde P_t(y;a,c,d\vert q^2)$ are orthonormal with respect to
the measure $dm(\cdot;a,0,c,d\vert q^2)$. The measure  $dm(\cdot;a,b,c,d\vert q^2)$
is defined by
\begin{equation}\label{E:meas1}
\int_{\mathbb{R}}f(y)dm(y;a,b,c,d\vert q^2)=\int_0^{\pi}f(\cos\theta)w(\cos
\theta)d\theta+\sum_kf(y_k)\,w_k
\end{equation}
with
\begin{multline*}
w(\cos\theta)=w(\cos\theta;a,b,c,d\vert q^2)\\
=\frac{1}{2\pi}\frac{(q^2,ab,ac,ad,bc,
bd,cd;q^2)_{\infty}}{(abcd;q^2)_{\infty}}\frac{(e^{2i\theta},e^{-2i\theta};q^2)_{
\infty}}{(ae^{i\theta},ae^{-i\theta},be^{i\theta},be^{-i\theta},
ce^{i\theta},ce^{-i\theta},de^{i\theta},de^{-i\theta};q^2)_{\infty}}
\end{multline*}
and $y_k=\tfrac{1}{2}(eq^{2k}+e^{-1}q^{-2k})$ for $e$ any of the parameters
$a,b,c,d$. The sum includes all $k\in\mathbb{N}_0$ with the property $\vert eq^{2k}
\vert>1$. For $e=b$, the coefficients are given by
\begin{multline}\label{E:meas1a}
w_k=w_k(b;a,c,d\vert q^2)=\\
\frac{1-b^2q^{4k}}{1-b^2}\frac{(b^{-2},ac,ad,cd;q^2)_{\infty}}{(a/b,c/b,d/b,
abcd;q^2)_{\infty}}\frac{(b^2,ab,bc,bd;q^2)_k}{(q^2,q^2b/a,q^2b/c,q^2b/d;q^2)_k}
\biggl(\frac{q^2}{abcd}\biggr)^k
\end{multline}
Throughout the following, the parameters $a,c,d$ satisfy the condition $max
\bigl(\vert a\vert,\vert c\vert,\vert d\vert\bigr)<1$. Thus the measure
$dm(\cdot;a,0,c,d\vert q^2)$ contains only the continuous part.
In \eqref{E:meas1} and \eqref{E:meas1a}, the notation for the $q$-shifted
factorials is adopted from \cite{gr}:
\begin{alignat}{2}
(a_1,\ldots,a_s;q^2)_{\infty}&=(a_1;q^2)_{\infty}\ldots(a_s;q^2)_{\infty},\qquad
&(a;q^2)_{\infty}&=\prod_{j=0}^{\infty}\bigl(1-aq^{2j}\bigr)\\
(a_1,\ldots,a_s;q^2)_n&=(a_1;q^2)_n\ldots(a_s;q^2)_n,\qquad&(a;q^2)_n&=
\prod_{j=0}^{n-1}\bigl(1-aq^{2j}\bigr)\\
(a_1,\ldots,a_s;q^2)_0&=1&&
\end{alignat}
The measure defined in \eqref{E:meas1} provides an orthogonality
measure for the Askey-Wilson polynomials $p_t(y;a,b,c,d\vert q^2)$
introduced by
\begin{multline}\label{E:askwidef}
p_t(y;a,b,c,d\vert q^2)=\\
a^{-t}(ab,ac,ad;q^2)_t\,{}_4\phi_3\bigl(q^{-2t},
abcdq^{2(t-1)},a\gamma,a\gamma^{-1};ab,ac,ad;q^2,q^2\bigr)
\end{multline}
with $y=\tfrac{1}{2}(\gamma+\gamma^{-1})$.
Here the basic hypergeometric series \cite{gr} is defined by 
\begin{equation}
_{n+1}\phi_n(a_1,a_1,\ldots,a_{n+1};b_1,\ldots,b_n;q^2,\alpha)=\sum_{t=0}^{\infty}
\frac{(a_1,a_2,\ldots,a_{n+1};q^2)_t}{(q^2,b_1,\ldots,b_n;q^2)_t}\alpha^t
\end{equation}
Due to Sears' transformation (\cite{gr}, eq. 2.10.4), the polynomials
$p_t(y;a,b,c,d\vert q^2)$ are symmetric in $a,b,c,d$.
The orthonormal continuous dual $q^2$-Hahn polynomials
and the Askey-Wilson polynomials are related by
\begin{equation}\label{E:hahndef}
\tilde P_t(y;a,b,c,d\vert q^2)=\frac{1}{(q^2,ac;q^2)_t}\,p_t(
y;a,0,c,d\vert q^2)
\end{equation}
The formal intertwining conditions on the R-matrix elements introduced on
$V_z\otimes V^*_w$ or $V_z^*\otimes V_w$ are given by \eqref{E:rdef} and
\eqref{E:rel} with $V^{(1)}=V$, $V^{(2)}=V^*$ or $V^{(1)}=V^*$, $V^{(2)}=V$.
%\begin{equation}\label{E:rdef2}
%\begin{split}
%R_{VV^*}\bigl(\tfrac{z}{w}\bigr)\Delta(a)&=\Delta'(a)R_{VV^*}\bigl(\tfrac{z}{w}
%\bigr)\\
%R_{V^*V}\bigl(\tfrac{w}{z}\bigr)\Delta(a)&=\Delta'(a)R_{V^*V}\bigl(\tfrac{w}{z}
%\bigr)\qquad\forall\,a\in U_q\bigl(\widehat{sl}(2\vert2)\bigr)
%\end{split}
%\end{equation}
%with $\Delta(a)$, $\Delta'(a)$ defined as in \eqref{E:rdef}. 
With the $U_q\bigl(
gl(2\vert2)\bigr)$-weight structure of $V$ and $V^*$ specified by \eqref{E:mod1}
and \eqref{E:dualmod}, these conditions imply
\begin{equation}\label{E:rpart}
\begin{split}
\check R_{VV^*}\bigl(\tfrac{z}{w}\bigr)\,(v_{2,t_1}\otimes v^*_{0,t_2})&=\sum_{t_3,t_4}
\check R_{2,t_1;\,0,t_2^*}^{0,t_4^*;\,2,t_3}\bigl(\tfrac{z}{w}\bigr)\,
v^*_{0,t_4}\otimes v_{2,t_3}\\
\check R_{V^*V}\bigl(\tfrac{w}{z}\bigr)\,(v^*_{0,t_2}\otimes v_{2,t_1})&=\sum_{t_3,t_4}
\check R_{0,t_2^*;\,2,t_1}^{2,t_3;\,0,t_4^*}\bigl(\tfrac{w}{z}\bigr)\,
v_{2,t_3}\otimes v^*_{0,t_4}
\end{split}
\end{equation}
where $t_1-t_2=t_3-t_4$.
Equations
\eqref{E:rprop1} and \eqref{E:relrel} imply
\begin{equation}\label{E:rmatcomp}
\begin{split}
\check R_{0,t_2^*;\,2,t_2+r}^{2,t_1+r;\,0,t_1^*}(z)&=q^{2(t_1-t_2)(2r-1)}\Biggl(
\frac{(q^2;q^2)_{t_2+r}(q^2;q^2)_{t_1+1}}{(q^2;q^2)_{t_1+r}(q^2;q^2)_{t_2+1}}\Biggr)^2
\,\check R_{2,t_1+r;\,0,t_1^*}^{0,t_2^*;\,2,t_2+r}(z)\\
\check R_{0,t_2+r^*;\,2,t_2}^{2,t_1;\,0,t_1+r^*}(z)&=q^{2(t_2-t_1)(2r+1)}\Biggl(
\frac{(q^2;q^2)_{t_1+r+1}(q^2;q^2)_{t_2}}{(q^2;q^2)_{t_2+r+1}(q^2;q^2)_{t_1}}\Biggr)^2
\,\check R_{2,t_1;\,0,t_1+r^*}^{0,t_2+r^*;\,2,t_2}(z)
\end{split}
\end{equation}
The matrix elements on the rhs of \eqref{E:rpart} may be viewed as the $\check R$-matrix
elements associated with two evaluation modules of $U_q\bigl(\widehat{sl}(2)\bigr)$.
A convenient choice of $U_q\bigl(\widehat{sl}(2)\bigr)$-generators is
\begin{equation}\label{E:U2def}
\begin{split}
\hat h_1&=h_2+h_3\\
\hat h_0&=h_0+h_1\\
\hat e_1&=\bigl(E^{3,-}_0E^{2,-}_0+q^{-1}E^{2,-}_0E^{3,-}_0\bigr)q^{h_2+h_3}\\
\hat f_1&=-q^{-h_2-h_3}\bigl(E^{2,+}_0E^{3,+}_0+qE^{3,+}_0E^{2,+}_0\bigr)\\
S(\hat e_0)&=-q^{h_2+h_3}\bigl(
E^{2,+}_0E^{3,+}_{-1}+q^{-1}E^{3,+}_{-1}E^{2,+}_0\bigr)\\
S(\hat f_0)&=
\bigl(E^{3,-}_1E^{2,-}_0+qE^{2,-}_0E^{3,-}_1\bigr)q^{-h_2-h_3}
\end{split}
\end{equation}
The defining relations obeyed by the generators $\hat e_0,\hat e_1,\hat f_0,
\hat f_1,\hat h_0,\hat h_1$ (see \cite{jm}, for example) are inherited from the
defining relations of $U_q\bigl(\widehat{sl}(2\vert2)\bigr)$ \cite{gade1}.
Two infinite-dimensional $U_q\bigl(sl(2)\bigr)$-modules $\hat V$ and $\hat V^*$
with basis $\{\hat v_j\}_{j\in\mathbb N_0}$ and $\{\hat v^*_j\}_{j\in\mathbb N_0}$
are introduced by
\begin{equation}
\hat v_t=q^{-\frac{1}{2}t(2t+1)}\frac{(q^2;q^2)_{t+1}}{(1-q^2)^{t+1}}\,v^*_{0,t}\qquad\;\;
\hat v^*_t=q^{\frac{1}{
2}t(2t-1)}\frac{(1-q^2)^t}{(q^2;q^2)_t}\,v_{2,t}
\end{equation}
Then \eqref{E:U2def} and \eqref{E:mod1}, \eqref{E:mod3} yield 
$U_q\bigl(\widehat{sl}(2)\bigr)$-structures on $\hat V$ and $\hat V^*$. With
these structures, $\{v_{2,t}\otimes z^m\}_{t\in\mathbb{N}_0,m\in\mathbb{Z}}$ and
$\{v^*_{0,t}\otimes w^m\}_{t\in\mathbb{N}_0,m\in\mathbb{Z}}$ can be regarded
as $U_q\bigl(\widehat{sl}(2)\bigr)$-evaluation modules $\hat V_{z'}$ and
$\hat V^*_{w'}$. With the definitions of the $U_q\bigl(\widehat{sl}(2)\bigr)$-structures
on $\hat V_{z'}$ and $\hat V_{w'}$ chosen analogous to \eqref{E:evmod},
the spectral parameters are related by $z'=q^{-3}z^{-1}$, $w'=q^{-3}w^{-1}$.
The $U_q\bigl(\widehat{sl}(2)\bigr)$-structures of $\hat V$ and $\hat V^*$ read
\begin{alignat}{2}
\hat e_0\hat v_t&=q^{t-\frac{3}{2}}[t+1]\,\hat v_{t+1}\qquad&\hat e_0\hat v^*_t
&=-q^{-t-\frac{5}{2}}[t]\hat v^*_{t-1}\\
\hat f_0\hat v_t&=-q^{-t+\frac{5}{2}}[t]\,\hat v_{t-1}\qquad&\hat f_0\hat v^*_t
&=q^{t+\frac{7}{2}}[t+1]\,\hat v^*_{t+1}\notag\\
\hat h_0\hat v_t&=(2t+1)\,\hat v_t\qquad&\hat h_0\hat v^*_t&=-(2t+1)\,\hat v^*_t
\notag\\
\hat e_1\hat v_t&=-q^{-t-\frac{1}{2}}[t]\,\hat v_{t-1}\qquad&\hat e_1\hat v^*_t
&=q^{t+\frac{1}{2}}[t+1]\hat v^*_{t+1}\notag\\
\hat f_1\hat v_t&=q^{t+\frac{3}{2}}[t+1]\,\hat v_{t+1}\qquad&\hat f_1\hat v^*_t
&=-q^{-t+\frac{1}{2}}[t]\hat v^*_{t-1}\notag\\
\hat h_1\hat v_t&=-(2t+1)\,\hat v_t\qquad&\hat h_1\hat v^*_t&=(2t+1)\,\hat v^*_t
\notag
\end{alignat}
Apart from simple factors depending on $r$, the vectors \eqref{E:ceig1},
\eqref{E:ceig2} can be rewritten by
\begin{equation}\label{E:ceig3}
\sum_{t=0}^{\infty}\tilde P_t\bigl(y;q^{2r+1},q,q\vert q^2\bigr)\,\hat v^*_t\otimes
\hat v_{t+r},\qquad\;\;
\sum_{t=0}^{\infty}\tilde P_t\bigl(y;q^{2r+1},q,q\vert q^2\bigr)\,\hat v^*_{t+r}
\otimes \hat v_t
\end{equation}
and
\begin{equation}\label{E:ceig4}
\sum_{t=0}^{\infty}\tilde P_t\bigl(y;q^{-2r-1},q^{-1},q^{-1}\vert q^{-2}\bigr)\,
\hat v_{t+r}\otimes \hat v^*_t,\;\;\;
\sum_{t=0}^{\infty}\tilde P_t\bigl(y;q^{-2r-1},q^{-1},q^{-1}\vert q^{-2}\bigr)\,
\hat v_t\otimes\hat v^*_{t+r}
\end{equation}
They are generalised
eigenvectors of the Casimir element of $U_q\bigl(sl(2)\bigr)$ given by
\begin{equation}
C=(q-q^{-1})^2\hat f_1\hat e_1+q^{\hat h_1+1}+q^{-\hat h_1-1}
\end{equation}
with eigenvalue $2y$.
For fixed $y=\cos\theta$, the vectors \eqref{E:ceig3} with $r=0,1,2,\ldots$
constitute an infinite-dimensional $U_q\bigl(sl(2)\bigr)$-module $\hat V^{(y)}$
corresponding to the principal unitary series representation $\pi^P_{\rho,0}$ of
$U_q\bigl(su(1,1)\bigr)$ with $\cos\theta=\frac{1}{2}\bigl(q^{2i
\rho}+q^{-2i\rho}\bigr)$ \cite{kalnins}, \cite{groene}. Each representation
$\pi^P_{\rho,0}$ is irreducible \cite{burban}. The tensor product $\hat V^*
\otimes \hat V$ decomposes into an integral over of all modules $\hat V^{(\cos\theta)}$ with
$0\leq \theta\leq\pi$ \cite{kalnins}, \cite{groene}.

The vectors in \eqref{E:ceig1} have the $U_q\bigl(gl(2\vert2)\bigr)$-weights
$(-r+1,r+1,r-1,-r+1)$ with $r\in\mathbb{Z}$. Acting with the coproducts of the
$U_q\bigl(gl(2\vert2)\bigr)$-generators on them produces further
generalised eigenvectors of
the quadratic Casimir $C^{(2)}$. Taking into account the defining relations
of $U_q\bigl(gl(2\vert2)\bigr)$ (see \cite{gade1}) and the explicite expression
\eqref{E:cform} for $C^{(2)}$, the set of all resulting linear independent vectors
for fixed
$y$ can be specified. They provide an irreducible $U_q\bigl(gl(2\vert2)
\bigr)$-module $V^{(y)}$. The weights of the vectors composing $V^{(y)}$
are listed by
\begin{equation}\label{E:weights}
\begin{split}
&(-r+1,r+1,r-1,-r+1),\;(-r+1,r-1,r-1,-r+3),\\
&(-r+1,r-3,r-1,-r+1),\;(-r+1,r-1,r-1,-r-1),\\
&(-r+1,r,r-1,-r+2)_2,\;(-r+1,r-2,r-1,-r+2)_2,\\
&(-r+1,r-2,r-1,-r)_2,\;(-r+1,r,r-1,-r)_2,\\
&(-r+1,r-1,r-1,-r+1)_4.
\end{split}
\end{equation}
Here the subscripts denotes the multiplicity of generalised eigenvectors at the given
weight.

A decomposition of the tensor product $V\otimes V^*$ contains all
$U_q\bigl(gl(2\vert2)\bigr)$-modules $V^{(\cos\theta)}$ with $0\leq\theta\leq\pi$
since the eigenvectors related to the first set of weights in
\eqref{E:weights} can be regarded as the set of vectors spanning the $U_q\bigl(
sl(2)\bigr)$-module $\hat V^{(y)}$.
Inspection of $U_q\bigl(gl(2\vert2)\bigr)$-weight structure of $V\otimes V^*$
and the list \eqref{E:weights} shows that no further modules occur in
the decomposition of $V\otimes V^*$.

The intertwining condition \eqref{E:rdef} should apply with the particular
choice $a=C^{(2)}$.
Thus, if the action of $\check R_{VV^*}(z)$ is well defined on
the irreducible components of $V\otimes V^*$, it is expected to take the form
\begin{multline}\label{E:rform1}
\check R_{VV^*}(z)\Biggl(\sum_{t=0}^{\infty}q^{-t(2r+1)}\frac{(q^2;q^2)_{t+r+1}}{
(q^2;q^2)_t(q^2;q^2)_{r+1}}
\tilde P_t\bigl(y;q^{2r+1},
q,q\vert q^2\bigr)v_{2,t}\otimes v^*_{0,t+r}\Biggr)=\\
r(y,z)\sum_{t=0}^{\infty}q^{-t(2r+1)}\frac{(q^2;q^2)_{t+r+1}}{(q^2;q^2)_t(q^2;q^2)_{r+1}}
\tilde P_t\bigl(y;q^{-2r-1},q^{-1},q^{-1}\vert q^{-2}\bigr)
v^*_{0,t+r}\otimes v_{2,t}
\end{multline}
or
\begin{multline}\label{E:rform2}
\check R_{VV^*}(z)\Biggl(\sum_{t=0}^{\infty}q^{t(2r-1)}\frac{(q^2;q^2)_{t+1}(q^2;
q^2)_r}{(q^2;q^2)_{t+r}(q^2;q^2)_1}
\tilde P_t\bigl(y;q^{2r+1},
q,q\vert q^2\bigr)v_{2,t+r}\otimes v^*_{0,t}\Biggr)=\\
r(y,z)\sum_{t=0}^{\infty}q^{t(2r-1)}\frac{(q^2;q^2)_{t+1}(q^2;q^2)_r}{(q^2;q^2)_{t+r}(q^2;q^2)_1}
\tilde P_t\bigl(y;q^{-2r-1},q^{-1},q^{-1}\vert q^{-2}\bigr)
v^*_{0,t}\otimes v_{2,t+r}
\end{multline}
with $\vert y\vert=1$ and
a suitable normalisation of the polynomials $\tilde P_t\bigl(y;q^{-2r-1},q^{-1},
q^{-1}\vert q^{-2}\bigr)$. Within a certain range of spectral parameters $z$,
the function $r(y,z)$ can be expressed as an infinite
sum over elements of $R_{VV}(q^2z^{-1})$ with the continuous dual $q^2$-Hahn polynomials
as coefficients. If $\vert qz^{-1}\vert<1$, the sum is absolutely convergent
and can be evaluated by means of a sum formula established in \cite{groene}
(see Appendix \ref{A:rmode}):
\begin{equation}\label{E:rmode1}
r(\cos\theta,z)=\bigl(1-q^2z^{-1}\bigr)\cdot\frac{\bigl(q^2z^{-1},q^2z^{-1};q^2
\bigr)_{\infty}}{\bigl(qz^{-1}e^{i\theta},qz^{-1}e^{-i\theta};q^2\bigr)_{
\infty}},\qquad\;\;\vert qz^{-1}\vert<1
\end{equation}
where the normalisation $\tilde P_0\bigl(y;q^{-2r-1},q^{-1},
q^{-1}\vert q^{-2}\bigr)=q^r$ has been chosen.
Within the range of validity, the rhs of \eqref{E:rmode1} does not possess
zeros or poles. For $\vert qz^{-1}\vert<1$,
the inverse transformations read:
\begin{multline}\label{E:trans1}
\check R_{2,t_1+r;\,0,t_1^*}^{0,t_2^*;\,2,t_2+r}(z)=
q^{(t_2-t_1)(2r-1)}\frac{(q^2;q^2)_{t_1+r}(q^2;q^2)_{t_2+1}}{(q^2;q^2)_{t_2+r}(q^2;
q^2)_{t_1+1}}\cdot\\
\int^{\pi}_0\tilde P_{t_1}\bigl(
\cos\theta;q^{2r+1},q,q\vert q^2\bigr)\tilde P_{t_2}\bigl(\cos\theta;q^{-2r-1},
q^{-1},q^{-1}\vert q^{-2}\bigr)\,r(\cos\theta,z)\,w(\cos\theta)d\theta
\end{multline}
and
\begin{multline}\label{E:trans2}
\check R_{2,t_1;\,0,t_1+r^*}^{0,t_2+r^*;\,2,t_2}(z)=
q^{(t_1-t_2)(2r+1)}\frac{(q^2;q^2)_{t_2+r+1}(q^2;q^2)_{t_1}}{(q^2;q^2)_{t_1+r+1}(
q^2;q^2)_{t_2}}\cdot\\
\int^{\pi}_0\tilde P_{t_1}\bigl(
\cos\theta;q^{2r+1},q,q\vert q^2\bigr)\tilde P_{t_2}\bigl(\cos\theta;q^{-2r-1},
q^{-1},q^{-1}\vert q^{-2}\bigr)\,r(\cos\theta,z)\,w(\cos\theta)d\theta
\end{multline}
with 
$r\geq0$, $r(\cos\theta,z)$ given by \eqref{E:rmode1} 
and $w(\cos\theta)=w\bigl(\cos\theta;q^{2r+1},0,q,q\vert q^2)$ defined by
\eqref{E:meas1}. The last two factors in \eqref{E:trans1} and \eqref{E:trans2} 
may be rewritten as
\begin{multline}\label{E:chmeas1}
r(\cos\theta,z)w\bigl(\cos\theta;q^{2r+1},0,q,q\vert q^2)=\\
\bigl(1-q^2z^{-1}\bigr)
(q^2z^{-1},q^2z^{-1};q^2)_{\infty}w(\cos\theta;q^{2r+1},qz^{-1},q,q\vert q^2)
\end{multline}
According to \eqref{E:meas1}, the function $w(\cos\theta;q^{2r+1},qz^{-1},q,
q\vert q^2)$
enters the orthogonality measure for the Askey-Wilson polynomials $p_t\bigl(
\cos\theta;q^{2r+1},qz^{-1},q,q\vert q^2\bigr)$. For $\vert qz^{-1}\vert<1$,
the orthogonality relation reads (\cite{gr}, eqs. 7.5.15-7.5.17)
\begin{multline}\label{E:orthog}
\int^{\pi}_0p_t\bigl(\cos\theta;q^{2r+1},qz^{-1},q,q\vert q^2\bigr)
p_s\bigl(\cos\theta;q^{2r+1},qz^{-1},q,q\vert q^2\bigr)\,
w_z(\cos\theta)d\theta=\\
\frac{\delta_{t,s}}{(q^2z^{-1},q^2z^{-1};q^2)_{\infty}h_t(q^{2r+1},qz^{-1},
q,q\vert q^2)}
\end{multline}
where $w_z(\cos\theta)=w\bigl(\cos\theta;q^{2r+1},qz^{-1},q,q\vert q^2\bigr)$
and
\begin{equation}
h_t(q^{2r+1},qz^{-1},q,q\vert q^2)=
\frac{1-q^{2(2t+r+1)}z^{-1}}{(q^2,q^2,q^{2(r+1)},q^{2(r+1)},
q^2z^{-1},q^2z^{-1};q^2)_t}
\end{equation}
The continuous dual Hahn polynomials $\tilde P_t(y;a,0,c,d\vert q^2)$ and
$\tilde P_t(y;a^{-1},0,c^{-1},d^{-1}\vert q^{-2})$
may be expressed in terms of the polynomials
$p_t\bigl(y;q^{2r+1},qz^{-1},q,q\vert q^2)$ making use of \eqref{E:hahndef}
and the connection
coefficients between Askey-Wilson polynomials established in \cite{askwi}
(see also section 7.6 in \cite{gr}). The required relation is a special case of
eqs. 7.6.2 and 7.6.3 in \cite{gr}:
\begin{equation}\label{E:con1}
\tilde P_t(y;a,c,d\vert q^2)=\sum_{s=0}^tc_{s,t}\,p_s(y;
a,b,c,d)
\end{equation}
with
\begin{equation}\label{E:con2}
c_{s,t}=b^{t-s}\frac{(cd;q^2)_t(acq^{2s};q^2)_{t-s}}{(q^2,ad,cd,abcdq^{2(s-1)};
q^2)_s(q^2,abcdq^{4s};q^2)_{t-s}}
\end{equation} 
Here the parameters $a,b,c,d$ are only restricted by the requirement that the
denominators in the polynomials and coefficients never vanish. With $a=q^{2r+1}$,
$c=d=q$ and $b=qz^{-1}$, the relation reads
\begin{equation}\label{E:con3}
\tilde P_t\bigl(y;q^{2r+1},qz^{-1},q,q\vert q^2)=\sum_{s=0}^tc_{s,t}(q,z)
\,p_s\bigl(y;q^{2r+1},qz^{-1},q,q\vert q^2\bigr)
\end{equation}
with
\begin{equation}\label{E:con4}
c_{s,t}(q,z)=\Bigl(\frac{q}{z}\Bigr)^{t-s}\frac{(q^2;q^2)_t\bigl(q^{2(r+s+1)};
q^2\bigr)_{t-s}}{\bigl(q^{2(r+1)},q^2,q^2,q^{2(r+s+1)}z^{-1};q^2\bigr)_s\bigl(
q^2,q^{2(r+2s+2)}z^{-1};q^2\bigr)_{t-s}}
\end{equation}
Replacing $q$ by $q^{-1}$ and $z$ by $z^{-1}$ in \eqref{E:con3}, \eqref{E:con4}
and changing from base $q^{-2}$ to base $q^2$ in
$p_s(y;q^{-2r-1},q^{-1}z,q^{-1},q^{-1}\vert q^{-2})$ yields
\begin{equation}\label{E:con5}
\frac{\tilde P_t\bigl(y;q^{-2r-1},q^{-1},q^{-1}\vert q^{-2}\bigr)}{
\tilde P_0\bigl(y;q^{-2r-1},q^{-1},q^{-1}\vert q^{-2}\bigr)}=
\sum_{s=0}^tc'_{s,t}(q,z)\,p_s\bigl(y;q^{2r+1},qz^{-1},q,q\vert q^2\bigr)
\end{equation}
with
\begin{equation}\label{E:con6}
c'_{s,t}(q,z)=z^sq^{-s(3s+2r+1)}(-1)^s\,c_{s,t}(q^{-1},z^{-1})
\end{equation}
Here the normalisation of the polynomials $\tilde P_t\bigl(y;q^{-2r-1},
q^{-1},q^{-1}\vert q^{-2}\bigr)$ is left arbitrary. Insertion of \eqref{E:orthog},
\eqref{E:con3} and \eqref{E:con5} into equation \eqref{E:trans2} allows to
express the R-matrix elements in terms of connection coefficients and
the function specifying the corresponding orthogonality relation:
\begin{multline}\label{E:rcon}
\check R_{2,t_1;\,0,t_1+r^*}^{0,t_2+r^*;\,2,t_2}(z)=q^{(t_1-t_2)(2r+1)}\frac{(q^2;
q^2)_{t_2+r+1}(q^2;q^2)_{t_1}}{(q^2;q^2)_{t_1+r+1}(q^2;q^2)_{t_2}}\tilde P_0\bigl(
y;q^{-2r-1},q^{-1},q^{-1}\vert q^{-2}\bigr)\cdot\\
(1-q^2z^{-1})\sum_{t=0}^{min(t_1,t_2)}c_{t,t_1}(q,z)c'_{t,t_2}(q,z)\frac{1}{
h_t(q^{2r+1},qz^{-1},q,q\vert q^2)}
\end{multline}
Equation \eqref{E:rcon} is valid for all values of $z$ provided that the R-matrix
element on the lhs is well defined. The orthogonality relations \eqref{E:orthog}
take a different form for $\vert qz^{-1}\vert>1$ (see chapter 6 of \cite{gr}).
Therefore the expression \eqref{E:rmode1} does not give rise to inverse
transformations of the form
\eqref{E:trans1} or \eqref{E:trans2} if the condition $\vert qz^{-1}\vert<1$ is
not satisfied.

Within the region of validity, the expression \eqref{E:rmode1} for the action
of $R_{VV^*}(z)$ has a well-defined limit as $q$ tends to zero. The latter
does not show a structure which could distinguish the
different irreducible components in the limit $q\to0$.
Hence in this region the action does not indicate an inconsistency in
the quasi-particle description of the space of infinite configurations in $\Omega^{(1)}_A$
in terms of the two type II vertex operators $\Phi^{\mathring V\,V_{I}}_{V_{I}}(z)$ and
$\Phi^{\mathring V^*\,V_{J}}_{V_{J}}(w)$.

In contrast, the action of $\check R_{V^*V}(z)$ on the irreducible components
\eqref{E:ceig2}
is well-defined for all $z\neq q^{2s}$, $s\in\mathbb Z$. The eigenvalue is finite
at $z=q^{2s-1}e^{\pm i\theta}$ for all $s\in\mathbb Z$. Details will be given
in a forthcoming publication.

\appendix

\section{The mapping for vertical border stripes}
\label{A:map}

A generator $\phi_{j,t;r}$ with suitable $j,t$ is assigned to the $r$-th
box of a vertical border strip if this box carries the number $0$ in the
reference labelling defined in section \ref{S:bs}. Hence, the normal forms
corresponding to a border strip with parameters $r_0$ and $(p_1,\ldots,p_{R-1};
\bar p_1,\ldots,p_R)$ or $(\emptyset;\bar p_1)$ can be written
\begin{equation}\label{E:nform3}
\begin{split}
&\phi_{j_{\bar p_1},t_{\bar p_1};\,s_1+\bar p_1-1}\ldots\phi_{j_2,t_2;\,s_1+1}
\phi_{j_1,t_1;\,s_1}\\
&\cdot\phi_{j_{\bar p_1+\bar p_2},t_{\bar p_1+\bar p_2};\,s_2+\bar p_2-1}\ldots
\phi_{j_{\bar p_1+2},
t_{\bar p_1+2};s_2+1}\phi_{j_{\bar p_1+1},t_{\bar p_1+1};\,s_2}\\
&\cdot\phi_{j_{\bar p_1+\bar p_2+\bar p_3},t_{\bar p_1+\bar p_2+\bar p_3};\,s_3+
\bar p_3-1}\ldots
\phi_{j_{\bar p_1+\bar p_2+2},t_{\bar p_1+\bar p_2+2};\,s_3+1}\phi_{j_{\bar p_1+
\bar p_2+1},t_{\bar p_1+\bar p_2+1};,s_3}\\
&\ldots\\
&\cdot\phi_{j_{\bar N},t_{\bar N};s_R+\bar p_R-1}\ldots\phi_{j_{\bar p_1
+\ldots+\bar p_{R-1}+2},t_{\bar p_1+\ldots+\bar p_{R-1}+2};s_R+1}
\phi_{j_{\bar p_1+\ldots+\bar p_{R-1}+1},t_{\bar p_1+\ldots+\bar p_{R-1}+1};s_R}
\end{split}
\end{equation}
where $\bar N=\sum_{i=1}^R\bar p_i$ and
the numbers $s_i$ are defined in  \eqref{E:sdef}.
Prescriptions for the indices $j,t$ are obtained from modifications
of figures \ref{FF:map1}-\ref{FF:map3}.
Each section of a border strip drawn in figure \ref{FF:map1} is transposed such
that the uppermost box of the column becomes the leftmost box of a row.
In figures \ref{FF:map2} and \ref{FF:map3},
each section is transposed such that the leftmost box of the row becomes the
uppermost box of a column.
Throughout the figures, the $p_i$ is replaced by $\bar p_i$
for all $i=1,2,\ldots R$ and $\bar p_i$ by $p_i$ for $i=1,2,\ldots,R-1$.
In the right part of the figures, each generator $\phi^*_{j,t;m}$ is replaced
by $\phi_{j,t;m}$. Then in figure \ref{FF:map1}
the order of the pairs $j_i,t_i$ in the product is reversed. The modes $m_i$
are kept unchanged. For example,
the lowest part of figure \ref{FF:map1} is changed to

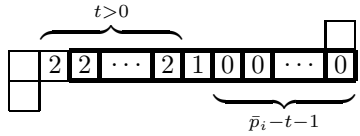
\begin{figure}[h]
\begin{center}
\setlength{\unitlength}{1cm}
\begin{picture}(16,2)
\put(0.4,1.2){\line(1,0){0.8}}
\put(0.4,0.8){\line(1,0){0.8}}
\multiput(0.4,0.8)(0.4,0){4}{\line(0,1){0.4}}
\multiput(2.3,0.8)(0.4,0){5}{\line(0,1){0.4}}
\put(0.4,0.4){\line(0,1){0.4}}
\put(0.8,0.4){\line(0,1){0.4}}
\put(0.4,0.4){\line(1,0){0.4}}
\put(1.75,1){$\ldots$}
\put(4.05,1){$\ldots$}
\put(0.92,0.9){$2$}
\put(1.32,0.9){$2$}
\put(2.42,0.9){$2$}
\put(2.82,0.9){$1$}
\put(3.22,0.9){$0$}
\put(3.62,0.9){$0$}
\put(4.72,0.9){$0$}
\put(0.82,1.1){$\overbrace{\phantom{3333333333\,\,}}^{t>0}$}
\put(3.12,0.65){$\underbrace{\phantom{3333333333\,\,}}_{\bar p_{i}-t-1}$}
\put(4.6,1.2){\line(0,1){0.4}}
\put(5,1.2){\line(0,1){0.4}}
\put(4.6,1.6){\line(1,0){0.4}}

\linethickness{1.5pt}
\multiput(1.2,0.8)(0.4,0){2}{\line(0,1){0.4}}
\multiput(2.3,0.8)(0.4,0){5}{\line(0,1){0.4}}
\put(1.2,1.2){\line(1,0){3.8}}
\put(1.2,0.8){\line(1,0){3.8}}
\put(4.6,0.8){\line(0,1){0.4}}
\put(5,0.8){\line(0,1){0.4}}
\end{picture}\par
\end{center}
\end{figure}

\begin{equation*}
\longleftrightarrow\begin{cases}
\bar p_i=t+1:&\phi_{0,0;s_i+\bar p_i-1}\ldots\phi_{0,0;s_i+2}
\phi_{0,0;s_i+1}\\
\bar p_i>t+1:&\phi_{0,0;s_i+\bar p_i-1}\ldots\phi_{0,0;s_i+\bar p_i-t+1}
\phi_{0,0;s_i+\bar p_i-t}\\
&\qquad\quad
\cdot\phi_{2,0;s_i+\bar p_i-t-1}\ldots\phi_{2,0;s_i+2}
\phi_{2,0;s_i+1}
\end{cases}
\end{equation*}

\vspace{0.3cm}

The above changes transform figures \ref{FF:map1}, \ref{FF:map2} and \ref{FF:map3}
to figures \ref{FF:map1}', \ref{FF:map2}' and \ref{FF:map3}', respectively.
Then the factors in the $i$-th
line in \eqref{E:nform2} are related to the numbers given to the boxes in the
$i$-th row and the left neighbouring column by the following three steps.

\begin{enumerate}

\item For $1\leq i\leq R$ and $\bar p_i\geq2$, the factors
\begin{multline}
\qquad\;\;\phi_{j_{\bar p_1+\ldots+\bar p_i-1},t_{\bar p_1+\ldots+\bar p_i-1};
s_i+\bar p_i-2}\ldots
\phi_{j_{\bar p_1+\ldots+\bar p_{i-1}+2},t_{\bar p_1+\ldots+\bar p_{i-1}+2};s_i+1}\notag\\
\cdot\phi_{j_{\bar p_1+\ldots+\bar p_{i-1}+1},t_{\bar p_1+\ldots+\bar p_{i-1}+1};s_i}
\end{multline}
are determined according to the numbers attributed to the $\bar p_i$ rightmost
boxes of the $i$-th row. Figure \ref{FF:map1}' specifies these factors 
for all cases admitted by the rules \eqref{E:label1} and \eqref{E:label2}
for the tableaux of vertical strips.
\label{R:step4}

\item To obtain the factor $\phi_{j_{\bar p_1+\ldots+\bar p_{i}},t_{\bar p_1+\ldots+
\bar p_{i}};\,s_i+\bar p_i}$  
with $i<R$ and $\bar p_i\geq1$, the $i$-th row and the $(i+1)$-th finite column with length
$p_{i+1}+1$ is taken into account. Figure \ref{FF:map2}' gives all possible
cases  together with the associated factors.
\label{R:step5}

\item Depending on the numbers given to the boxes of the left infinite column,
the leftmost factor $\phi_{j_{\bar N},t_{\bar N};\,s_R+\bar p_R-1}$ is determined.
All cases are found in figure \ref{FF:map3}'. 
\label{R:step6}

\end{enumerate}

For a semi-standard super tableau of the border strip $(\emptyset,\bar p_1)$ with
$\bar p_1>1$, the normal form \eqref{E:phiset1} is determined by the second
and third step. In the case $(\emptyset;1)$, only the last step is needed.
The set of all normal forms \eqref{E:nform3}
may be called $B(p_1,\ldots,p_{R-1};\bar p_1,\ldots,\bar p_R;r_0)$ if $R>1$
and $B(\emptyset;\bar p_1)$ for $R=1$.
Steps \ref{R:step4}-\ref{R:step6} describe the one-to-one correspondence
claimed in Result \ref{RR:r2}.

\section{The R-matrix $\check R_{VV}(z)$}
\label{A:rmatrix}

The intertwiner $\check R:V_{z}\otimes V_{w}\longrightarrow V_{w}\otimes
V_{z}$ satisfies the conditions
\begin{equation}\label{E:intc}
\check R\Delta(a)=\Delta(a)\check R\qquad\;\;\forall a\in U_q\bigl(\widehat{gl}(2
\vert2)\bigr).
\end{equation}
This set of linear equations has a solution
\begin{equation}
\check R_{VV}\bigl(\tfrac{z}{w}\bigr)=P^{gr} R_{VV}\bigl(\tfrac{z}{w}\bigr)
\end{equation}
where $P^{gr}$ is the graded transposition defined in
section \ref{S:vo} and $R_{VV}\bigl(\tfrac{z}{w}\bigr)$ denotes the R-matrix
in the first line of \eqref{E:vocom1}. 
In the following, the nonvanishing
matrix elements $\check R_{j_1,t_1;\,j_2,t_2}^{j_3,t_3;\,j_4,t_4}(z)$
of $\check R_{\mathring{V}\mathring{V}}(z)$ with $t\geq0$
and the basis of $\mathring{V}$ specified by \eqref{E:mod2}, \eqref{E:mod3}
are collected.
They are related by
\begin{multline}\label{E:rprop1}
\check R_{j_1,t_1;\,j_2,t_2}^{j_3,t_3;\,j_4,t_4}(z)=
q^{\frac{1}{8}(1-(-1)^{j_3})(1-(-1)^{j_4})-\frac{1}{8}
(1-(-1)^{j_1})(1-(-1)^{j_2})}\\
\cdot q^{-(t_1+\frac{1}{2}\delta_{j_1,1}+\frac{1}{2}\delta_{j_1,3})^2
-(t_2+\frac{1}{2}\delta_{j_2,1}+\frac{1}{2}\delta_{j_2,3})^2
+(t_3+\frac{1}{2}\delta_{j_3,1}+\frac{1}{2}\delta_{j_3,3})^2
+(t_4+\frac{1}{2}\delta_{j_4,1}+\frac{1}{2}\delta_{j_4,3})^2}
\\
\cdot\left(\frac{[t_1]![t_2]!}{[t_3]![t_4]!}\right)^2\frac{[t_1+1]^{2-2\delta_{
j_1,2}+\delta_{j_1,1}-\delta_{j_1,3}}[t_2+1]^{2-2\delta_{j_2,2}+\delta_{j_2,1}
-\delta_{j_2,3}}}
{[t_3+1]^{2-2\delta_{j_3,2}+\delta_{j_3,
1}-\delta_{j_3,3}}[t_4+1]^{2-2\delta_{j_4,2}+\delta_{j_4,1}-\delta_{j_4,3}}}
\check R_{j_3,t_3;\,j_4,t_4}^{j_1,t_1;\,j_2,t_2}(z)
\end{multline}
and
\begin{multline}\label{E:rprop2}
\check R_{j_1,t_1;\,j_2,t_2}^{j_3,t_3;\,j_4,t_4}(z)=\\
z^{t_4-t_1+\delta_{j_4,1}-\delta_{j_1,1}}\frac{[t_1+1]^{\delta_{j_1,0}-\delta_{
j_1,2}}[t_2+1]^{\delta_{j_2,0}-\delta_{j_2,2}}}{[t_3+1]^{\delta_{j_3,0}-\delta_{
j_3,2}}[t_4+1]^{\delta_{j_4,0}-\delta_{j_4,2}}}\check R_{\sigma(j_2),t_2;\,\sigma
(j_1),t_1}^{\sigma(j_4),t_4;\,\sigma(j_3),t_3}(z)
\end{multline}
with $\sigma(0)=2$, $\sigma(2)=0$, $\sigma(1)=1$, $\sigma(3)=3$ and
\begin{multline}\label{E:rprop3}
\check R_{j_1,t_1;\,j_2,t_2}^{j_3,t_3;\,j_4,t_4}(z)=\\z^{t_4-t_1-\delta_{j_4,2}
+\delta_{j_1,2}}\frac{[t_1+1]^{\delta_{j_1,1}-\delta_{j_1,3}}
[t_2+1]^{\delta_{j_2,1}-\delta_{j_2,3}}}{[t_3+1]^{\delta_{j_3,1}
-\delta_{j_3,3}}[t_4+1]^{\delta_{j_4,1}-\delta_{j_4,3}}}
\check R_{\tau(j_2),t_2;\,\tau(j_1),t_1}^{\tau(j_4),t_4;\,\tau(j_3),t_3}(z)
\end{multline}
with $\tau(0)=0$, $\tau(2)=2$, $\tau(1)=3$ and $\tau(3)=1$.
Here the notation $[t]!=[t][t-1]\ldots[1]$ is used.

The matrix elements $\check R_{j,t_1;\,j,t_2}^{j,t_3;\,j,t_4}(z)$ are
nonvanishing for $t_1+t_2=t_3+t_4$. The solution of the intertwining condition
\eqref{E:intc} is unique up to a scalar.
  Normalising $\check R_{\mathring{V}\mathring{V}}(z)$ by
\begin{equation}\label{E:rnorm2}
\check R_{2,0;\,2,0}^{2,0;\,2,0}(z)=1,
\end{equation}
explicit solution of the intertwining condition \eqref{E:intc} gives
\begin{equation}\label{E:r1}
\check R_{2,t_1;\,2,t_2}^{2,0;\,2,t_1+t_2}(z)=z^{t_2}q^{t_1(2t_2+1)}
(q^2;q^2)_{t_2}\frac{z-1}{q^2-1}
\prod_{r=0}^{t_2}\frac{q^2-1}{q^{2(t_1+r)}z-1}
\end{equation}
and
\begin{equation}\label{E:r2}
\begin{split}
&\check R_{2,t_1;\,2,t_2}^{2,s;\,2,t_1+t_2-s}(z)=\frac{1}{(q^2;q^2)_s}
\frac{q^{2t_1}z-1}{z-q^{2(t_1
-s)}}q^{(t_1-s)(2t_2-2s+1)-t_1(2t_2+1)}\bar R_{2,t_1;\,2,t_2}^{2,0;\,2,t_1+t_2}(z)\\
&\cdot\sum_{r=0}^sz^{r-s}q^{(s-r)(s-r+1)}
\frac{1}{(q^2;q^2)_{r}(q^2;q^2)_{s-r}}\frac{(q^2;q^2)_{t_2+r}}{(q^2;q^2)_{
t_2-s+r}}\left(\frac{(q^2;q^2)_{t_1}}{(q^2;q^2)_{t_1-r}}\right)^2\\
&\cdot\prod_{r_1=0}^{s-r}\bigl(z-q^{2(t_1-r-r_1)}\bigr)\prod_{r_2=0}^r
\frac{1}{q^{2(t_1-r_2)}z-1}
\end{split}
\end{equation}
for $t_1\geq s>0$ and $t_2-s\geq0$.
In \eqref{E:r1}, \eqref{E:r2} and the remainder, 
the notation $(a;q^2)_n=\bigl(1-q^{2(n-1)}a\bigr)\ldots(1-q^2a)(1-a)$ is used.
The remaining elements $\check R_{2,t_1;
\,2,t_2}^{2,s;\,2,t_1+t_2-s}(z)$ are obtained from the relations
\begin{equation}\label{E:rel1}
\check R_{2,t_1;\,2,t_2}^{2,t_3;\,2,t_4}(z)=z^{t_4-t_1}\check R_{2,t_2;\,2,t_1}^{2,t_4;\,
2,t_3}(z)
\end{equation}
and
\begin{equation}\label{E:rel2}
\check R_{2,t_1;\,2,t_2}^{2,t_3;\,2,t_4}(z)=q^{-2(t_1^2+t_2^2-t_3^2
-t_4^2)}\left(\frac{(q^2;q^2)_{t_1}(q^2;q^2)_{t_2}}{(q^2;q^2)_{t_3}(q^2;q^2)_{t_4}}
\right)^2\check R_{2,t_3;\,2,t_4}^{2,t_1;\,2,t_2}(z)
\end{equation}
which are special cases of \eqref{E:rprop1}-\eqref{E:rprop3}.
All matrix elements $\check R_{j,t_1;\,j,t_2}^{j,t_3;\,j,t_4}(z)$ can be expressed
in terms of $\check R_{2,t_1;\,2,t_2}^{2,t_3;\,2,t_4}(z)$:
\begin{multline}\label{E:r3}
\check R_{3,t_1;\,3,t_2}^{3,s;\,3,t_1+t_2-s}(z)=
\frac{1}{(1-z)\bigl(q^{2(t_2+1)}-1\bigr)}\cdot\\
\shoveleft{
\biggl\{q^{s-t_1}\bigl(q^{2(s-t_1)}-1\bigr)\bigl(q^{2(t_1+t_2-s+1)}-1\bigr)
\check R_{2,t_1;\,2,t_2+1}^{2,s;\,2,t_1+t_2-s+1}(z)}\\
+q^{t_1+2t_2-3s-1}\bigl(z-
q^{2(s-t_1+1)}\bigr)\bigl(q^{2(s+1)}-1\bigr)\check R_{2,t_1;\,2,t_2+1}^{2,
s+1;\,2,t_1+t_2-s}(z)\biggr\}
\end{multline}

\begin{equation}\label{E:rel3}
\begin{split}
\check R_{0,t_1;\,0,t_2}^{0,t_3;\,0,t_4}(z)&=q^{t_3+t_4-t_1-t_2}\frac{\bigl(
q^{2(t_1+1)}-1\bigr)\bigl(q^{2(t_2+1)}-1\bigr)}{\bigl(q^{2(t_3+1)}-1\bigr)\bigl(
q^{2(t_4+1)}-1\bigr)}\check R_{2,t_1;\,2,t_2}^{2,t_3;\,2,t_4}(z)\\
\check R_{1,t_1;\,1,t_2}^{1,t_3;\,1,t_4}(z)&=q^{t_3+t_4-t_1-t_2}\frac{\bigl(
q^{2(t_1+1)}-1\bigr)\bigl(q^{2(t_2+1)}-1\bigr)}{\bigl(q^{2(t_3+1)}-1\bigr)\bigl(
q^{2(t_4+1)}-1\bigr)}\check R_{3,t_1;\,3,t_2}^{3,t_3;\,3,t_4}(z)
\end{split}
\end{equation}
The R-matrix elements $\check R_{j_1,t_1;\,j_2,t_2}^{j_3,t_3;\,j_4,t_4}(z)$
with $j_1=j_3\neq j_2=j_4$ or $j_1=j_4\neq j_2=j_3$ are given by
\begin{multline}\label{E:r4}
\check R_{2,t_1;\,3,t_2}^{2,s;\,3,t_1+t_2-s}(z)=\frac{q^{t_1+1}}{q^{2(
t_1+t_2+2)}-1}\\
\cdot\Bigl\{q^{t_1+1}\bigl(q^{2(t_2+1)}-1\bigr)\check R_{2,t_1;\,2,t_2}^{2,s;\,
2,t_1+t_2-s}(z)+q^{-s-1}\bigl(q^{2(s+1)}-1\bigr)
\check R_{3,t_1;\,3,t_2}^{3,s;\,3,t_1+t_2-s}(z)\Bigr\}
\end{multline}

\begin{multline}\label{E:r5}
\check R_{3,t_1;\,2,t_2}^{3,s;\,2,t_1+t_2-s}(z)=\frac{q^{s+1}}{q^{2(t_1+t_2+2)}-1}\\
\cdot
\Bigl\{q^{-t_1-1}\bigl(q^{2(t_1+1)}-1\bigr)\check R_{2,t_1;\,2,t_2}^{2,s_1;
\,2,s_2}(z)+q^{s+1}\bigl(q^{2(t_1+t_2-s+1)}-1\bigr)
\check R_{3,t_1;\,3,t_2}^{3,s;\,3,t_1+t_2-s}(z)\Bigr\}
\end{multline}

\begin{multline}\label{E:r6}
\check R_{2,t_1;\,3,t_2}^{3,s;\,2,t_1+t_2-s}(z)=\frac{q^{s-t_2
+t_1+1}}{q^{2(t_1+t_2+2)}-1}\\
\cdot\Bigl\{\bigl(q^{2(t_2+1)}-1\bigr)\check R_{2,t_1;\,2,t_2}^{2,
s;\,2,t_1+t_2-s}(z)-q^{s-t_1}\big(q^{2(t_1+t_2-s+1)}-1\bigr)
\check R_{3,t_1;\,3,t_2}^{3,s;\,3,t_1+t_2-s}(z)\Bigr\}
\end{multline}

\begin{equation}\label{E:rel4}
\check R_{3,t_1;\,2,t_2}^{2,s;\,3,t_1+t_2-s}(z)=\Bigl(\frac{z}{q}\Bigr)^{t_2-s}
\check R_{2,t_2;\,3,t_1}^{3,t_1+t_2-s;\,2,s}(z)
\end{equation}

\begin{multline}\label{E:rel5}
\check R_{2,t_1;\,3,t_2}^{3,s;\,2,t_1+t_2-s}(z)=\\
\frac{q^{2(s+1)}-1}{q^{2(t_2
+1)}-1}\check R_{2,t_1;\,1,t_2}^{1,s;\,2,t_1+t_2-s}(z)=q^{2(s-t_2)}
\frac{q^{2(t_1+t_2-s+1)}
-1}{q^{2(t_1+1)}-1}\check R_{0,t_1;\,3,t_2}^{3,s;\,0,t_1+t_2-s}(z)=\\
\frac{\bigl(q^{2(
s+1)}-1\bigr)\bigl(q^{2(t_1+t_2-s+1)}-1\bigr)}{\bigl(q^{2(t_1+1)}-1\bigr)
\bigl(q^{2(t_2+1)}-1\bigr)}\check R_{0,t_1;\,1,t_2}^{1,s;\,0,t_1+t_2-s}(z)
\end{multline}

\begin{multline}\label{E:rel6}
\check R_{3,t_1;\,2,t_2}^{2,s;\,3,t_1+t_2-s}(z)=\\
\frac{q^{2(s+1)}-1}{q^{2(t_2
+1)}-1}\check R_{3,t_1;\,0,t_2}^{0,s;\,3,t_1+t_2-s}(z)=q^{2(s-t_2)}
\frac{q^{2(t_1+t_2-s+1)}
-1}{q^{2(t_1+1)}-1}\check R_{1,t_1;\,2,t_2}^{2,s;\,1,t_1+t_2-s}(z)=\\
\frac{\bigl(q^{2(
s+1)}-1\bigr)\bigl(q^{2(t_1+t_2-s+1)}-1\bigr)}{\bigl(q^{2(t_1+1)}-1\bigr)
\bigl(q^{2(t_2+1)}-1\bigr)}\check R_{1,t_1;\,0,t_2}^{0,s;\,1,t_1+t_2-s}(z)
\end{multline}

\begin{multline}\label{E:rel7}
\check R_{2,t_1;\,3,t_2}^{2,s;\,3,t_1+t_2-s}(z)=
\frac{\bigl(q^{2(s+1)}-1\bigr)\bigl(q^{2(t_1+t_2-s+1)}-1\bigr)}{q^{2(t_1+1)}
-1\bigr)\bigl(q^{2(t_2+1)}-1\bigr)}\check R_{0,t_1;\,1,t_2}^{0,s;\,1,t_1+t_2
-s}(z)=\\
q^{s-t_1}z^{t_2-s+1}\frac{q^{2(t_1+t_2-s+1)}-1}{q^{2(t_2+1)}-1}
\check R_{1,t_2;\,2,t_1}^{1,t_1+t_2-s;\,2,s}(z)=\\
q^{t_1-s}z^{t_2-s}\frac{q^{2(s+1)}-1}{q^{2(t_1+1)}-1}\check R_{3,t_2;\,0,t_1}^{
3,t_1+t_2-s;\,0,s}(z)
\end{multline}

\begin{multline}\label{E:rel8}
\check R_{3,t_1;\,2,t_2}^{3,s;\,2,t_1+t_2-s}(z)=
\frac{\bigl(q^{2(s+1)}-1\bigr)\bigl(q^{2(t_1+t_2-s+1)}-1\bigr)}{q^{2(t_1+1)}
-1\bigr)\bigl(q^{2(t_2+1)}-1\bigr)}\check R_{1,t_1;\,0,t_2}^{1,s;\,0,t_1+t_2
-s}(z)=\\
q^{t_1-s}z^{t_2-s-1}\frac{q^{2(s+1)}-1}{q^{2(t_1+1)}-1}
\check R_{2,t_2;\,1,t_1}^{2,t_1+t_2-s;\,1,s}(z)=\\
q^{s-t_1}z^{t_2-s}\frac{q^{2(t_1+t_2-s+1)}-1}{q^{2(t_2+1)}-1}\check R_{0,t_2;\,3,t_1}^{
0,t_1+t_2-s;\,3,s}(z)
\end{multline}

\begin{multline}\label{E:r7}
\check R_{2,t_2;\,0,t_1}^{0,t_1+t_2-s;\,2,s}(z)=z^{s-t_2}
\check R_{0,t_1;\,2,t_2}^{2,s;\,0,
t_1+t_2-s}(z)=\\
\shoveleft{\frac{q^{t_2}z^{s-t_2}}{q^{2(t_1+t_2+1)}-1}
\biggl\{q^{t_1-s+1}\frac{\bigl(
q^{2(t_1+1)}-1\bigr)\bigl(q^{2t_2}-1\bigr)}{q^{2(t_1+t_2-s+1)}-1}\check R_{2,t_1;\,3,
t_2-1}^{3,s-1;\,2,t_1+t_2-s}(z)}\\
+\bigl(q^{2(t_1+1)}-1\bigr)\check R_{3,t_1;\,2,t_2}^{2,s;\,3,t_1+t_2-s}(z)\biggr\}
\end{multline}

\begin{multline}\label{E:r8}
\check R_{3,t_2;\,1,t_1}^{1,t_1+t_2-s;\,3,s}(z)=z^{s-t_2}\check R_{1,t_1;\,3,t_2}^{
3,s;\,1,t_1+t_2-s}(z)=\\
\shoveleft{-\frac{q^{t_2+1}z^{s-t_2}}{q^{2(t_1+t_2+2)}-1}\biggl\{q^{t_1-s}
\frac{\bigl(q^{2(t_1+1)}-1\bigr)\bigl(q^{2(t_2+1)}-1\bigr)}{q^{2(t_1+t_2-s+1)}
-1}\check R_{3,t_1;\,2,t_2}^{2,s;\,3,t_1+t_2-s}(z)}\\
+\bigl(q^{2(t_1+1)}-1\bigr)
\check R_{2,t_1+1;\,3,t_2}^{3,s;\,2,t_1+t_2-s+1}(z)\biggr\}
\end{multline}

\begin{multline}\label{E:r9}
\check R_{2,t_2;\,0,t_1}^{2,t_1+t_2-s;\,0,s}(z)=z^{s-t_2+1}\check R_{0,t_1;\,2,t_2}^{
0,s;\,2,t_1+t_2-s}(z)\\
\shoveleft{=z^{s-t_2+1}\frac{q^{2(t_1+1)}-1}{q^{2(t_1+t_2+1)}-1}\biggl\{q^{t_1+2t_2-s}
\check R_{3,t_1;\,2,t_2}^{3,s;\,2,t_1+t_2-s}}\\
+\frac{q^{2t_2}-1}{q^{2(s+1)}-1}
\check R_{2,t_1;\,3,t_2-1}^{2,s;\,3,t_1+t_1-s-1}(z)\biggr\}
\end{multline}

\begin{equation}\label{E:rel9}
\check R_{3,t_2;\,1,t_1}^{3,t_1+t_2-s;\,1,s}(z)=z^{s-t_2}q^{2(s-t_1)}
\frac{\bigl(q^{2(t_1+1)}-1\bigr)\bigl(q^{2(t_1+t_2-s+1)}-1\bigr)}
{\bigl(q^{2(s+1)}-1\bigr)\bigl(q^{2(t_2+1)}-1\bigr)}
\check R_{3,t_1;\,1,t_2}^{3,s;\,1,t_1+t_2-s}(z)
\end{equation}
\begin{multline}\label{E:r10}
\check R_{3,t_1;\,1,t_2}^{3,s;\,1,t_1+t_2-s}(z)=
\frac{q^{2(t_2+1)}-1}{q^{2(t_1
+t_2+2)}-1}\\
\cdot\biggl\{\check R_{3,t_1;\,2,t_2+1}^{3,s;\,2,t_1+t_2-s+1}(z)+q^{t_1+2t_2-s+2}
\frac{q^{2(t_1+1)}-1}{q^{2(t_1+t_2-s+1)}-1}\check R_{2,t_1;\,3,t_2}^{2,s;\,3,
t_1+t_2-s}(z)\biggr\}
\end{multline}

\begin{equation}\label{E:rel10}
\check R_{1,t_2;\,3,t_1}^{1,t_1+t_2-s;\,3,s}(z)=z^{s-t_2}q^{2(t_1-s)}
\frac{\bigl(q^{2(s+1)}-1\bigr)\bigl(q^{2(t_2+1)}-1\bigr)}
{\bigl(q^{2(t_1+1)}-1\bigr)\bigl(q^{2(t_1+t_2-s+1)}-1\bigr)}
\check R_{1,t_1;\,3,t_2}^{1,s;\,3,t_1+t_2-s}(z)
\end{equation}
\begin{multline}\label{E:r11}
\check R_{1,t_1;\,3,t_2}^{1,s;\,3,t_1+t_2-s}(z)=\\
\shoveleft{\frac{1}{z}\frac{q^{-t_1+2t_2+s+1}}
{q^{2(t_1+t_2+2)}-1}\frac{q^{2(t_1+1)}-1}{q^{2(s+1)}-1}
\cdot\biggl\{\bigl(q^{2(t_1+1)}-1\bigr)\check R_{2,t_1;\,3,t_2}^{2,s;\,3,
t_1+t_2-s}(z)}\\+q^{-t_1-2t_2+s-1}
\bigl(q^{2(t_1+t_2-s+1)}-1\bigr)\check R_{3,t_1;\,2,t_2+1}^{
3,s;\,2,t_1+t_2-s+1}(z)\biggr\}
\end{multline}
The above expressions are valid for $0\leq s\leq t_1+t_2$. In \eqref{E:r9},
the second contribution on the rhs is dropped for $s=t_1+t_2$.
The remaining nonvanishing R-matrix elements are 
\begin{multline}\label{E:rel11}
\check R_{3,t_2;\,1,t_1}^{0,t_1+t_2-s+1;\,2,s}(z)=\\
q^{2(t_2-s)+1}\frac{\bigl(
q^{2(t_1+1)}-1\bigr)\bigl(q^{2(s+1)}-1\bigr)}{\bigl(q^{2(t_2+1)}-1\bigr)\bigl(
q^{2(t_1+t_2-s+2)}-1\bigr)}\check R_{1,t_2;\,3,t_1}^{2,t_1+t_2-s+1;\,0,s}(z)=\\
z^{s-t_2-1}q^{t_2-t_1}\frac{q^{2(t_1+1)}-1}{q^{2(t_2+1)}-1}\check R_{3,t_1;\,1,t_2}^{
2,s;\,0,t_1+t_2-s+1}(z)=\\
z^{s-t_2}q^{t_1+t_2-2s+1}\frac{q^{2(s+1)}-1}{q^{2(t_1
+t_2-s+2)}-1}\check R_{1,t_1;\,3,t_2}^{0,s;\,2,t_1+t_2-s+1}(z)
\end{multline}

\begin{multline}\label{E:r12}
\check R_{1,t_1;\,3,t_2}^{0,s;\,2,t_1+t_2-s+1}(z)=
\frac{q^{2(t_1+1)}-1}{q^{2(t_1
+t_2+2)}-1}\\
\cdot\biggl\{q^{t_1+2t_2-s+2}\check R_{2,t_1+1;\,3,t_2}^{3,s;\,2,t_1+t_2
-s+1}(z)-\frac{q^{2(t_2+1)}-1}{q^{2(s+1)}-1}\check R_{3,t_1;\,2,t_2}^{2,s;\,3,
t_1+t_2-s}(z)\biggr\}
\end{multline}
with $0\leq s\leq t_1+t_2+1$ and        
\begin{multline}\label{E:rel12}
\check R_{2,t_2;\,0,t_1}^{1,t_1+t_2-s;\,3,s-1}(z)=\\
q^{2(t_2-s)+1}\frac{\bigl(q^{2(t_1+1)}-1\bigr)\bigl(q^{2s}-1\bigr)}{\bigl(q^{2(
t_2+1)}-1\bigr)\bigl(q^{2(t_1+t_2-s+1)}-1}\check R_{0,t_2;\,2,t_1}^{3,t_1+t_2-s;\,
1,s-1}(z)=\\
z^{s-t_2}q^{t_1+t_2-2s+1}\frac{q^{2s}-1}{q^{2(t_1+t_2-s+1)}-1}\check R_{0,t_1;\,
2,t_2}^{1,s-1;\,3,t_1+t_2-s}(z)=\\
z^{s-t_2-1}q^{t_2-t_1}\frac{q^{2(t_1+1)}-1}{q^{2(t_2+1)}-1}\check R_{2,t_1;\,0,
t_2}^{3,s-1;\,1,t_1+t_2-s}(z)
\end{multline}

\begin{multline}\label{E:r13}
\check R_{2,t_1;\,0,t_2}^{3,s-1;\,1,t_1+t_2-s}(z)=
\frac{q^{2(t_2+1)}-1}{q^{2(t_1+t_2+1)}-1}\\
\cdot\biggl\{q^{t_1+2t_2-s+2}\frac{q^{2t_1}
-1}{q^{2(t_1+t_2-s+1)}-1}\check R_{3,t_1-1;\,2,t_2}^{2,s-1;\,3,t_1+t_2-s}(z)-
\check R_{2,t_1;\,3,t_2}^{3,s-1;\,2,t_1+t_2-s+1}(z)\biggr\}
\end{multline}
with $t_1+t_2>0$ and $1\leq s\leq t_1+t_2$. In the limit $q\to0$,
well-defined expressions are found for the R-matrix elements with
respect to a different basis $\{\tilde v_{j,t}\}_{0\leq j\leq3,t\in
\mathbb{N}_0}$ given by
\begin{equation}\label{E:newbas}
\tilde v_{j,t}=q^{t(t+\alpha_j)}v_{j,t}
\end{equation}
with $\alpha_{j}=\tfrac{1}{2}+\delta_{j,1}-\delta_{j,2}$.
The matrix elements $\tilde R_{j_1,t_1;\,j_2,t_2}^{j_3,t_3;\,j_4,t_4}(z)$
of the corresponding R-matrix are related to the above matrix elements by
\begin{equation}\label{E:r14}
\tilde R_{j_1,t_1;\,j_2,t_2}^{j_3,t_3;\,j_4,t_4}(z)=q^{t_1(t_1+\alpha_{j_1})
+t_2(t_2+\alpha_{j_2})-t_3(t_3+\alpha_{j_3})-t_4(t_4+\alpha_{j_4})}
\bar R_{j_1,t_1;\,j_2,t_2}^{j_3,t_3;\,j_4,t_4}(z)
\end{equation}
Straightforward analysis of \eqref{E:r1}-\eqref{E:r13} with properties
\eqref{E:rprop1}-\eqref{E:rprop3} yields
\begin{equation}\label{E:r15}
\lim_{q\to0}\tilde R_{j_1,t_1;\,j_2,t_2}^{j_3,t_3;\,j_4,t_4}(z)=
\delta_{j_1,j_4}\delta_{j_2,j_3}\delta_{t_1,t_4}\delta_{t_2,t_3}\,z^{t_2+
\theta_{j_2,j_1}}
\end{equation}

\section{Evaluation of $r(y,z)$}
\label{A:rmode}

The function $r(y,z)$ in \eqref{E:rform1} and \eqref{E:rform2} can be obtained
by collecting all contributions to the term $v^*_{0,0}\otimes v_{2,r}$ on
the rhs of \eqref{E:rform2}:
\begin{multline}\label{E:rmodex1}
r(y,z)=
\frac{1}{\tilde P_0\bigl(y;q^{-2r-1},q^{-1},q^{-1}\vert q^{-2}\bigr)}\cdot\\
\sum_{t=0}^{\infty}q^{t(2r+1)}\frac{(q^2;q^2)_{t+1}(q^2;q^2)_r}{(q^2;q^2)_{t+r}(q^2;
q^2)_1}\tilde P_t\bigl(y;q^{2r+1},
q,q\vert q^2\bigr)\,\check R_{2,t+r;\,0,t^*}^{0,0^*;\,2,r}(z)
\end{multline}
A choice of $\tilde P_0\bigl(y;q^{-2r-1},q^{-1},q^{-1}\vert q^{-2}\bigr)$ sets
the normalisation of the polynomials $\tilde P_t\bigl(y;q^{-2r-1},q^{-1},q^{-1}
\vert q^{-2}\bigr)$. Its value may depend on $r$.
Comparison of the intertwining conditions yields a relation between the matrix elements
$\check R_{2,t_1;\,0,t_2^*}^{0,t_3^*;\,2,t_1-t_2+t_3}(z)$ or $\check R_{0,t^*_2;\,
2,t_1}^{2,t_1-t_2+t_3;\,0,t^*_3}(z)$ 
and matrix elements of $\check R_{VV}(q^2z^{-1})$:
\begin{equation}\label{E:relrel}
\begin{split}
\check R_{2,t_1;\,0,t_2^*}^{0,t_3^*,\,2,t_1-t_2+t_3}(z)&=q^{t_3-t_2}\frac{[t_3+1]}{[
t_2+1]}\,\check R_{2,t_3;\,2,t_1}^{2,t_1-t_2+t_3;\,2,t_2}(q^2z^{-1})\\
\check R_{0,t_2^*;\,2,t_1}^{2,t_1-t_2+t_3;\,0,t^*_3}(z)&=q^{t_2-t_3}\frac{[t_3+1]}{[
t_2+1]}\,\check R_{2,t_1;\,2,t_3}^{2,t_2;\,2,t_1-t_2+t_3}(q^2z^{-1})
\end{split}
\end{equation}
Here an overall normalisation is fixed by equating the R-matrix elements on the
lhs with those on rhs for $t_1=t_2=t_3=0$.
Using \eqref{E:relrel} with \eqref{E:rprop1}, \eqref{E:r1}, the sum in \eqref{E:rmodex1}
can be written
\begin{equation}\label{E:rmodex2}
q^r\frac{1-q^2z^{-1}}{1-q^{2(r+1)}z^{-1}}\cdot\sum_{t=0}^{\infty}\tilde P_t\bigl(y;
q^{2r+1},q,q\vert q^2\bigr)\frac{\bigl(q^{2(r+1)};q^2\bigr)_{t}}{\bigl(
q^{2(r+2)}z^{-1};q^2\bigr)_{t}}\bigl(qz^{-1}\bigr)^t
\end{equation}
The infinite sum in \eqref{E:rmodex2} is related to a generating function for
Askey-Wilson polynomials derived in \cite{groene}.
In base $q^2$, eq. 2.10 in \cite{groene} reads
\begin{multline}\label{E:genfun}
\sum_{t=0}^{\infty}\frac{(abcd;q^2)_{2t}p_t(\cos\theta;a,b,c,d\vert q^2)}{(q^2,
ab,ac,bc,abcdq^{2(t-1)};q^2)_t}\frac{(f/g,abc/f;q^2)_t}{(abcdg/f,fd;q^2)_t}
g^t=\\
\frac{(abcd,dg,abcdge^{i\theta}/f,fe^{i\theta};q^2)_{\infty}}{(abcdg/f,df,abce^{i
\theta},ge^{i\theta};q^2)_{\infty}}{}
_8W_7\bigl(abce^{i\theta}/q^2;ae^{i\theta},be^{i\theta},ce^{i\theta},f/g,abc/f;q^2,
ge^{-i\theta}\bigr)
\end{multline}
for $\vert g\vert<1$. For the definitions of the
very-well-poised basic hypergeometric series
$_8W_7(a_1;a_4,a_5,a_6,a_7,a_8;q^2,\alpha)$, the reader is referred to \cite{gr}.
The orthonormal continuous dual $q^2$-Hahn polynomials $\tilde P_t\bigl(\cos
\theta;a,c,d\vert q^2\bigr)$ are related to the Askey-Wilson polynomials by
\begin{equation}
\tilde P_t(\cos\theta;a,c,d\vert q^2)=\frac{p_t(\cos\theta;a,0,c,d
\vert q^2)}{(q^2,ac;q^2)_t}
\end{equation}
For $b=0$, equation \eqref{E:genfun} simplifies:
\begin{multline}\label{E:rmodex3}
\sum_{t=0}^{\infty}\tilde P_t(\cos\theta;a,c,d\vert q^2)\frac{(f/g;q^2)_t}{(
df;q^2)_t}g^t=\\
\frac{(dg,fe^{i\theta};q^2)_{\infty}}{(df,ge^{i\theta};q^2)_{\infty}}
{}_3\phi_2\bigl(ae^{i\theta},ce^{i\theta},f/g;ac,fe^{i\theta};q^2,ge^{-i\theta}
\bigr),\qquad\;\;\vert g\vert<1
\end{multline}
With $a=q^{2r+1}$, $c=d=q$, $f=q^{2r+3}z^{-1}$ and $g=qz^{-1}$,
the rhs of \eqref{E:rmodex3} coincides with the sum \eqref{E:rmodex2}
while the rhs becomes
\begin{equation}\label{E:rmodex5}
\frac{(qg,fe^{i\theta};q^2)_{\infty}}{(qf,ge^{i\theta};q^2)_{\infty}}
{}_2\phi_1\bigl(q^{2r+1}e^{i\theta},qe^{i\theta};q^{2r+3}z^{-1}e^{i\theta};
q^2,qz^{-1}e^{-i\theta}\bigr)
\end{equation}
The basic hypergeometric series in \eqref{E:rmodex5} has the form $_2\phi_1
(A,B;C;q^2,C/AB)$. If $\vert C/AB\vert<1$, the latter can be summed
according to eq. 1.5.1 in \cite{gr}:
\begin{equation}\label{E:gauss}
_2\phi_1(A,B;C;q^2,C/AB)=\frac{(C/A,C/B;q^2)_{\infty}}{(C,C/AB;q^2)_{\infty}}
\end{equation}
Use of \eqref{E:gauss} in \eqref{E:rmodex5} and collecting terms yields
\begin{multline}\label{E:rmodex6}
r(y,z)=\frac{1}{\tilde P_0\bigl(y;q^{-2r-1},q^{-1},q^{-1}\vert q^{-2}\bigr)}
q^r\bigl(1-q^2z^{-1}\bigr)\frac{(q^2z^{-1},q^2z^{-1};q^2)_{\infty}}{(qz^{-1}
e^{i\theta},qz^{-1}e^{-i\theta};q^2)_{\infty}}\,,\\
\vert qz^{-1}\vert<1
\end{multline}
Setting  $\tilde P_0\bigl(y;q^{-2r-1},q^{-1},q^{-1}\vert q^{-2}\bigr)=q^r$
leads to \eqref{E:rmode1}. Equations \eqref{E:rmodex3} and \eqref{E:gauss}
allow to handle
the contributions to the terms $v_{0,s}^*\otimes v_{2,r+s}$ on the rhs of
\eqref{E:rform2} or $v_{0,r+s}^*\otimes v_{2,s}$ on the rhs of \eqref{E:rform1}
 in a similar way. 
The condition $qz^{-1}<1$ proves sufficient to reproduce the result \eqref{E:rmode1}
in all cases.


\begin{thebibliography}{9}

\bibitem{gade1}{R.M. Gade, \emph{An integrable $U_q\bigl(\widehat{gl}(2\vert2)
\bigr)_1$-Model: Corner Transfer Matrices and Young Skew Diagrams},
Nucl.Phys. B \textbf{694}~(2004) 354-404}


\bibitem{jm}{M. Jimbo, T. Miwa, Algebraic Analysis of Solvable Lattice Models,
Regional Conference Series in Mathematics, AMS, Vol. \textbf{85}~(1995)}



\bibitem{dav}{B. Davies, O. Foda, M. Jimbo, T. Miwa and A. Nakayashiki, Diagonalization
of the $XXZ$ Hamiltonian by vertex operators, Comm. Math. Phys, 151 (1993) 89-153}



\bibitem{idz}{M. Idzumi, K. Iohara, M. Jimbo, T. Miwa, T. Nakashima and T.
Tokihiro, \emph{Quantum affine symmetry in vertex models}, Int. J. Mod.
Phys. A \textbf{8}~(1993) 1479-1511}


\bibitem{naka}{A. Nakayashiki and Y. Yamada, \emph{Crystallizing the spinon basis},
Comm. Math. Phys. \textbf{178}~(1996) 179}


\bibitem{aff}{S.-J. Kang, M. Kashiwara, K. Misra, T. Miwa, T. Nakashima and
A. Nakayashiki, \emph{Affine crystals and vertex models}, Int. J. Mod. Phys.
A \textbf{7}(Suppl. 1A)~(1992) 449-484}

\bibitem{kash}{M. Kashiwara, \emph{On crystal bases of the $q$-analogue of
universal enveloping algebras}, Duke Math. J. \textbf{63}~(1991) 465-516}

\bibitem{kac1}{V.G. Kac and M. Wakimoto, \emph{Integrable highest weight modules over affine
superalgebras and Appell's function}, Comm. Math. Phys. \textbf{215}~(2001), 631-682}

\bibitem{jimbo}{M. Jimbo, \emph{A $q$-Analogue of $U\bigl(gl(N+1)\bigr)$, Hecke
Algebra, and the Yang Baxter Equation}, Lett. Math. Phys.
\textbf{10}~(1985) 63-68}

\bibitem{zhang1}{R.B. Zhang, \emph{Universal $L$-operator and invariants of the quantum
supergroup $U_q\bigl(gl(m\vert n)\bigr)$}, J. Math. Phys. \textbf{33}~(1992) 1970-1979}


\bibitem{gade2}{R.M. Gade, \emph{On the quantum affine superalgebra $U_q\bigl(
\widehat{gl}(2\vert2)\bigr)$ at level one}, Nucl. Phys. B \textbf{500}~(1997)
547-564}

\bibitem{kuniba}{A. Kirillov, A. Kuniba and T. Nakanishi, \emph{Skew Young Diagram
method in spectral decomposition of integrable lattice models}, Comm. Math.
Phys. \textbf{185}~(1997) 441-465}



\bibitem{zhang2}{W.-L. Yang and Y.-Z. Zhang, \emph{Vertex operators of $U_q\bigl(
\widehat{gl}(N\vert N)\bigr)$ and highest weight representations of
$U_q\bigl(\widehat{gl}(2\vert2)\bigr)$}, J. Math. Phys. \textbf{41}~(2000)
2460-2481}


\bibitem{gade3}{R.M. Gade, \emph{Universal $R$-matrix and graded Hopf algebra
structure of $U_q\bigl(\widehat{gl}(2\vert2)\bigr)$}, J. Phys. A
\textbf{31}~(1998) 4909-4925}




\bibitem{kato}{A. Kato, Y.-H. Quano and J. Shiraishi, \emph{Free Boson Representation
of $q$-Vertex Operators and their Correlation Functions}, Comm. Math. Phys.
\textbf{157}~(1993) 119-137}

\bibitem{konno}{H. Konno, \emph{Free Field Representation of Quantum Affine
Algebra $U_q(\widehat{sl}_2)$ and Form Factors in Higher Spin XXZ
Model}, Nucl. Phys. B \textbf{432}~(1994) 457-486}

\bibitem{koyama}{Y. Koyama, \emph{Staggered Polarization of Vertex Models with
$U_q\bigl(\widehat{sl}(n)\bigr)$-Symmetry}, Comm. Math. Phys.
\textbf{164}~(1994) 277-291}
\bibitem{idzumi}{M. Idzumi, K. Iohara, M. Jimbo, T. Miwa, T. Nakashima and
T. Tokihiro, \emph{Quantum affine symmetry in vertex models}, Int. J. Mod.
Phys. A \textbf{8}~(1993) 1479-1511}

\bibitem{askwi}{R. Askey and J. Wilson, \emph{Some basic hypergeometric orthogonal
polynomials}, Mem. Amer. Math. Soc. \textbf{319}~(1985)}

\bibitem{gr}{G. Gasper and M. Rahman, \emph{Basic Hypergeometric Series}, 2nd. ed.,
Cambridge University Press, Cambridge, 2004}

\bibitem{kalnins}{E.G. Kalnins and W. Miller Jr., \emph{A note on tensor products
of q-algebra representations and orthogonal polynomials}, J. Comp. Appl. Math.
\textbf{68}~(1996), 197-207}

\bibitem{groene}{W. Groenevelt, \emph{Bilinear summation formulas from quantum
algebra representations}, Ramanujan J. \textbf{8}~(2004), 383-416}

\bibitem{burban}{I.M. Burban and A.U. Klimyk, \emph{Representations of the
quantum algebra $U_q\bigl(\mathfrak{su}(1,1)\bigr)$}, J. Phys. A: Math. Gen.
\textbf{26}~(1993), 2139-2151}



\end{thebibliography}
\end{document}